\documentclass[final,5p,times,twocolumn]{elsarticle}

\usepackage{amssymb}

\usepackage{lineno}

\usepackage{graphicx}
\usepackage{multirow}
\usepackage{float}
\usepackage{amsmath}
\usepackage{gensymb}
\biboptions{numbers,sort&compress}

\usepackage[labelformat=simple, justification=centering]{subcaption}

\journal{Nuclear Instruments and Methods A}

\begin{document}

\begin{frontmatter}



\title{Model-based Design Evaluation of a Compact, High-Efficiency Neutron Scatter Camera}


\author[1]{Kyle Weinfurther}
\author[1]{John Mattingly}
\author[2]{Erik Brubaker}
\author[2]{John Steele}
\address[1]{North Carolina State University, Raleigh NC 27695}
\address[2]{Sandia National Laboratories, Livermore CA 94550}

\begin{abstract}
This paper presents the model-based design and evaluation of an instrument that estimates incident neutron direction using the kinematics of  neutron scattering by hydrogen-1 nuclei in an organic scintillator.  The instrument design uses a single, nearly contiguous volume of organic scintillator that is internally subdivided only as necessary to create optically isolated pillars, i.e., long, narrow parallelepipeds of organic scintillator.  Scintillation light emitted in a given pillar is confined to that pillar by a combination of total internal reflection and a specular reflector applied to the four sides of the pillar transverse to its long axis.  The scintillation light is collected at each end of the pillar using a photodetector, e.g., a microchannel plate photomultiplier (MCP-PM) or a silicon photomultiplier (SiPM).  In this optically segmented design, the (x, y) position of scintillation light emission (where the x and y coordinates are transverse to the long axis of the pillars) is estimated as the pillar's (x, y) position in the scintillator "block", and the z-position (the position along the pillar's long axis) is estimated from the amplitude and relative timing of the signals produced by the photodetectors at each end of the pillar.  The neutron's incident direction and energy is estimated from the (x, y, z)-positions of two sequential neutron-proton scattering interactions in the scintillator block using elastic scatter kinematics. For proton recoils greater than 1 MeV, we show that the (x, y, z)-position of neutron-proton scattering can be estimated with $<$ 1 cm root-mean-squared [RMS] error and the proton recoil energy can be estimated with $<$ 50 keV RMS error by fitting the photodetectors' response time history to models of optical photon transport within the scintillator pillars. Finally, we evaluate several alternative designs of this proposed single-volume scatter camera made of pillars of plastic scintillator (SVSC-PiPS), studying the effect of pillar dimensions, scintillator material (EJ-204, EJ-232Q and stilbene), and photodetector (MCP-PM vs.\ SiPM) response vs.\ time.  We demonstrate that the most precise estimates of incident neutron direction and energy can be obtained using a combination of scintillator material with high luminosity and a photodetector with a narrow impulse response.  Specifically, we conclude that an SVSC-PiPS constructed using EJ-204 (a high luminosity plastic scintillator) and an MCP-PM will produce the most precise estimates of incident neutron direction and energy.

\end{abstract}

\begin{keyword}
Neutron \sep Scatter Camera \sep Scintillator  \sep Photodetector \sep Imaging



\end{keyword}

\end{frontmatter}


\graphicspath{ {.}}

\section{Introduction}
\label{intro}

This paper will start with a brief overview of the operational principles of neutron scatter cameras. Then we will discuss  previous neutron scatter camera designs and the advantages of our proposed design. Next, we will discuss the reconstruction method used to estimate scintillation position, proton recoil energy and scintillation time. Subsequently, we will detail the factors used in our expected response, such as the scintillator, photodetector and pillar impulse response. We will then detail the methodology of simulating observed responses to compare to expected responses. Using the observed and expected responses, we will show the precision of scintillation position, proton recoil energy and scintillation time estimation. This paper will conclude by illustrating back-projected image resolution of a compact, high-efficiency neutron scatter camera.

\subsection{Operational Principles}

\begin{figure}[h!!]
\centering
  \includegraphics[width=\columnwidth]{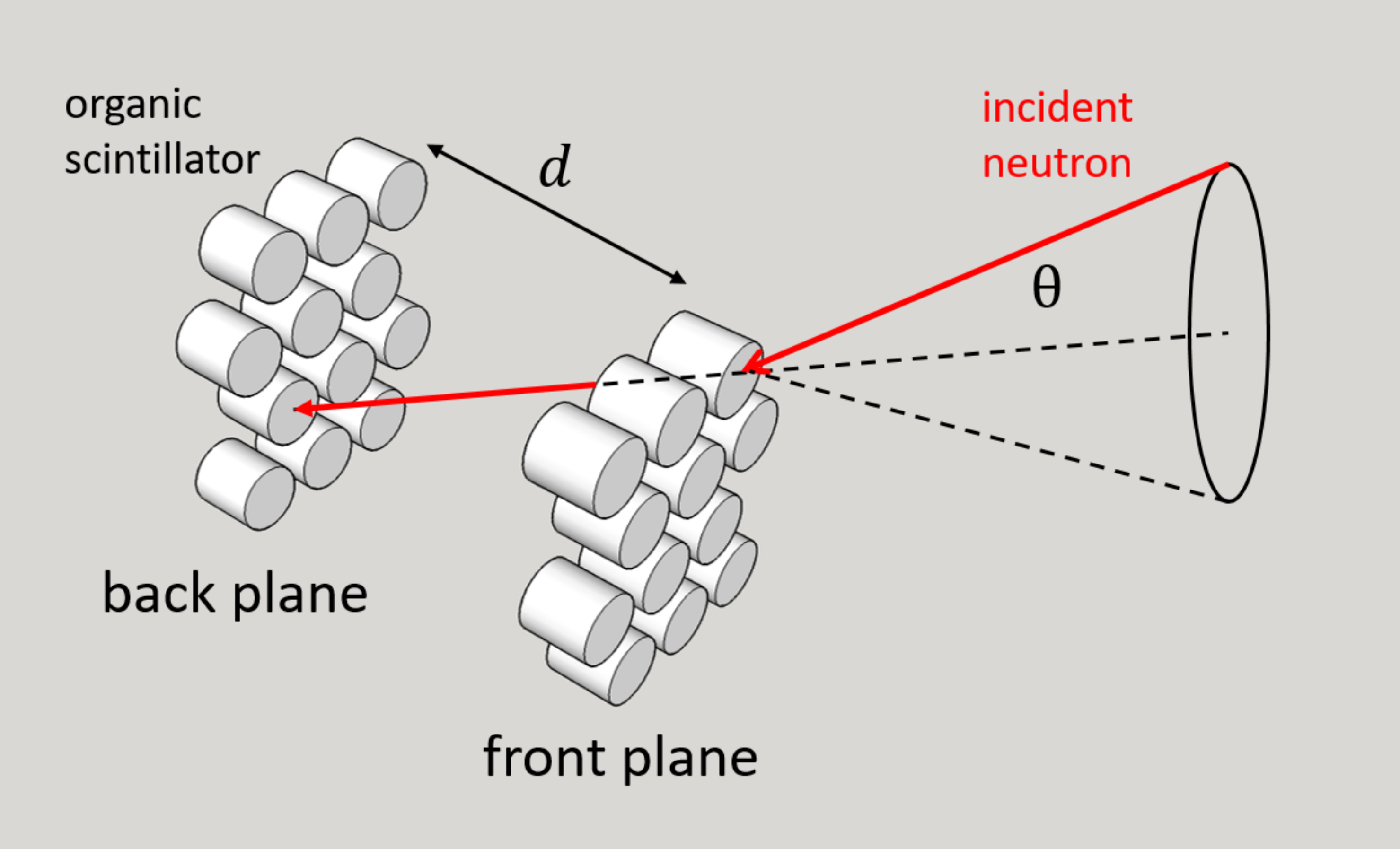}
  \caption{Neutron Scatter camera operational principles. Incident neutron cone angles are created using the time of flight of the neutron between the two planes and the scintillation brightness from neutron elastic scatter in the front plane}
  \label{fig:NSC_cartoon}
\end{figure}

Neutron scatter imagers estimate incident neutron direction using the kinematics of neutron scattering off hydrogen-1 in an organic scintillator. A neutron must scatter twice within the active volume of the detector to estimate incident neutron direction. The location of both scatters, the time between scatters and the energy deposited in the first scatter may be used to describe a cone of possible incident neutron directions whose axis is aligned with the vector connecting the two scatters. A graphic illustrating this method is shown in Figure \ref{fig:NSC_cartoon}.

\subsection{Theory}
\label{theory}
Neutrons transfer some or all of their energy to the detection medium during elastic scatter\cite{knoll_textbook}. The amount of energy transferred is dependent upon recoil nucleus mass and the angle of scatter given by

\begin{align}
\label{eqn:scatterNeutronEnergy}
{E_{n^{'}}} &= \dfrac{(1+\alpha) + (1-\alpha) \cos \theta_{CM}}{2} E_n
\end{align}

\noindent
where $\alpha = \left(\dfrac{A-1}{A+1}\right)^2$, $A$ is the mass of the target nucleus, $E_n$ is the incident energy of the neutron, $E_{n^{'}}$ is the energy of the neutron after elastic scatter and $\theta_{CM}$ is the scatter angle in the center-of-mass (CM) coordinate frame. Scattering by light nuclei is isotropic in the center-of-mass coordinate frame. Neutrons transfer all of their energy to a proton ($A = 1$) in a head on collision when $\theta_{CM} = 180^\circ$. The center-of-mass scattering angle is related to angle in the lab frame by

\begin{align}
\label{CM->Lab}
\tan \theta_L &= \dfrac{\sin \theta_{CM}}{\dfrac{1}{A}+\cos \theta_{CM}}
\end{align}

\noindent
where $\theta_L$ is the lab frame scattering angle. We can simplify Equation \ref{eqn:scatterNeutronEnergy} to calculate the scattered neutron angle in the lab frame. If we use only scatters on hydrogen ($A = 1$), $\alpha = 0$  and Equation \ref{eqn:scatterNeutronEnergy} simplifies to

\begin{align}
\label{eqn:scatE over initE}
E_p &= E_n cos^{2} \theta_L
\end{align}

\noindent
where  $\theta_L$ is the angle between the incident neutron and the scattered neutron directions in the lab frame. We cannot directly measure the incident energy of the neutron; however, we can reconstruct it by summing the proton recoil energy in the first scatter and the energy of the scattered neutron shown in Equation \ref{eqn:test} 

\begin{align}
\label{eqn:test}
{E_n} &= E_p + E_{n^{'}} 
\end{align}

\noindent
where $E_p$ is the proton recoil energy. We can estimate the scattered neutron energy $E_{n^{'}}$ using neutron time of flight between two scatters using 

\begin{align}
\label{eqn:KE}
{E_n}^{'} &= \frac{1}{2}m_n v^{2} = \frac{1}{2}m_n\left(\dfrac{d}{\Delta t}\right)^{2} 
\end{align}

\noindent
where $m_n$ is the mass of a neutron, $v$ is the speed of the scattered neutron, $d$ is the distance between the first and second neutron elastic scatter, and $\Delta t$ is the time between the two scatters. A second neutron scatter must occur, otherwise scattered neutron energy cannot be estimated and cone back-projection is impossible.

For the proposed design, we estimated the proton recoil energy using the intensity of light emitted in the first neutron elastic scatter. Concurrently, we estimated the scintillation position along the pillar using photodetectors' signal amplitude and relative timing. A neutron must interact in different pillars to reconstruct the scintillation position for both scatter events. We have all the information needed to back-project a cone of incident neutron angles using Equation \ref{eqn:coneAngle}.

\begin{align}
\label{eqn:coneAngle}
\theta_L &= tan^{-1} \left(\sqrt{\frac{E_p}{E_n^{'}}} \right)
\end{align}

\subsection{Imager Design}
\begin{figure}[h!!]
\centering
  \includegraphics[width=\columnwidth]{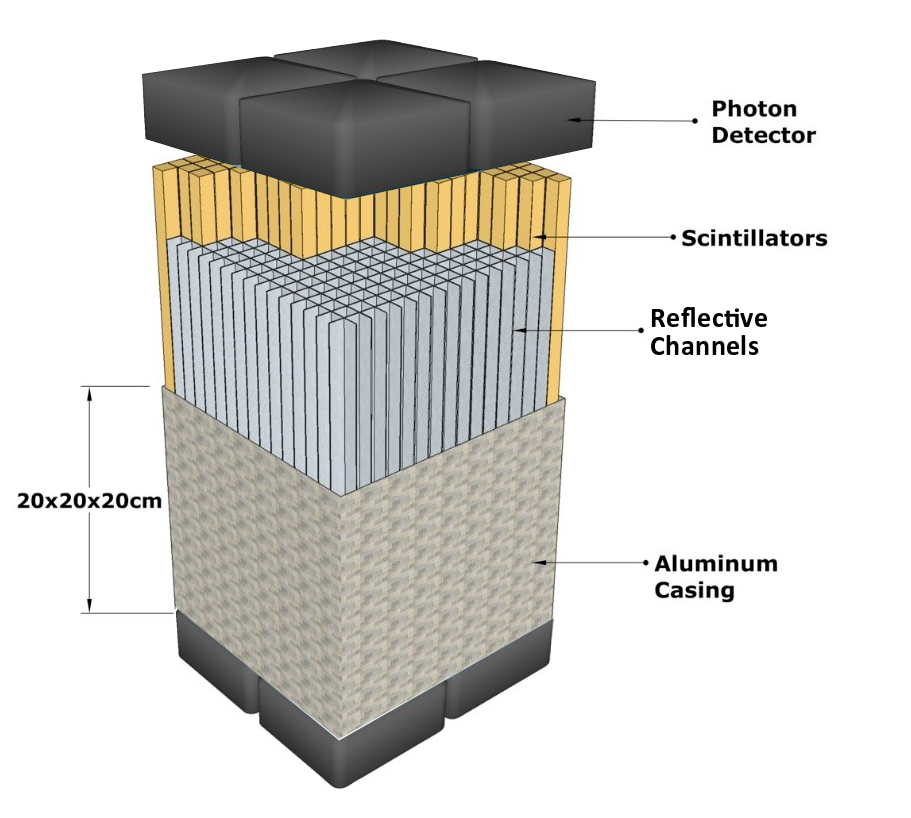}
  \caption{Handheld neutron scatter camera made of optically segmented pillars of scintillator and reflective channels}
  \label{fig:svscPips}
\end{figure}

In this paper, we propose a high efficiency imager design to localize neutron emitting material. The instrument design uses a semi-contiguous volume of organic scintillator that is subdivided into optically isolated pillars. Each scintillator pillar is surrounded by a 1 mm air gap; this air gap allows scintillation light to undergo total internal reflection (TIR) to increase light collection efficiency. Each channel is lined with a reflective film/paint to reflect photons escaping back into the pillar. Orthogonal to each pillar and attached to opposing ends are photodetectors. Opposing photodetectors enable two waveforms to be recorded for each neutron elastic scatter. The device consists of a scintillating volume of approximately 8000 cm$^3$.  A depiction of a single volume scatter camera made of pillars of plastic scintillator (SVSC-PiPS) is shown in Figure \ref{fig:svscPips}.

\section{Previous Work}
\label{previousWorks}

One of the first known neutron scatter telescopes was proposed in 1968\cite{Grannan1972}. The device was used to measure solar neutrons and the resulting albedo neutrons from earth at an altitude of 120,000 feet \cite{Preszler1972, Preszler1974, Zych1975}. The device consisted of two large planes of mineral oil liquid scintillator, optically separated into four cells per plane, located 1 m apart. In 1986, a double scatter fast neutron detector measured the neutron energy spectrum from a thermonuclear plasma source by utilizing the time of flight of neutrons between successive scatters\cite{Walker1986}. In 1992, researchers measured the energy spectrum (15 MeV to 100 MeV) and direction of neutrons discharged from solar flares using neutron double scatter events\cite{Ryan1992, Ryan1993}.

In 2005, the Fast Neutron Imaging Telescope (FNIT) was constructed to measure solar flare neutrons from 2 MeV to 20 MeV\cite{Moser2005, Bravar2006, Bravar2006a}. Neutrons of these energies are highly attenuated by Earth's atmosphere;  therefore, the FNIT was designed to operate on spacecraft where weight, size, and power constraints are stringent. The original design of the FNIT consisted of eight stacked BC-404 (12 cm x 12 cm x 1.5 cm) detector layers. Wavelength shifting optical fibers were used to transport scintillation light from the detection layers to multianode photomultiplier tubes. The FNIT also had applications in homeland security measuring fission neutrons. Later iterations of the FNIT and a scaled down version - Solar Neutron Experiment (SONNE) - used liquid scintillator filled cylindrical pillars arranged in a radially symmetric pattern\cite{Bravar2009, Woolf2009}.

\subsection{Neutron Scatter Camera}

\begin{figure}[h!!]
\centering
  \includegraphics[width=\columnwidth]{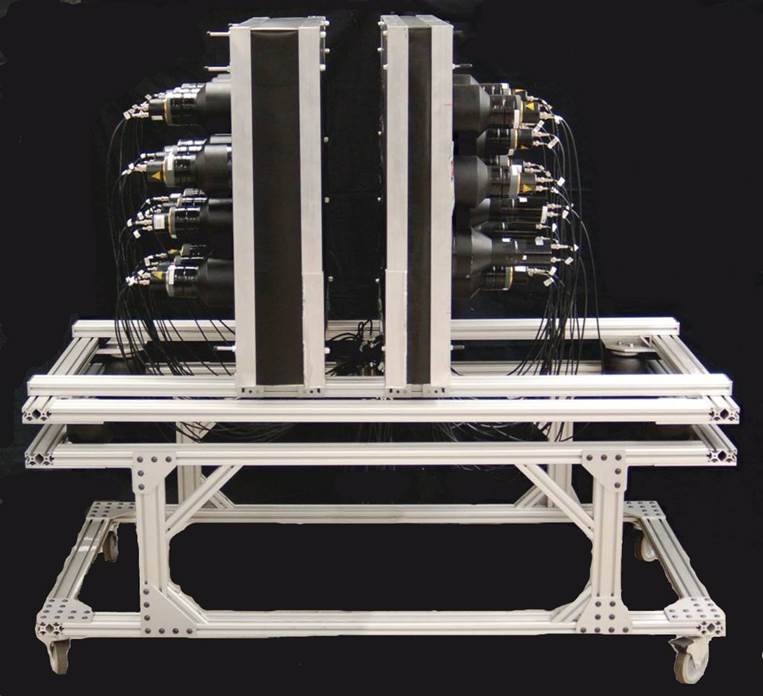}
  \caption{Neutron scatter camera developed at Sandia National Laboratories. The device consists of two planes of EJ-309 organic liquid scintillator. Incident neutron cone angles created using the time of flight of the neutron between the two planes and the scintillation brightness from neutron elastic scatter in the front plane. This version of the neutron scatter camera uses 12 detectors in each plane.}
  \label{fig:NSC_photo}
\end{figure}

Sandia National Laboratories developed a neutron scatter camera (NSC) in 2007 \cite{Marleau2007, Mascarenhas2008a}. A picture of this device is shown in Figure \ref{fig:NSC_photo}. Section \ref{theory} discusses operating methodology of a scatter camera. The front plane consists of sixteen 12.7 cm diameter x 5.1 cm thick EJ-309 organic liquid scintillators. The back plane uses 12.7 cm thick detectors to increase efficiency for double scatter. They used a smaller detector thickness in the front plane to reduce the probability of multiple scatters in the first detector.

The distance between the two detection planes is variable; 40 cm is the standard operating distance between detection planes for the NSC. Reducing detection plane distance increases detection efficiency at the cost of decreased angular resolution. Similarly, increasing detection plane separation reduces detection efficiency, but increases angular resolution. The neutron scatter camera has a angular resolution of 10$^{\circ}$ for normal operating conditions \cite{NSC_patent1}. A neutron must scatter once in the front plane and once in the back plane of the NSC to estimate source location. This requirement reduces the efficiency of the device. Similarly, A hybrid neutron scatter camera and Compton scatter camera (dual particle imager) is being developed at the University of Michigan which is able to image both neutrons and gamma rays simultaneously \cite{Poitrasson-Riviere2014}.

\subsection{Neutron Scatter Camera Efficiency}

\begin{figure}[h!!]
\centering
  \includegraphics[width=\columnwidth]{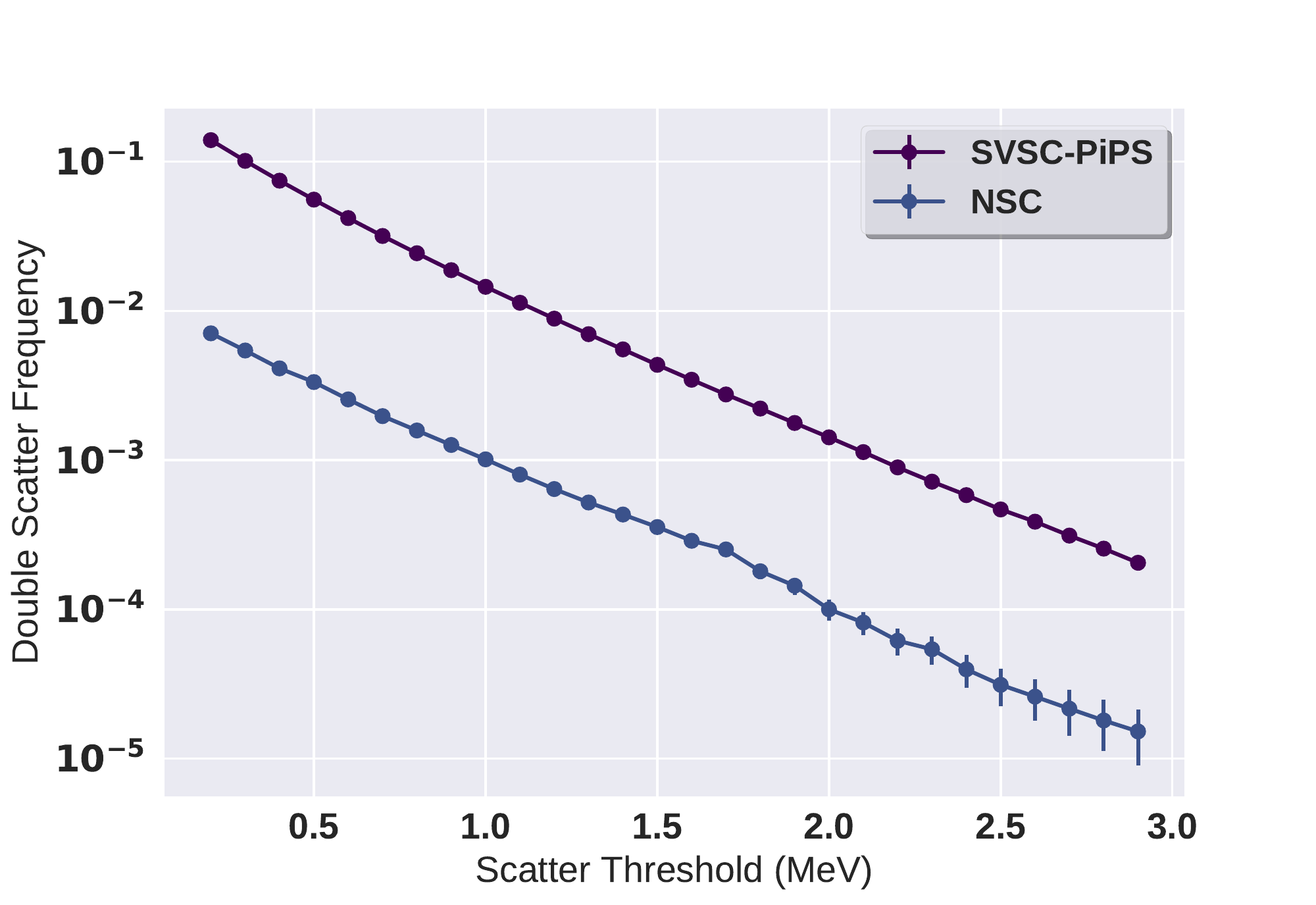}
  \caption{Event detection threshold comparison for a NSC and SVSC-PiPS for a pencil beam of Cf-252 fission spectrum neutrons aimed at the center of the respective device.}
  \label{fig:EffComparison}
\end{figure}

We compared the efficiency of the NSC and an SVSC-PiPS device using the simulation tool MCNPX-PoliMi \cite{Pelowitz2011}. MCNPX-PoliMi is a Monte Carlo particle transport code particularly suited to neutron problems. The code records state variables (e.g.\ position, direction, incident energy, and energy deposited) for particles that interact in user-specified detector cells. We calculated the efficiency of the two devices as a function of the proton recoil energy threshold shown in Figure \ref{fig:EffComparison}. The simulation consisted of a pencil beam of fission spectrum neutrons aimed at the center of both devices. We simulated a NSC at the normal operating distance of 40 cm between the two planes. For both simulations, we required the first two interactions to be elastic scatters off hydrogen-1. Neutron scatters off carbon are difficult to detect due to the low scintillation light emission. At a maximum, a carbon nucleus recoils with approximately 28\% of the incident neutron energy, and quenching of the recoil carbon nucleus causes much less scintillation light to be emitted compared to a recoil proton.

The lists of scattering events produced by the simulations required cuts to take into account operation of the respective systems' photodetection and signal acquisition schemes. For the SVSC-PiPS, the first two hydrogen-1 scatters could not occur in the same pillar; for the NSC, a neutron must scatter on hydrogen-1 for both the first scatter in the front plane and the second scatter in the back plane.

Figure \ref{fig:EffComparison} shows that the single volume design has significantly better efficiency (about an order of magnitude better). At a 200 keV threshold for both neutron scatters, the NSC has about 1\% efficiency. Using a contiguous volume of scintillator results in an order of magnitude efficiency increase. Even though the active volume of a SVSC-PiPS is less than that of the NSC, its efficiency to neutrons incident on the face of the detector is higher because the double scatter probability is higher in a contiguous volume than it is in multiple separated volumes.

\section{Scintillation Position and Proton Recoil Energy Reconstruction Method}
\label{reconstructionMethod}
Both interaction positions, interaction times and the proton recoil energy from the first scatter are needed to construct a cone of possible incident neutron directions using neutron kinematics. Photodetector responses on opposing ends of the pillar allow us to estimate scintillation location  along the elongated dimension of the pillar (z-direction) and proton recoil energy by using the the amplitude and relative timing of the signals produced by the photodetectors. Interaction position estimation in the (x,y) coordinates is not possible within a single pillar. When we back-project incident neutron cones, we set the estimated scintillation position in (x,y) at the center of the pillar illuminated with scintillation light. We will show simulation results of how we estimated position of interaction, proton recoil energy and time between scatters using neutron time of flight (TOF). We show results for various pillar lengths, pillar widths, scintillators and photodetectors for multiple proton recoil energies. Throughout the paper, pillar length will refer to pillar extent in the z-direction and pillar width will refer to the size in (x,y) dimension. Pillars have an aspect ratio of one in the (x,y) plane for simplicity, i.e.\ they have the same width in both the x and y direction.

There are four main steps to the simulation. 

\begin{enumerate}
	\item Simulate scintillation photon emission
	\begin{enumerate}
		\item Randomly select the number of scintillation photons from a Poisson distribution with a mean equal to proton recoil energy times the luminosity
		\item Scintillation photon timing is randomly sampled from the scintillator time response function
	\end{enumerate}
	
	\item Simulate light collection efficiency
	\begin{enumerate}
		\item Randomly select the number and time of arrival of scintillation photons on the photodetectors from the optical pillar response function
	\end{enumerate}
	
	\item Model charge carrier production in the photodetectors
	\begin{enumerate}
		\item Randomly sample the number of charge carriers generated at the photodetectors based on the MCP-PM quantum efficiency and SiPM photon detection efficiency
		\item Randomly sample the time of arrival of charge carriers from the photodetector transit time distribution
		\item Smear the charge carrier arrival time histogram by convolving with the photodetector impulse response
	\end{enumerate}
	
	\item Apply Poisson maximum likelihood estimation to fit the randomly sampled charge carrier arrival time history with the expected arrival time history where scintillation position and proton recoil energy are unknown variables
	
\end{enumerate}


The histograms shown in Figure \ref{fig:rmsErrorPosHist} and Figure \ref{fig:rmsErrorEnergyHist} result from analyzing 10,000 charge carrier waveforms simulated using the preceding steps for each photodetector/scintillator combination. The uncertainties in scintillation position and proton recoil energy are the sample standard deviations for those 10,000 simulations which account for most of the relevant effects, excluding variations in scintillation photon wavelength specific to each scintillator, wavelength dependent charge carrier production in the photodetectors, and anisotropy in luminosity and optical photon transport in stilbene (where the average behavior of stilbene was modeled).

We tabulated nominal responses using Equation \ref{eqn:fullResponse}

\begin{align}
\label{eqn:fullResponse}
R_{nom}(t) = R_{scint}(t) * R_{pil}(t) * R_{TTS}(t) * R_{imp}(t)
\end{align}

\begin{itemize}
  \item $R_{scint}(t)$ is the scintillator time response
  \item $R_{pil}(t)$ is the pillar response function
  \item $R_{TTS}(t)$ is the transit time spread of the photodetector
  \item $R_{imp}(t)$ is the photodetector impulse response
  \item $R_{nom}(t)$ is the nominal response resulting from a convolution of the preceding factors
\end{itemize}

\noindent
where $*$ is the convolution operator. Pillar response functions estimate the temporal spread of photon arrival times as photons traverse the pillar; they also provide collection efficiency as a function of scintillation position along the pillar.

The following sections will illustrate how we modeled each factor to produce nominal and observed responses for all alternative designs.

\section{Scintillator and Photodetector Impulse Response}
This section illustrates how we modeled the scintillator time response ($R_{scint}$) and photodetector impulse response ($R_{imp}$) to produce observed and nominal responses. We chose scintillators and photodetectors with different attributes  to determine desirable characteristics for our proposed imager design. 

\subsection{Scintillator Impulse Response}

\begin{figure}[h!!]
\centering
  \includegraphics[width=\columnwidth]{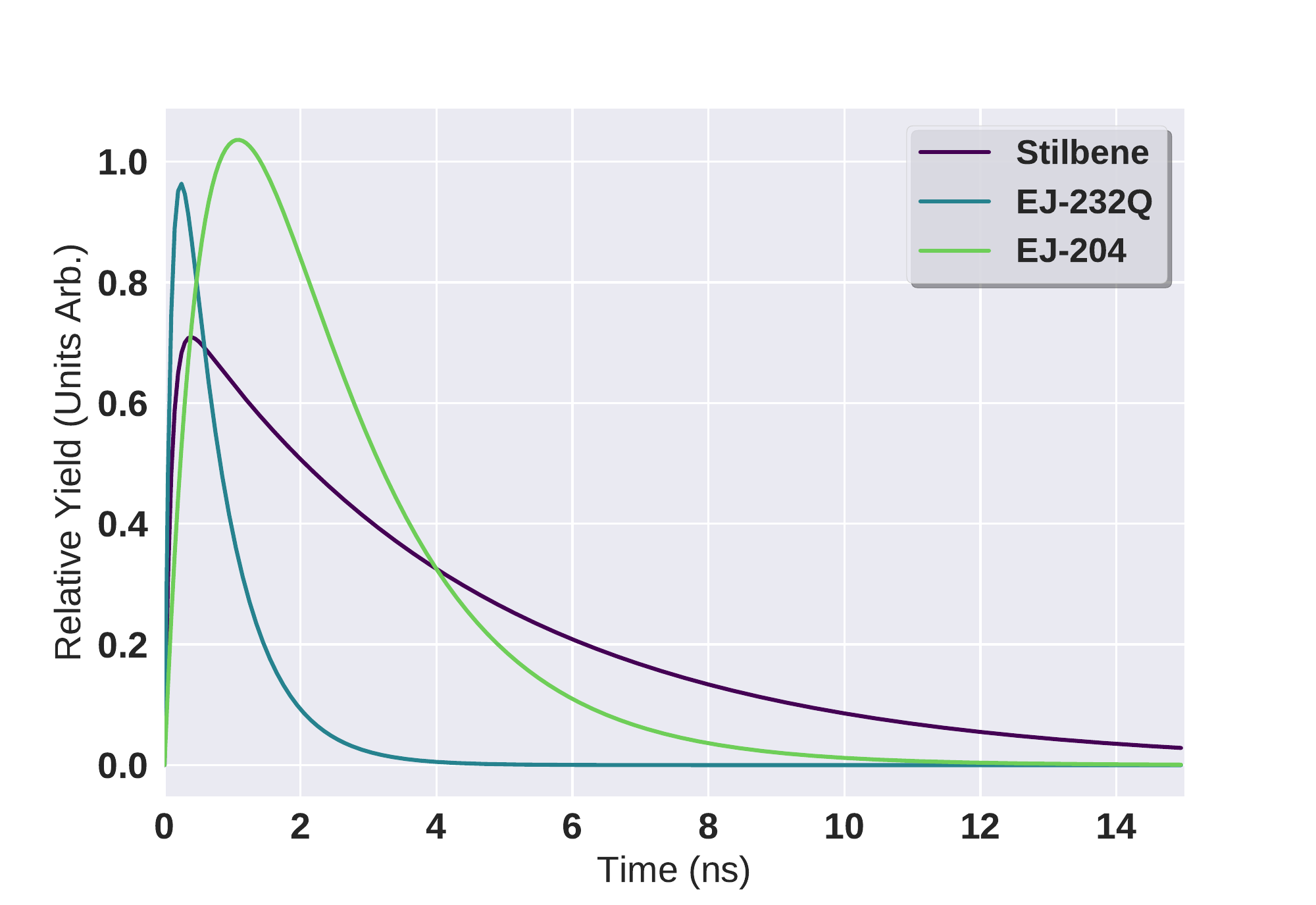}
  \caption{Relative time response of scintillators. Area under each curve is normalized to the luminosity of the scintillator.}
  \label{fig:scintillators}
\end{figure}

We investigated three different organic scintillators for use in an SVSC-PiPS device. We simulated a general purpose plastic scintillator EJ-204, a quenched plastic scintillator EJ-232Q (0.5\% Benzophenone), and a solution grown crystalline organic scintillator stilbene. We chose these three scintillators primarily due to the inherent differences in their time response. The time response of each scintillator can be seen in Figure \ref{fig:scintillators}. We modeled the scintillator time response using Equation \ref{eqn:scintTiming}

\begin{align}
\label{eqn:scintTiming}
R_{scint}(t) = \exp\left(\dfrac{-t}{\tau_f}\right) - \exp\left(\dfrac{-t}{\tau_r}\right)
\end{align}

\noindent
where $t$ is the time elapsed after the neutron and hydrogen-1 scatter, $\tau_f$ is the fall time, and $\tau_r$ is the rise time. Each waveform is normalized to the number of scintillation photons produced per 1 MeVee energy deposition. EJ-204 and stilbene have similar brightnesses at approximately 68\% anthracene luminosity\cite{ej204, Zaitseva2015}. The quenched plastic has a much lower brightness at 19\% anthracene luminosity\cite{ej232q}. There are also significant differences in the time responses of the scintillators evident in Table \ref{table:scintProps}. The light attenuation length for both plastic scintillators are known to be 8 cm for EJ-232Q \cite{knoll_textbook} and 160 cm for EJ-204 \cite{ej204}. The attenuation length for stilbene is unknown; we used an assumed length of 100 cm based on \cite{Zaitseva2015} (this is most likely an overestimate).

\begin{table*}[t]
\centering
\caption{Modeled scintillator properties \cite{ej232q, ej204, Kuchnir1968}}
\label{table:scintProps}
\begin{tabular}{lccccc}
Scintillator & EJ-204 & EJ-232Q  & Stilbene\\
Rise Time (ns) & 0.7  & 0.105 & 0.1\\
Fall Time (ns) & 1.8  & 0.7 & 4.5\\
Luminosity (\% Anthracene) & 68 & 19 & 67 \\
Wavelength of Maximum Emission (nm) & 408 & 370 & 382\\
Light Attenuation Length (cm) & 160 & 8 & 100\\

\end{tabular}
\end{table*}

Even though stilbene has the largest fall time of all three scintillators, it does have a 0.1 ns rise time comperable to EJ-232Q. Fast rise time should enable more precise estimatation of TOF between the two interactions. It may also enable more precise scintillation position reconstruction from the early arrival of scintillation photons. Scintillation light creation in stilbene is dependent on proton recoil direction in relation to the crystal structure\cite{Schuster2016}. This effect was ignored for the purposes of this paper as it is currently not fully characterized. The index of refraction of stilbene is dependent on the direction of the propagating optical photon as well. For the Geant4 simulation, we set the refractive index of stilbene to 1.757, the average of the all indices given in \cite{Stilbene1910}. Both EJ-204 and EJ-232Q have a refractive index of 1.58. Scintillators produce scintillation light over a range of wavelengths; the most intense wavelengths are 390 nm, 408 nm, 370 nm for stilbene, EJ-204 and EJ-232Q respectively. We fixed scintillation light wavelength at 400 nm for the simulation for all scintillators.

\subsection{Photodetector Impulse Response}
We simulated the responses of two different photodetectors of interest. We required photodetectors to have a small form factor and the ability to be used in arrays or to have pixelated anodes. The two photodetectors we chose to model were a micro-channel plate photomultiplier (MCP-PM) and a silicon photomultipliers (SiPM). These photodetectors use different mechanisms for detection. The MCP-PM was modeled after the Planacon XP85012 manufactured by Photonis\cite{planacon_xp85012} and the SiPM was modeled after the J-Series SiPMs manufactured by SensL\cite{sipm_Jseries}.

MCPs work in a fashion similar to standard photomultiplier tubes. First, scintillation photons impinge upon the photocathode. These photons dislodge electrons via the photoelectric effect. Electrons dislodged this way are referred to as photoelectrons (PE). This occurs in the photocathode of a Planacon at about 20\% quantum efficiency (QE) for photons with wavelengths between 325 and 425 nm\cite{planacon_xp85012}. These photoelectrons travel through the MCP's microchannels where each collision with the microchannel walls emits more electrons. The electrons continue to multiply until they reach anode pads at the back of the multiplying region. The anode pads are much larger than individual channels and their shape mostly creates the apparent ``pixels'' for the photodetector. 

SiPMs operate in a different manner. Scintillation photons directly interact in the silicon via photoelectric effect to create electron-hole pairs which propagate to the anode and cathode and create a signal. The photon detection efficiency (PDE)  is higher for SiPMs. PDE is defined as the statistical probability that a photon interacts with a microcell in the SiPM to produce an avalanche\cite{SensL2011}. SiPMs can reach up to 50\% PDE depending on photon wavelength and the overvoltage applied to the photodetector\cite{sipm_Jseries}. Increasing the overvoltage increases the PDE, but it also increases crosstalk between microcells of the SiPMs. Crosstalk can reach up to 22\% for SensL J-series SiPMs. For this analysis, we assumed an operating overvoltage of 2.5 volts which corresponds to a photon detection efficiency of 35\% and a cross talk percentage of 5\%. Cross talk and overvoltage were not modeled for these simulations.

\begin{figure}[h!!]
\centering
  \includegraphics[width=\columnwidth]{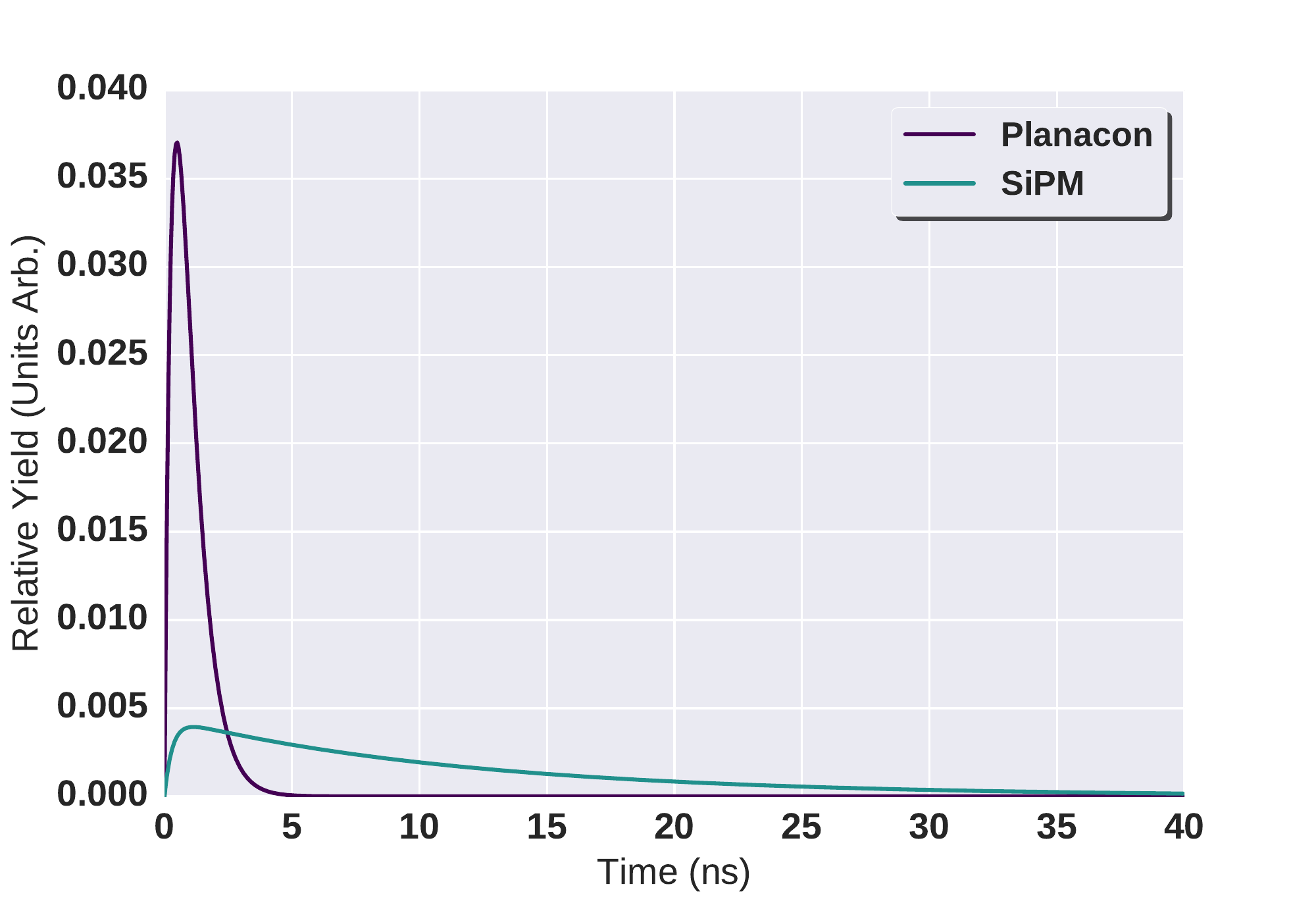}
  \caption{Photodetector impulse response to charge carriers. The area under each waveform is normalized to unity. The Planacon has a much more compact time response compared to the silicon photomultiplier.}
  \label{fig:photodetectors}
\end{figure}

One major difference between the two photodetectors is the impulse response; see Figure \ref{fig:photodetectors}. The Planacon has an impulse response with a rise time and fall time of 0.6 ns and 0.4 ns respectively\cite{Grigoryev2016a}; the SensL J-series SiPMs have in impulse response that includes a fast rise time of 0.3 ns and a slower fall time of 12 ns. We modeled the photodetector impulse response using Equation \ref{eqn:scintTiming}. The longer the photodetector impulse response, the larger the spread of charge carrier collection over time. Recent advancement in SiPM technology now provides a ``fast output'' which uses approximately 2\% of the energy deposited for the signal. This creates a faster rise and fall time of the output signal. However, with only 2\% of the signal being collected in the fast output, it will be subject to large fluctuations from small charge collections. So, we did not model the  fast output; all results are based on the standard SiPM output impulse response. 
In these simulations, no electron multiplication was modeled because it does not contribute to the random fluctuations in charge collection. Quantum efficiency and photon detection efficiency depend on scintillation photon wavelength. We fixed the QE for the MCP-PM at 20\% and the PDE at 35\% for an SiPM for all scintillation photons. 

\section{Models of Pillar Impulse Response}
In this section we examine propagation of scintillation photons through the pillar. We used the light transport module in Geant4\cite{Agostinelli2003a, Allison2006} to simulate the movement of scintillation photons through the pillar to the photodetectors. In this section, we also tabulated pillar response functions ($R_{pil}$). This section concludes with a comparison of different pillar designs.

\subsection{Optical Photon Transport Simulations}
\begin{figure}[h!!]
\centering
  \includegraphics[width=\columnwidth]{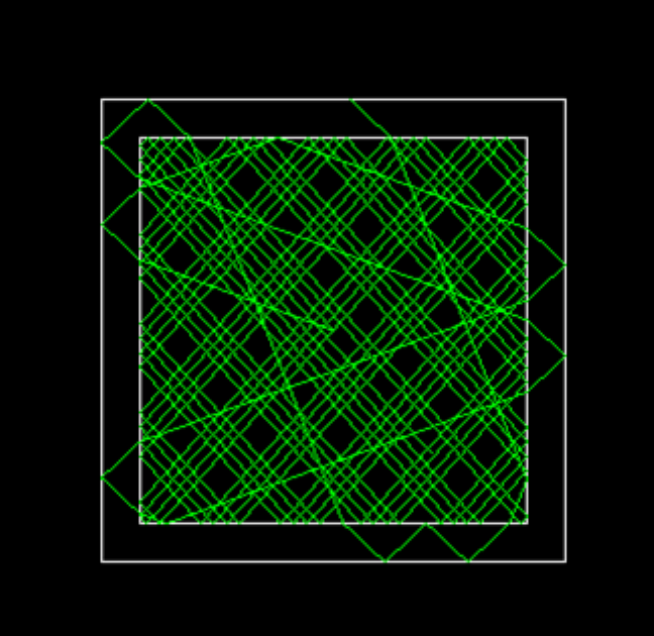}
  \caption{Two simulated photon trajectories through the pillar. One photon undergoes total internal reflection resulting in the ``grid'' pattern. The other photon does not undergo TIR and leaves the scintillator into the air and glances off the reflector lining the housing walls to ultimately be absorbed by the reflective film at the top of the figure.}
  \label{fig:geant1}
\end{figure}

We simulated light emission and transmission throughout the scintillator pillar to the photodetectors at each end of the pillar. Two example photon trajectories are shown from Geant4 in Figure \ref{fig:geant1}. The geometry of the simulation included a pillar of scintillator surrounded by a 1 mm air gap to allow total internal reflection. Outside the air gap, we lined the pillar housing walls with a reflector to contain light that did not undergo total internal reflection in the scintillator pillar. 

In Figure \ref{fig:geant1}, both photons originate at the center of the pillar. One photon undergoes total internal reflection (shown in the ``grid'' pattern). Photons emitted from scintillation with little to no velocity in the z-direction exhibit long transit times and a large number of reflections. Total internal reflection caused photons to travel large distances (over 100 cm) before arriving at the photodetector. TIR can occur when light is traveling from a material with a higher index of refraction to a material with lower index of refraction. Light will undergo TIR if the angle between the incident direction of the light and the normal to the interface is larger than the critical angle, where the critical angle is defined by Snell's law:

\begin{align}
\label{eqn:SnellsLaw}
\theta_c &= \arcsin \left( \frac{n_2}{n_1} \right)
\end{align}

\noindent
In Equation \ref{eqn:SnellsLaw}, $\theta_c$ is the critical angle, $n_1$ is the index of refraction of the material where the light originated and $n_2$ is the index of refraction of the adjacent material. As an example, we simulated an EJ-204 pillar ($n_1=1.58$) surrounded by air ($n_2=1$). Using Equation \ref{eqn:SnellsLaw}, the critical angle is 39$^\circ$ fo an EJ-204/air interface. Therefore, photons emitted at an angle above 39$^\circ$ relative to the scintillator surface normal will undergo TIR. Total internal reflection does not occur when light is transmitting to a medium with a higher index of refraction.

Figure \ref{fig:geant1} shows that the TIR photon reflected off the walls of the scintillator approximately 100 times and traveled a distance over 100 cm before reaching the photodetector. The downside to TIR is that photon travel distances can be large depending on the angle of emission. The loss mechanism for TIR photons is absorption in the scintillator. 

The second photon path originating at the center of Figure \ref{fig:geant1} does not experience TIR. It Fresnel refracts out of the pillar into the air and reflects off the enhanced specular reflector (ESR) film lining the housing wall, then it refracts back into the scintillator pillar. Subsequently, the photon Fresnel reflects to stay inside the pillar. This path continues until photon is absorbed in the ESR film at the top of the figure.

\begin{figure}[h!!]
\centering
  \includegraphics[width=\columnwidth]{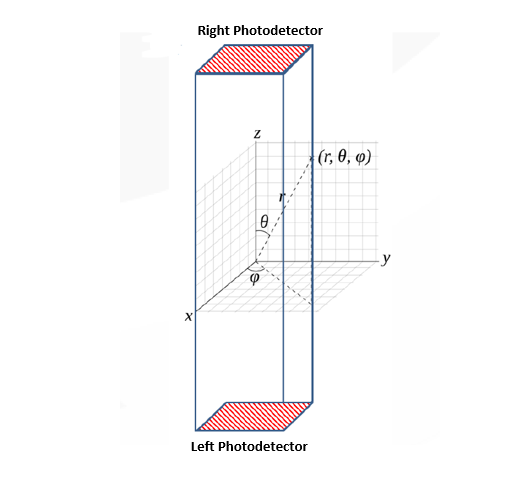}
  \caption{Polar ($\theta$) and azimuth ($\phi$) angles with respect to the pillar geometry.}
  \label{fig:angleDefinition}
\end{figure}

We tabulated Geant4 photon travel times, number of reflections and photon emission angles for a 1 cm x 1 cm x 20 cm EJ-204 plastic scintillator surrounded by 1 mm of air with ESR film lining the pillar housing walls. The simulation isotropically emitted $10^7$ photons at a distance of 6 cm from the photodetector. For these plots, a reflection is defined as the photon altering its direction for any reason; this includes a reflection from TIR, a reflection from the ESR film, and a Fresnel reflection or refraction. Polar and azimuth angle definitions are shown in Figure \ref{fig:angleDefinition}.

\begin{figure}[h!!]
\centering
  \includegraphics[width=\columnwidth]{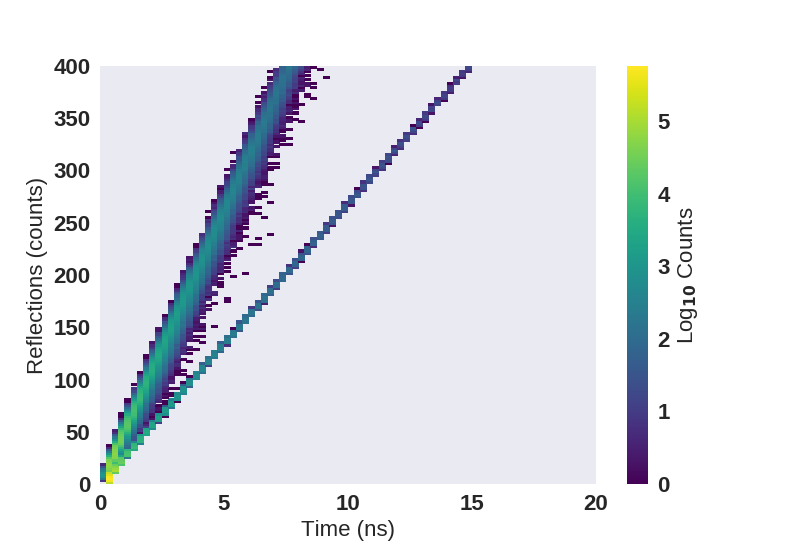}
  \caption{A comparison of the time of arrival of photons at the photodetector vs. the number of reflections endured. 10$^7$ optical photons were simulated at a 6 cm distance from the photodetector. Reflections are defined as any change in photon direction. The thicker line on the left are photons that escaped the pillar to reflect off the ESR film lining the pillar housing walls. The thinner line on the right are photons that undergo TIR.}
  \label{fig:timeVsRef}
\end{figure}

We will first examine the number of reflections and time of arrival of photons at the photodetector shown in Figure \ref{fig:timeVsRef}. We simulated all photons starting at time zero. The 2-D histogram displays two distinct bands of photon arrival times. Most photons arrive between 0 and 2 ns. 
The thicker band of counts at shorter arrival times are from photons that escape the pillar and travel in the air gap surrounding the pillar. Due to Snell's law, when a photon escapes the pillar, the light will angle towards the photodetector from refraction. Light travels faster in air than in EJ-204 due to the lower refractive index of air. This contributes to shorter arrival times for photons escaping the pillar. Photons escaping the pillar can travel a tortuous path though the air gap, off the reflector and back into the scintillator multiple times. Transmitting through the pillar increases travel time before arriving at the photodetector. The amount of lag is indicative of the number of times the photon passes through the scintillator. Photons that undergo total internal reflection create the small thin band at longer arrival times in the figure. Emission angles close to $\theta = 90^\circ$ and $\phi = 45^\circ$ will allow photons to undergo TIR, but they do not travel quickly towards the photodetector with each reflection. These emission angles lead to arrival times of $\geq$10 ns due to numerous reflections. 

\begin{figure}[h!!]
\centering
  \includegraphics[width=\columnwidth]{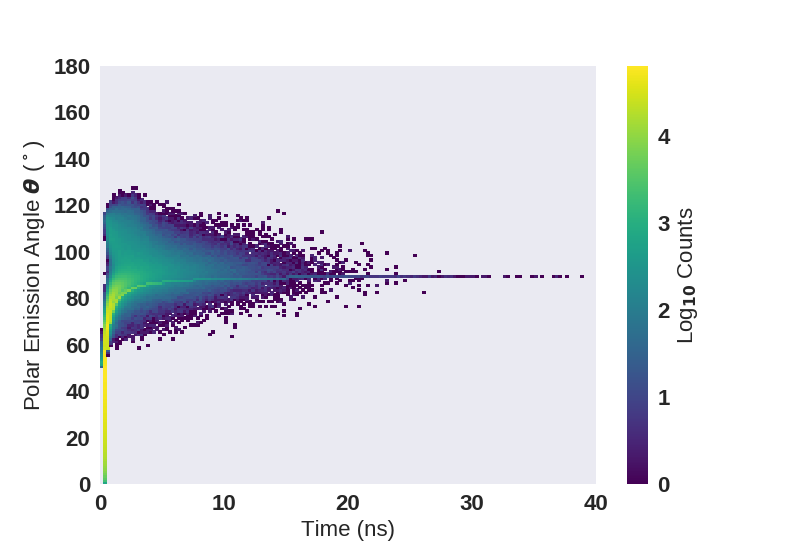}
  \caption{Arrival time of photons at the photocathode as a function of their polar emission angle $\theta$. EJ-204 has a critical angle $\theta_c \approx 39^\circ$ from the surface normal at $\theta = 51^\circ$. Only TIR occurs for emission angles below this angle. TIR can still occur afterwards, but it is dependent on the azimuth angle of emission $\phi$. A majority of the photons undergo TIR, which is clearly visible as a distinct line starting at short time and reaching an asymptote near 90$^\circ$.}
  \label{fig:timeVsTheta}
\end{figure}

We tabulated the time of arrival of photons and binned them with respect to initial polar emission angle $\theta$ shown in Figure \ref{fig:timeVsTheta}. There is a distinct line starting at low emission angles which exhibits an asymptote near 90$^\circ$. This results from photons that undergo TIR. As the polar emission angle $\theta$ nears 90$^\circ$, photon velocity in the z-direction (the direction along the long axis of the pillar) approaches zero, leading to a very large transit time before arriving at the photodetector. Outside the TIR line are photons that exit the pillar and reflect in the ESR film. Emission angles over 90$^\circ$ are initially directed towards the opposing photodetector. There is a possibility that the photons can change direction over multiple reflections and move in the opposite z-direction from slight deviations in the reflected photons exit angle. The critical angle of 39$^\circ$ is measured with respect to the normal of the scintillator surface. This is why we do not see photons escaping the pillar below $\theta \approx 51^\circ = 90^\circ - 39^\circ$.

\begin{figure}[h!!]
\centering
  \includegraphics[width=\columnwidth]{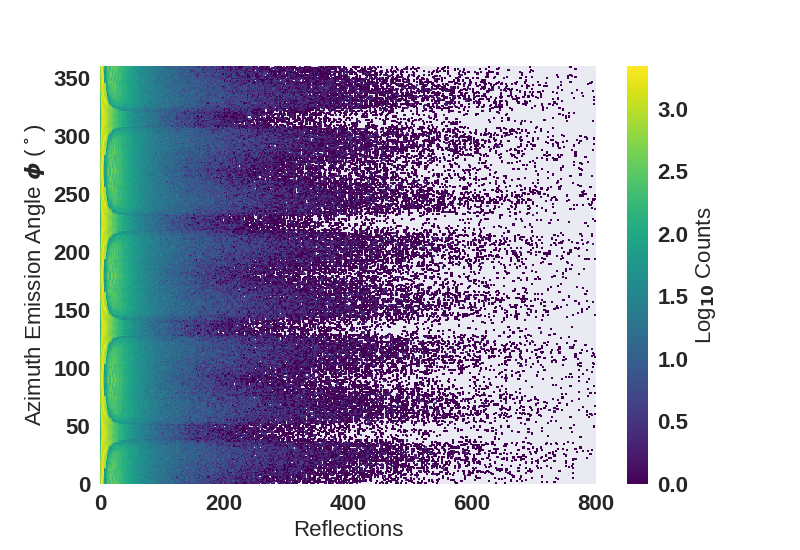}
  \caption{Number of reflections that scintillation photons underwent to arrive at the photodetector as a function of azimuth angle $\phi$. TIR occurs near azimuth angles $\phi = 45^\circ, 135^\circ, 225^\circ, 315^\circ$.}
  \label{fig:refVsPhi}
\end{figure}

We histogramed the azimuth emission angle $\phi$ against the number of reflections of photons as shown in Figure \ref{fig:refVsPhi}. At angles around $45^\circ$, $135^\circ$, $225^\circ$ and $315^\circ$, only TIR  occurs, regardless of polar emission angle $\theta$. However, the emission angle $\theta$ dictates how direct the path of the photon's travel is towards the photodetector. Most angles allow for quick travel to the photodetector indicated by the larger population of counts at a low number of reflections. Photons that escape the pillar produce counts outside the ``comb-like'' pattern seen in the figure. The tortuous path traversed by escaping photons increases the number of reflections before arriving at the photodetector.

\begin{figure}[h!!]
\centering
  \includegraphics[width=\columnwidth]{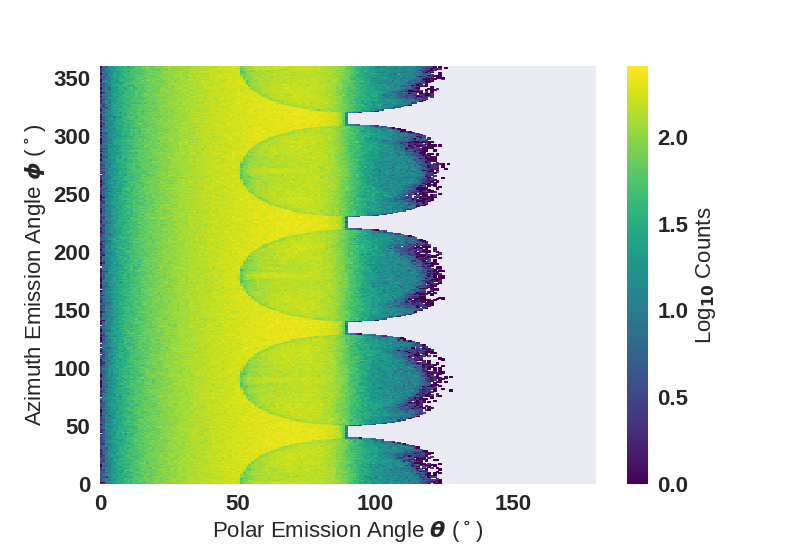}
  \caption{Histogram of the number of photons that arrive at the photodetector versus the polar and azimuth emission angle $\theta$ and $\phi$. TIR occurs near azimuth angles $\phi = 45^\circ, 135^\circ, 225^\circ, 315^\circ$. The ``ovals'' at polar emission angles $\theta \approx 50^\circ$ through $\theta \approx 130^\circ$ are scintillation photons escaping the pillar}
  \label{fig:thetaVsPhi}
\end{figure}

The last plot, in Figure \ref{fig:thetaVsPhi}, shows the number of photons arriving at the photodetector as a function of both the polar and azimuth emission angles $\theta$ and $\phi$, respectively. At small values of $\theta$, photons arrive at the photodetector at short times with a low number of reflections because those photons travel directly to the photodetector. Again, as $\theta$ increases towards $90^\circ$, the photon travels parallel to the pillar normal with little to no velocity component towards the photodetector. At these steep angles, photons are most likely to exit the pillar into the air and strike the ESR film reflector. The ``oval'' patterns seen in the figure are photons escaping the scintillator. This plot demonstrates what we saw in Figure \ref{fig:timeVsRef}, where photons initially emitted  at $\theta > 90^\circ$ can reverse direction in the ESR film and interact in the opposite photodetector. 

\subsection{Pillar Impulse Response}

\begin{figure}[h!!]
\centering
  \includegraphics[width=\columnwidth]{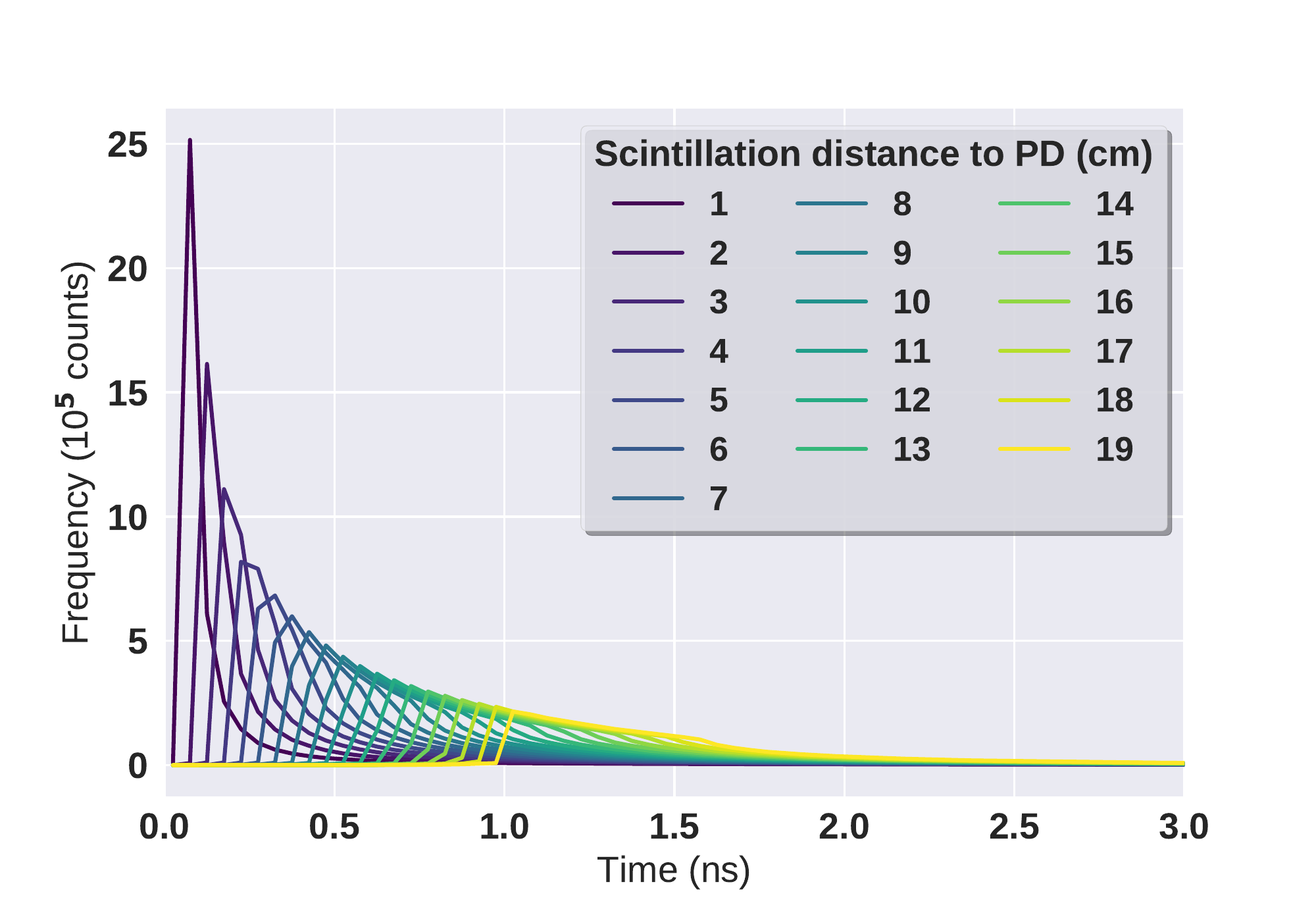}
  \caption{Time spread of scintillation photons as they arrive at the photodetector.}
  \label{fig:channelTimeSpread}
\end{figure}

As scintillation photons propagate down the pillar, the pillar's geometry induces a temporal spread of arrival times at the photodetector.  To estimate the distribution of arrival times, we simulated $10^7$ scintillation photons emitted isotropically at 0.5 cm increments along the pillar length. All scintillation photons were emitted at the center of the pillar in the (x,y) plane. We tabulated the arrival time of each photon that hit the photodetector and binned them to produce a pillar time spread histogram as a function of distance from the photodetector. These ``pillar response functions'' are shown in Figure \ref{fig:channelTimeSpread}. 

Photons produce a more compactly-supported pillar response function the closer they originate near the photodetector.  At 1 cm away, pillar geometry has a small effect and most incident photons arrive within 0.25 ns of scintillation. However, as the position moves farther away, we observe a much larger spread of arrival times up to 1 ns at the farthest distance. Note that the area under each curve decreases as the distance increases. This is due to self-absorption of scintillation light in addition to losses in the ESR reflector lining the housing walls. 

\begin{figure}[h!!]
\centering
  \includegraphics[width=\columnwidth]{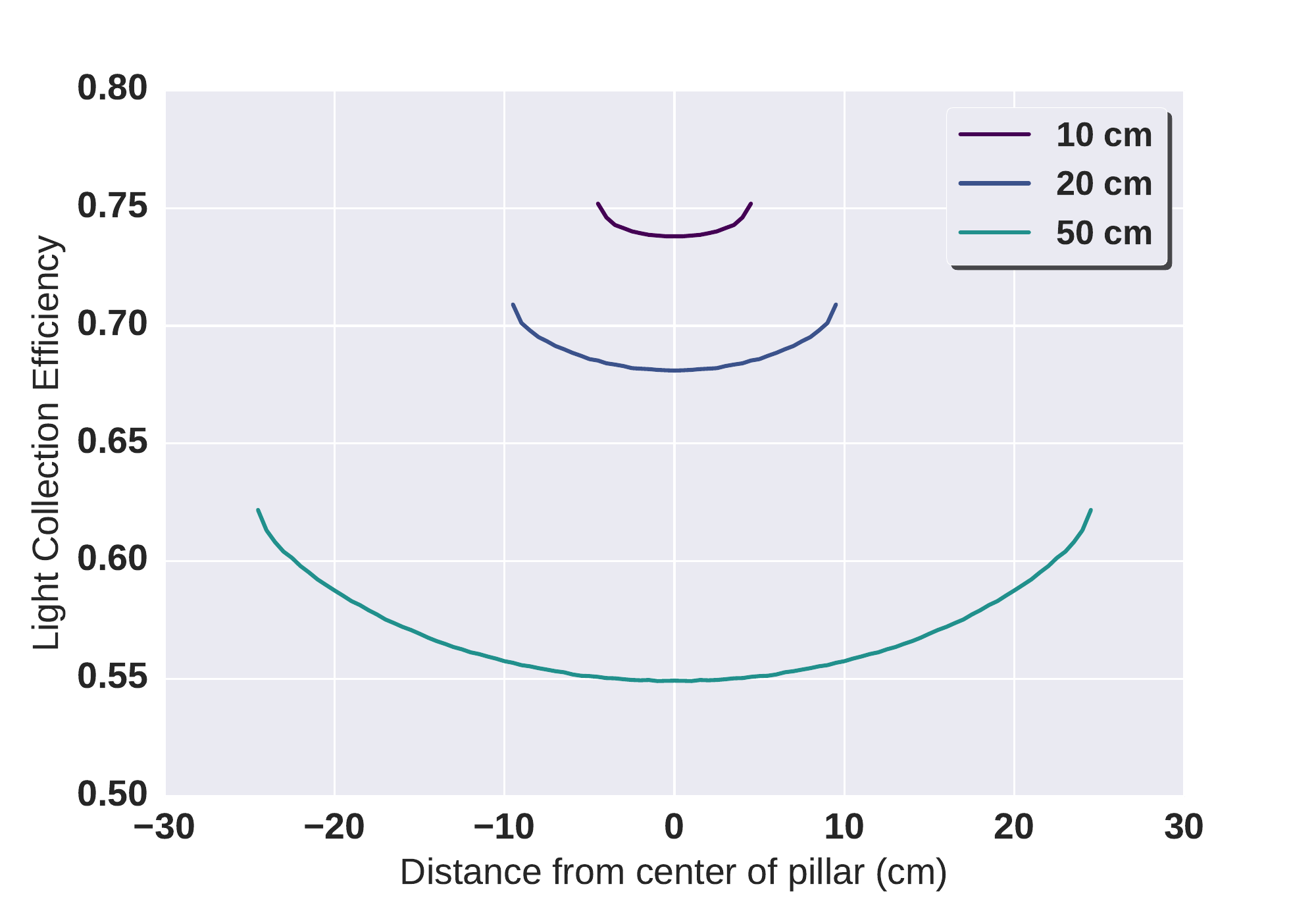}
  \caption{Light collection efficiency using both photodetectors for 10 cm, 20 cm and 50 cm pillars. Higher collection efficiencies are seen for shorter pillars; there is less chance of self-absorption or losses in the reflective ESR film.}
  \label{fig:lightColletionEff}
\end{figure}

Using the area under the curves at each distance, we calculated the collection efficiency as a function of position in the pillar. We summed the light collected on both photodetectors to obtain a light collection efficiency curve for the pillar. Figure \ref{fig:lightColletionEff} shows the efficiency for three different length pillars. As expected, the highest efficiency is seen for a 10 cm pillar and the lowest is seen for a 50 cm pillar. The larger the distance from scintillation to the photodetector, the more tortuous path a photon can travel before arrival at the end of the pillar; consequently, the probability of absorption increases.

\begin{figure}[h!!]
\centering
  \includegraphics[width=\columnwidth]{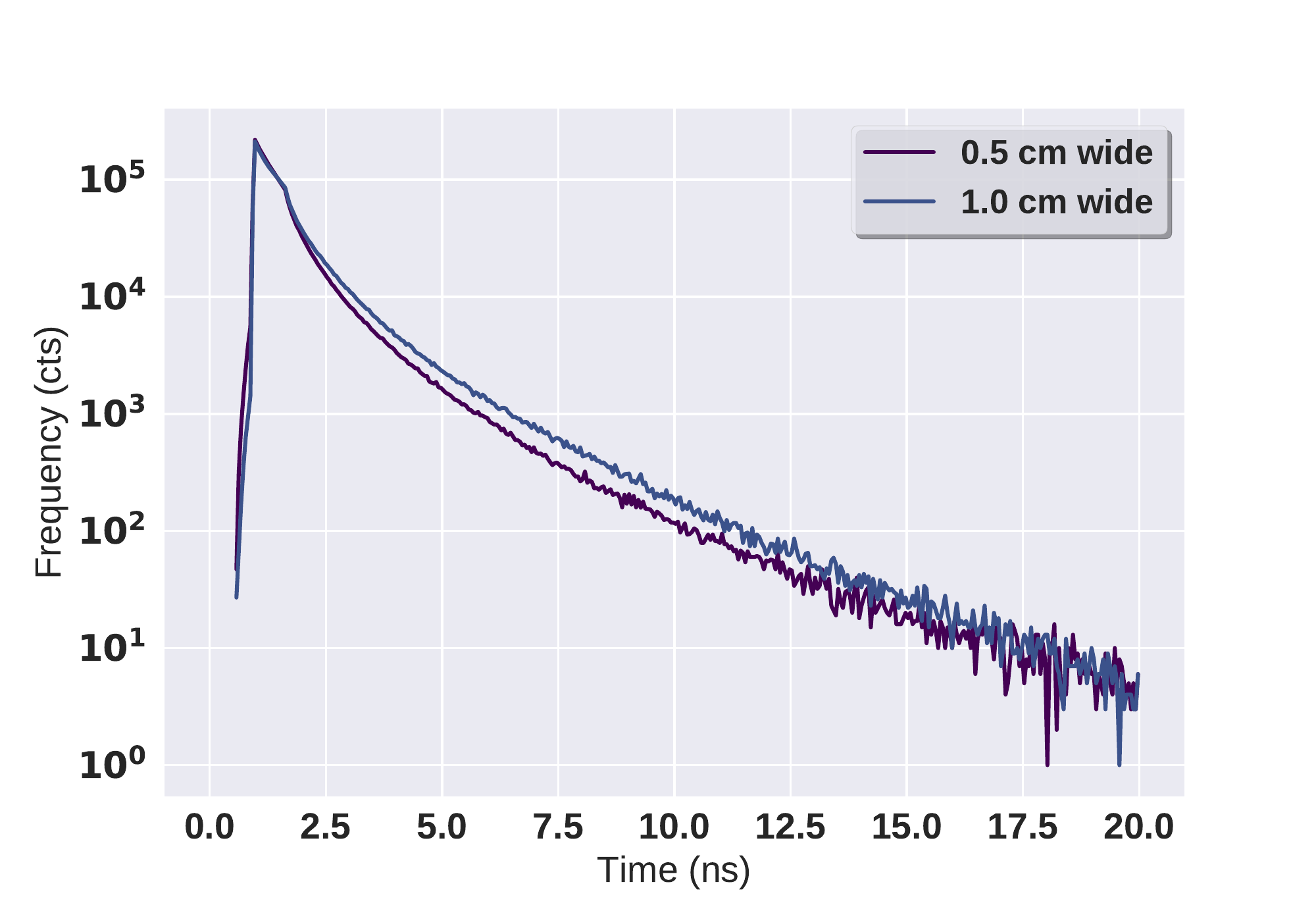}
  \caption{Photon arrival time histogram for a 20 cm long EJ-204 pillar with different widths. 1.0 cm width slightly increases pillar time spread of scintillation photons.}
  \label{fig:sideSizeChannelTimeSpread}
\end{figure}

We produced pillar impulse response functions for multiple scintillator types, lengths and widths. The width of the scintillator pillar has a slight effect on the pillar time spread of scintillation photons. A larger width results in a slightly larger temporal spread of photons shown in Figure \ref{fig:sideSizeChannelTimeSpread}. 

\subsection{Effect of Reflector}
We studied different types of reflective films to line the pillar housing walls. The two common types of reflectors are specular reflectors and diffuse reflectors \cite{Janecek2008, Janecek2012}. Specular reflectors reflect light analogous to mirrors where the reflected angle of the light is equal to the incident angle. Diffuse reflectors reflect incident light over a distribution of angles due to the rough reflector surface and photon scattering within the reflective material itself. An example of a diffuse reflector is a Lambertian reflector, where the surface appears equally bright at all viewing angles.

We simulated two types of reflector materials; titanium dioxide paint and enhanced specular reflector film. We simulated each reflector using built in look up tables (LUT) in Geant4. We used the \texttt{polishedvm2000air} for the ESR film LUT and \texttt{polishedtioair} for the TiO$_2$ LUT. The titanium dioxide paint and ESR film behave as a Lambertian and specular reflectors respectively. At incident photon angles over 50$^\circ$, TiO$_2$ paint transitions into a hybrid of a specular reflector and Lambertian reflector\cite{Janecek2008}. ESR film has a significant reduction in reflectivity for wavelengths below 395 nm\cite{Janecek2012, Loignon-houle2016}. This was not included in the LUT for ESR. This would decrease scintillation position and proton recoil energy reconstruction precision for scintillators who mostly emit scintillation photons below 395 nm (EJ-232Q and stilbene). We did not include the decrease in ESR reflectivity; we were interested in the scintillator and photodetector properties to give us the most precise estimate of scintillation position, proton recoil energy and scintillation time. Careful consideration of the scintillator emission spectrum must be taken to ensure maximum ESR reflectivity.

\begin{figure}[h!!]
\centering
  \includegraphics[width=\columnwidth]{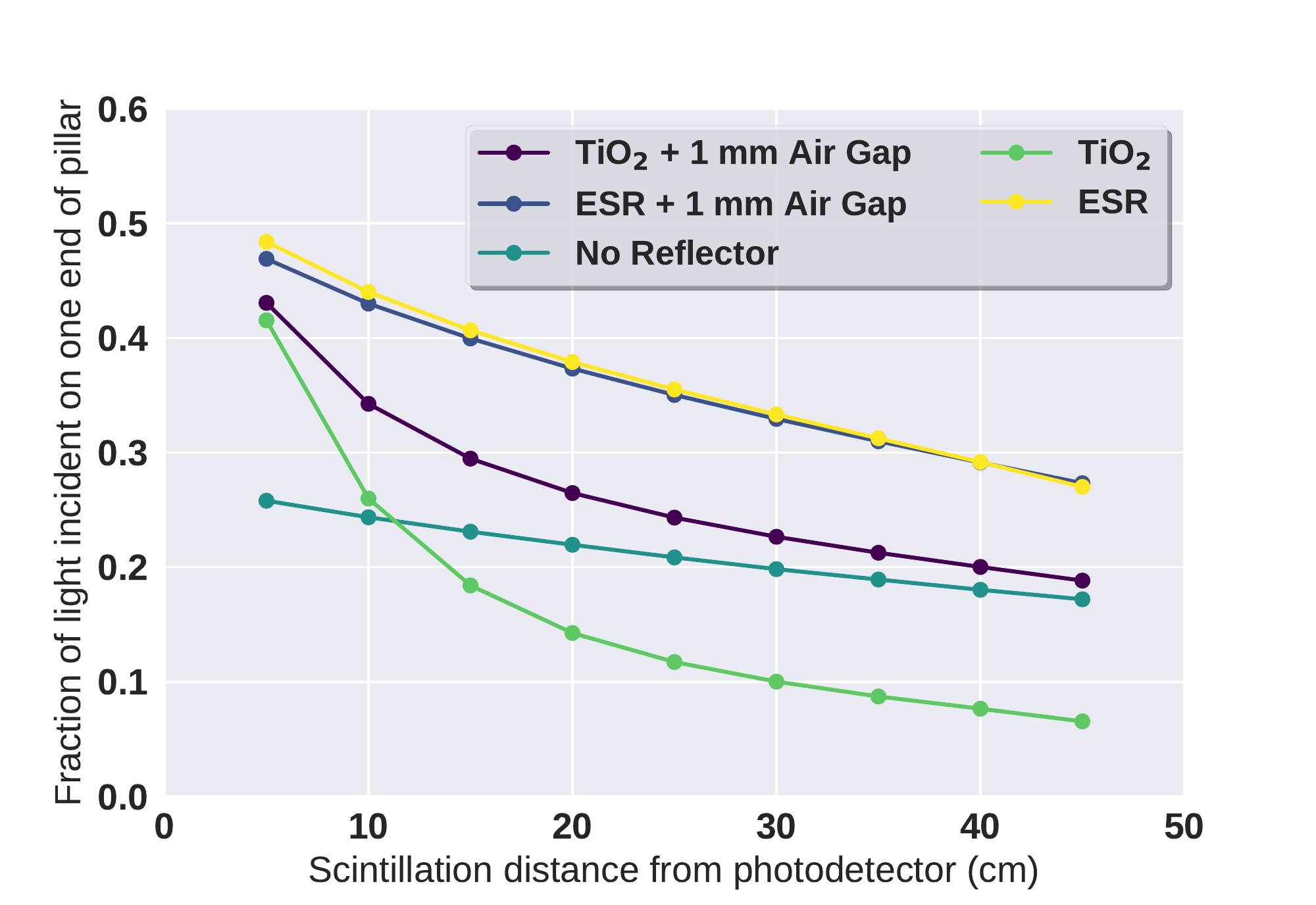}
  \caption{Reflector study with commonly used reflective materials in detector applications. TiO$_2$ is a diffuse reflector and ESR is a specular reflector. A 1 mm air gap was simulated to allow photons to undergo TIR. The `No Reflector' case consists of a scintillator surrounded by air.}
  \label{fig:reflectors}
\end{figure}

We performed simulations in Geant4 in two configurations. For one configuration, we directly  attached the reflector to the pillar. For the second configuration, the reflector lined the housing walls with a 1 mm gap of air between the scintillator and reflector. We estimated the collection efficiency of each arrangement from one photodetector. We compared each configuration to a base case consisting of a bare scintillator surrounded by air. We simulated a 1 cm x 1 cm x 50 cm EJ-204 pillar using $10^7$ photons emitted isotropically at each position. The results are shown in Figure \ref{fig:reflectors}.

For the base case with no reflector, we see a collection efficiency of 25\% at a 5 cm distance from the photodetector. Collection efficiency steadily decreases as distance to the photodetector increases. The collected photons undergo TIR while propagating down the pillar. All other photons that escape the pillar are not detected. We see a decrease in collection efficiency as distance increases due to self-absorption of scintillation light from the scintillator.

Titanium dioxide paint with a 1 mm air gap still allows for TIR of scintillation photons, but it diffusely reflects escaping photons back into the pillar that did not undergo TIR. The most likely angle of reflection in the titanium oxide paint is parallel to the surface normal of the housing wall. This explains the decreasing light collection efficiency as scintillation-detector distance increases. 

Directly painting the pillar with TiO$_2$ shows a reduction in collection efficiency after a 10 cm distance. TiO$_2$ has a higher index of refraction than the scintillator pillar which prohibits total internal reflection. This results in a very poor light collection efficiency for long travel distances. 

The ESR film exhibits high light collection efficiency both with and without an air gap. A pillar directly wrapped in ESR film slightly outperforms the the simulation that includes an air gap. We decided to use an ESR film as our reflector of choice. We included the air gap in the simulations to allow for a physical space between pillars.

\subsection{Scintillation Position Estimate Using Light Intensity}

Previous work \cite{Brubaker2009a} performed measurements using long pillar scintillators for double scatter imaging. This experiment used the intensity of light on opposing photodetectors to estimate the position of interaction within the pillar. The experiment utilized gamma rays interactions instead of neutron elastic scatters. The intensity of scintillation light produced by electrons is directly proportional to the amount of energy deposited by the electron \cite{Leo1993}. Protons do not exhibit a direct proportionality between energy deposited and light created in the scintillator. The denser population of excited states is more prone to non-radiative decays, thus the light conversion is not as efficient. Non-linearities are observed at proton recoil energies below 1 MeV. Scintillation position estimates are less precise for neutron scatter events due to the less scintillation light produced from a proton recoil compared to Compton scatter. One can estimate the position of interaction using the average light intensity on the photodetectors obtained from a multitude of scatter events in a similar fashion to \cite{Brubaker2009a}.

\begin{figure}[h!!]
\centering
  \includegraphics[width=\columnwidth]{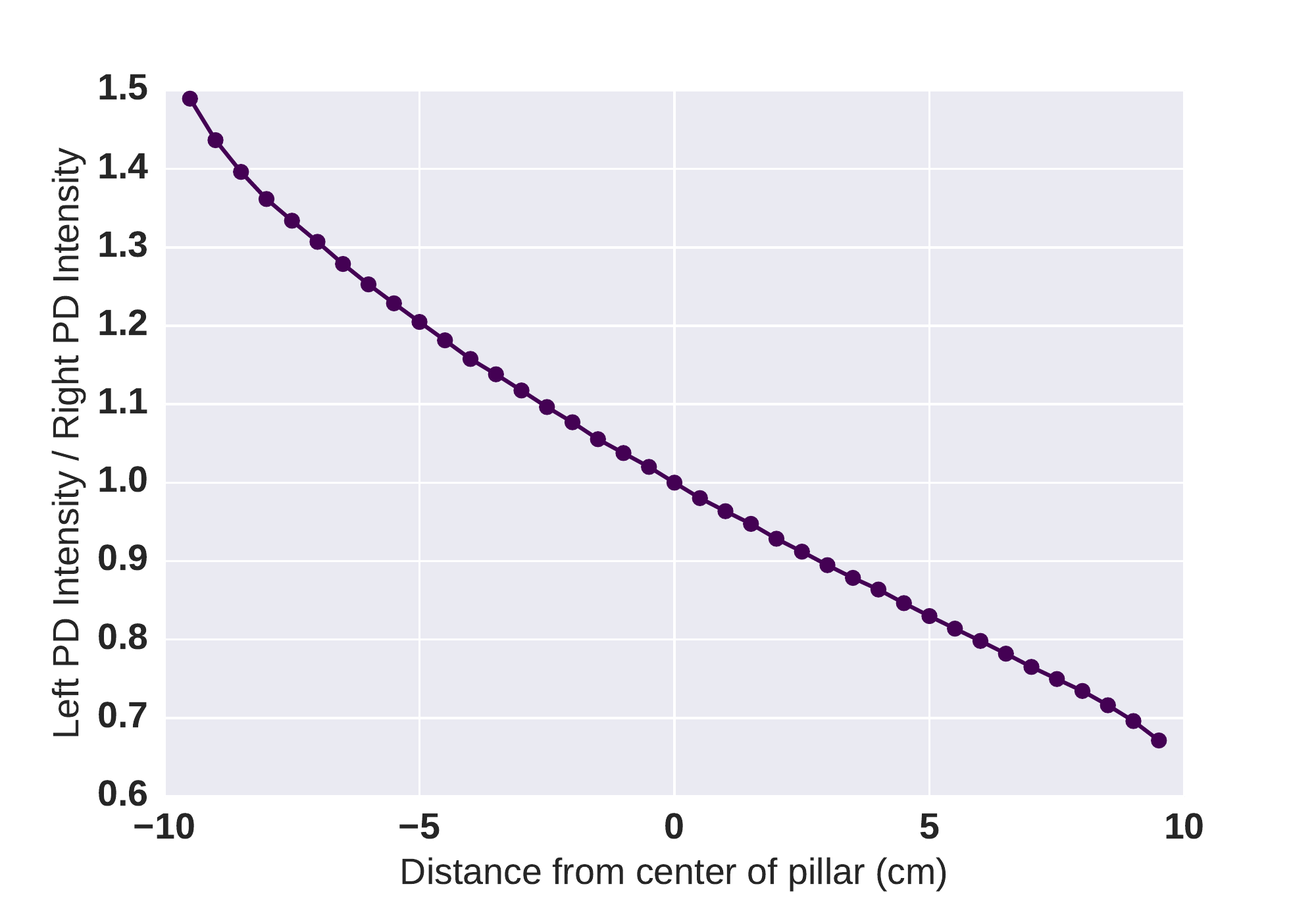}
  \caption{Light ratio curve created using $10^7$ photons at each position to get an average ratio of light at each position along the pillar. Uncertainties for the ratio of light intensity is smaller than the data points.}
  \label{fig:lightRatio}
\end{figure}

We used the same optical light transport model to simulate the transit of scintillation light throughout the pillar to obtain scintillation position estimates using the intensity of charge carrier created in the photodetector.  The simulation consisted of a 1 cm x 1 cm x 20 cm EJ-204 scintillator pillar with MCP-PMs attached on the left and right sides of the pillar. We simulated 1 mm of air surrounding the pillar to allow total internal reflection of scintillation photons. An enhanced specular reflector \citep{3M2010} film lined the housing walls to reflect escaping photons back into the scintillator. We simulated $10^7$ scintillation photons in 0.5 cm increments spanning the pillar. At each position, we calculated the intensity of light incident on the left and right photodetector. The ratio of light intensity on the left and right photodetectors is shown in Figure \ref{fig:lightRatio}. On average, scintillation events on the left side of the pillar will result in ratios greater than 1. Conversely, scintillation events on the right side of pillar have ratios lower than 1. 

\begin{figure}
\centering
  \includegraphics[width=\columnwidth]{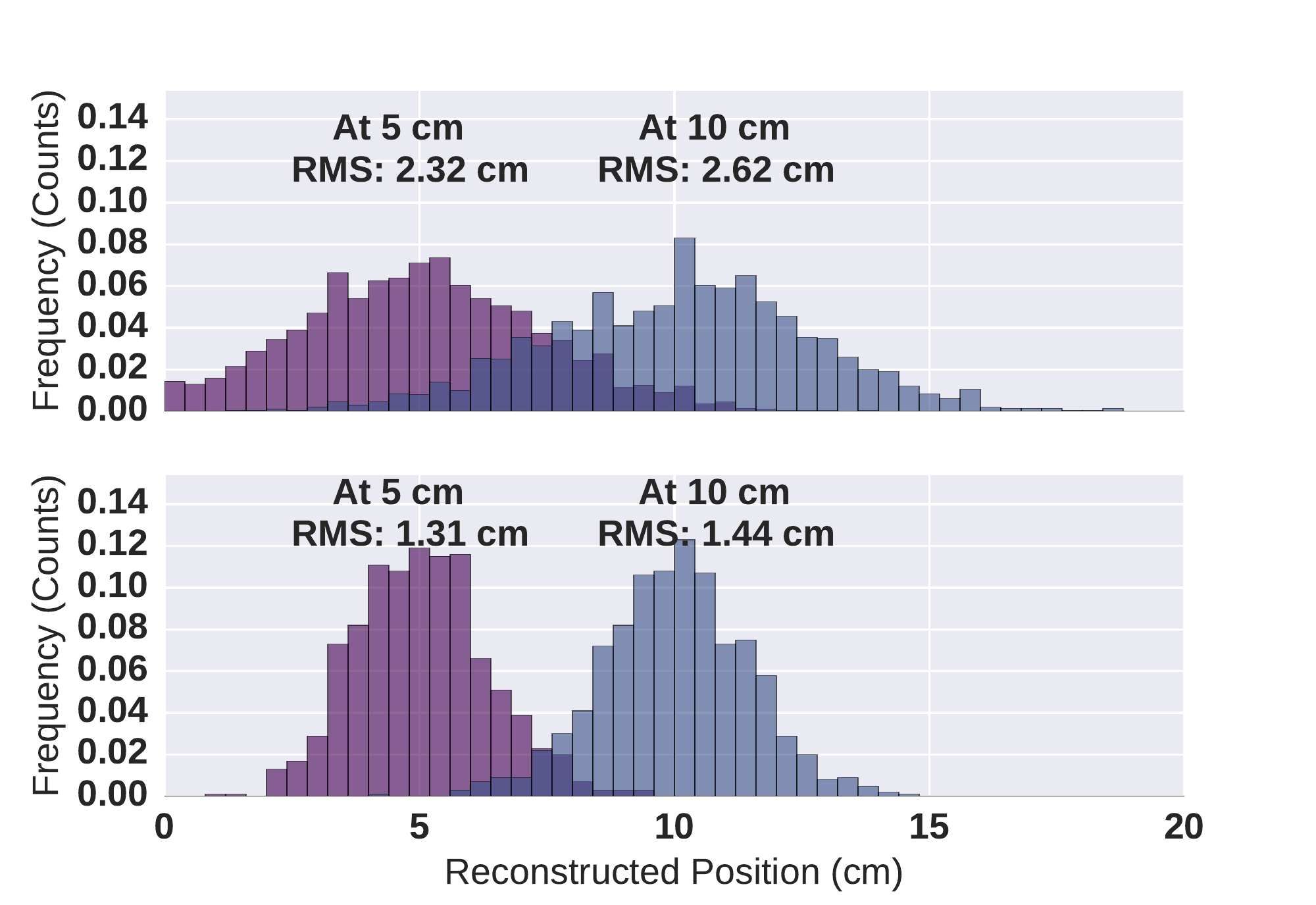}
  \caption{Scintillation position estimation using light intensity incident on the left photodetector to light incident on the right photodetector. Simulation performed using a 1 cm x 1 cm x 20 cm EJ-204 organic plastic scintillator. The top and bottom figure shows position reconstruction using a 1 MeV and 2 MeV proton recoil energy.}
  \label{fig:lightIntensityEstimate}
\end{figure}

We estimated interaction location using light intensity of photodetector waveforms from neutron elastic scatter. We simulated 1,000 proton recoil events at a location 5 cm along the pillar and in the center of the pillar. We fixed proton recoil energies at 1 MeV and 2 MeV. Using the average light ratio curve in Figure \ref{fig:lightRatio}, we estimated scintillation position (using only the observed waveform intensities) with results shown in Figure \ref{fig:lightIntensityEstimate}. The top of the figure shows results for 1 MeV proton recoils and the bottom shows the results for 2 MeV proton recoils. 1 MeV proton recoils have a larger root mean squared (RMS) error of approximately 2.5 cm compared to the RMS error of approximately 1.35 cm for 2 MeV proton recoils. Photostatistics dictate the accuracy of scintillation estimate when using light intensity. More photons are created for a higher proton recoil energy. This produces smaller uncertainties in scintillation position for higher proton recoils. Position reconstruction using the light intensity ratio for a 2 MeV proton recoil resulted in estimated scintillation positions that span about a quarter of the length of the pillar. When only tens of charge carriers are created in both photodetectors, the reconstruction is less precise due to random fluctuations in the number of charge carriers. Imprecise interaction position estimation produces incorrect pointing vectors. We cannot rely on average behavior to reconstruct scintillation position in the pillar. 

The optimal imager design to estimate scintillation position using light intensity on opposing photodetectors may not be the same as using photodetectors' signal amplitude and relative timing. These results hold for the as-simulated design using a 1 mm air gap and an ESR reflective film.

\section{Observed and Nominal Responses}
\label{peCreation}
In this section, we will show the process of simulating observed responses on an event-by-event basis for scintillation interactions within the pillar. We also illustrate the method to tabulate nominal  (expected) responses for comparison to the observed responses. 

\subsection{Observed Responses}
We simulated optical light transport and propagated an impulse of photons to opposing ends of the pillar to estimate the time spread photons undergo as a result of pillar geometry. These are tabulated in pillar response functions ($R_{pil}$). However, scintillations have a characteristic time response unique to each scintillator (see Figure \ref{fig:scintillators}). We convolved each pillar response function with the time response of the scintillator. This produced  ``scintillation pillar responses'' at 0.5 cm increments in the z-dimension along the pillar. We interpolated between the 0.5 cm increments to obtain scintillation pillar responses at any location along the pillar. 

Scintillation pillar responses estimate time of arrival history of photons at the photodetector. The next step is to determine the time and number of charge carriers created. Charge carriers are created when photons interact in either the photocathode of an MCP-PM or directly inside an SiPM. So, we calculated the number of photons incident on the photodetector to determine the number of charge carriers created.

In simulation, we know the proton recoil energy of interactions. Therefore, we can determine the expected number of scintillation photons produced in each interaction, given the scintillators luminosity. As previously mentioned, protons do not exhibit a linear relationship of proton recoil energy to light production due to quenching interactions. These quenching interactions reduce the amount of scintillation light created. To account for this, we used neutron light output (LO) functions for EJ-309 given in \cite{Naeem2013} and \cite{Enqvist2012}. The LO function predicts how much MeV electron equivalent (MeVee) the proton deposited in the scintillator. Each scintillator produces a characteristic amount of scintillation photons, on average, per MeVee; this is the luminosity. Using the luminosity, we sampled the actual number of scintillation photons produced in the scintillator from a Poisson distribution. This number is multiplied by the collection efficiency of each photodetector to estimate the number of photons that arrive at the face of the photodetector. Not all photons incident on the photodetector create a charge carrier. Therefore, the number of incident photons was multiplied by the QE or PDE of the photodetector to estimate the number of charge carriers created. The number of charge carriers created is shown in Equation \ref{eqn:createdPE}

\begin{align}
\label{eqn:createdPE}
n_{CC} = LO(E_p) \cdot L \cdot \epsilon_{LC}(z) \cdot \epsilon_{PD}
\end{align}

\noindent
where $n_{CC}$ is the number of charge carriers created, $LO()$ is the light output function (MeVee), $E_p$ is the proton recoil energy (MeV), $L$ is the luminosity of the scintillator (scintillation photons/MeVee), $\epsilon_{LC}(z)$ is the collection efficiency as a function of position in the scintillator (collected photons/emitted photons), and $\epsilon_{PD}$ is the quantum efficiency or photon detection efficiency (charge carriers/incident photon). Charge carrier arrival times were randomly sampled from the nominal pillar impulse response.

Photodetectors are positioned on opposite sides of the pillar. This provides a correlated pair of responses for each scintillation event. For example, consider a 20 cm length pillar. If scintillation occurs 5 cm away from the left photodetector, the same scintillation photons emerge 15 cm from the right photodetector. Therefore, both photodetector responses can be used to estimate the position of scintillation along the length of the pillar. 

\begin{figure}[h!!]
\centering
  \includegraphics[width=\columnwidth]{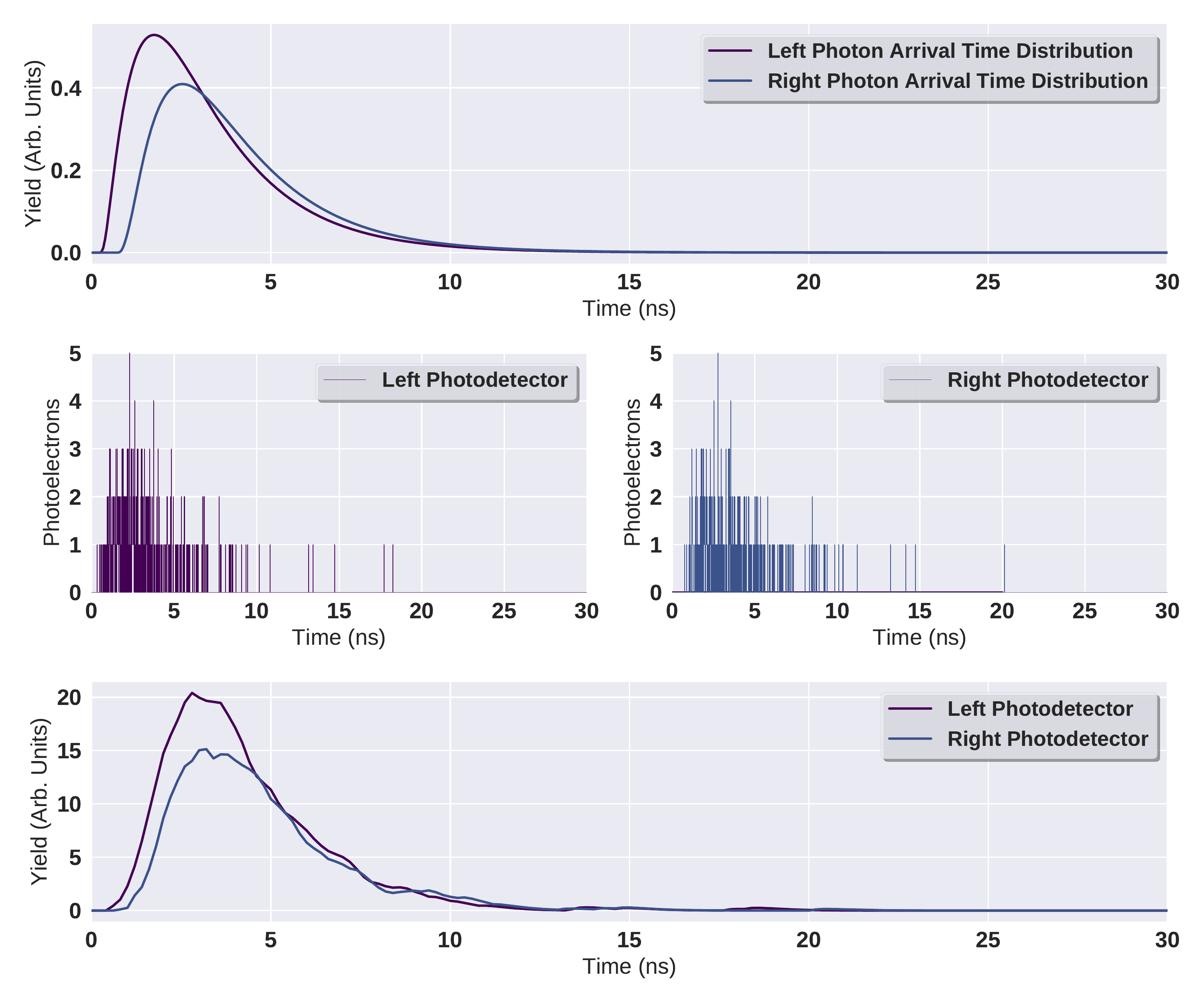}
  \caption{Simulated waveforms. Top) A pair of correlated pillar response functions for a scintillation closer to the left photodetector. We randomly sample charge carrier arrival times. Middle) Charge carrier arrival times after undergoing TTS. Bottom) Observed photodetector waveforms after convolving the charge carriers with the impulse response of the PD.}
  \label{fig:waveformCreation}
\end{figure}

Figure \ref{fig:waveformCreation} illustrates the waveform simulation process. The first step involves calculating the pair of scintillation pillar response functions at the actual distance to the left and right photodetector shown in the top of the figure. Using the pair of scintillation pillar responses, we randomly sample charge carrier arrival times. The number of charge carriers sampled is dictated by Equation \ref{eqn:createdPE} and shown in the middle of Figure \ref{fig:waveformCreation}. We smeared each charge carrier arrival by the transit time spread (TTS) characteristic to each photodetector. The transit time spread in both types of photodetector is small: 50 ps for the MCP-PM and 115 ps for the SiPM \cite{Grigoryev2016a, Nemallapudi2016}. Finally, to estimate the observed output waveform, we convolved the charge carrier arrival time histories shown in the middle of the figure with the photodetector impulse response shown in Figure \ref{fig:photodetectors}. This operation spreads out the charge carriers over multiple time bins. Commercially available fast digitizers based on switched capacitor arrays can sample waveforms every 200 ps\cite{Ritt2010, Bitossi2016, Oberla2014}. Therefore, we binned our observed waveform in 200 ps intervals. The observed photodetector output waveform is shown at the bottom of Figure \ref{fig:waveformCreation} for both the left and right photodetector. 

\subsection{Nominal Responses}

\begin{figure}[h!!]
\centering
  \includegraphics[width=\columnwidth]{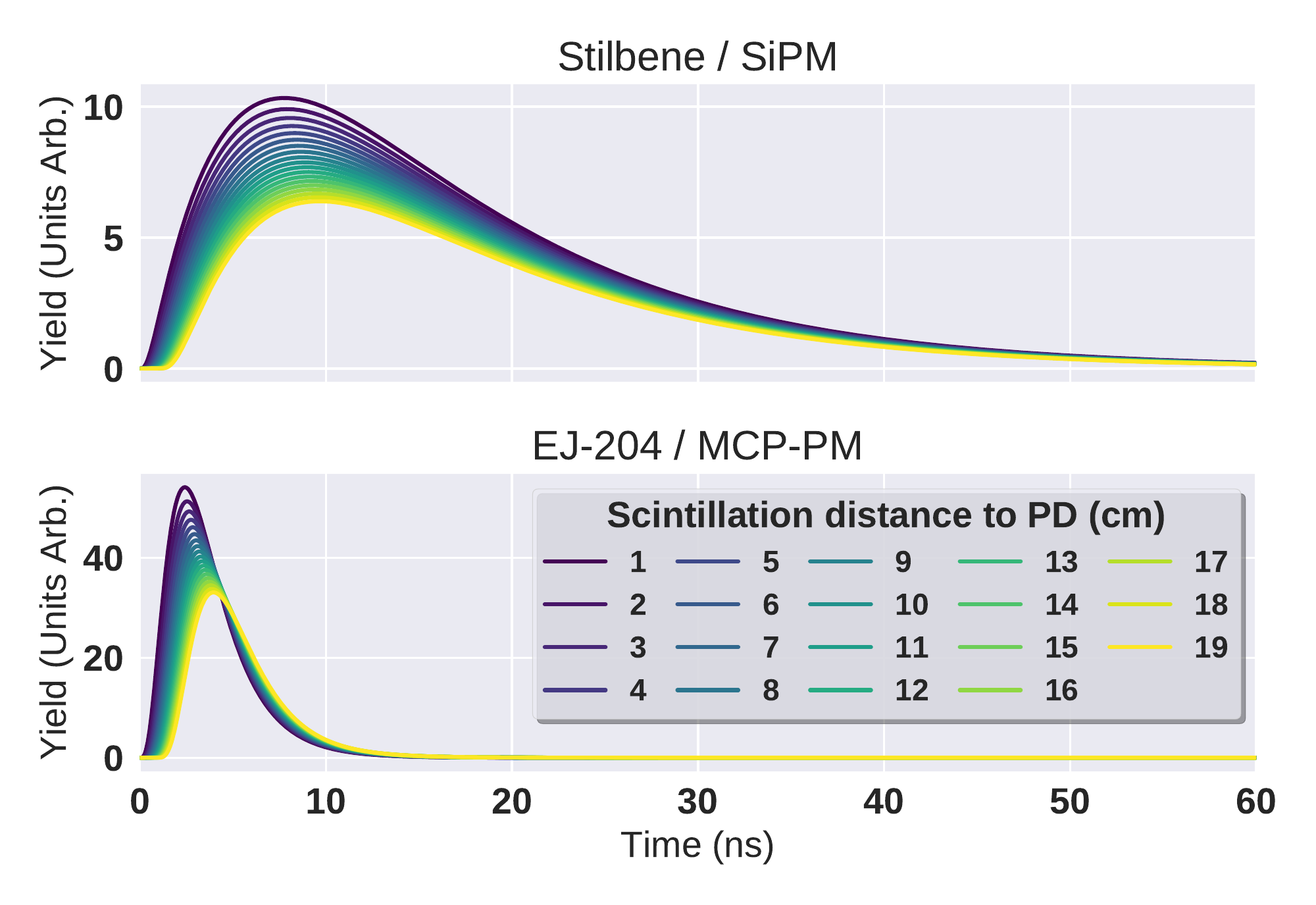}
  \caption{Nominal responses from a 1 cm x 1 cm x 20 cm scintillator pillar. Top shows stilbene with an SiPM and bottom shows EJ-204 with an MCP-PM. }
  \label{fig:idealResponses}
\end{figure}

We tabulated nominal responses for all combinations of scintillator, photodetector and pillar geometries using Equation \ref{eqn:fullResponse}. Recall that it convolves a scintillator time response, pillar response function, photodetector impulse response, and photodetector transit time spread. All responses were discussed in previous sections. Two examples of nominal responses are shown in Figure \ref{fig:idealResponses}. The top of the figure shows a stilbene pillar using an SiPM and the bottom of the figure shows an EJ-204 pillar using an MCP-PM. The slow fall time of both stilbene and the SiPM result in an elongated decay time for the nominal response.

\section{Estimation of Scintillation Position, Proton Recoil Energy, and Scintillation Time}
\label{positionReconstruction}
In this section, we show the best scintillator and photodetector combination is EJ-204/MCP-PM. The compact charge carrier distribution in the MCP-PM and the fast, bright scintillation of EJ-204 allows for the most precise position estimate ($<$1 cm RMS error) and proton recoil energy estimate ($<$50 keV RMS error) for proton recoil energies above 1 MeV; these results are described in detail in Section \ref{RMS description} and illustrated in Figures \ref{fig:rmsErrorPosHist} and \ref{fig:rmsErrorEnergyHist}.  The precision of position reconstruction is dependent upon where in the pillar the scintillation occurred; it is less precise in the center of the pillar due to the low collection efficiency at that location.  
  We examine the scintillation position and proton recoil energy reconstruction precision from 0.2 to 2.9 MeV proton recoil energy using the EJ-204/MCP-PM combination. 

\subsection{Analysis Method}
In the previous section, we discussed how we estimated observed photodetector waveforms. We used the observed photodetector output waveforms (shown in the bottom of Figure \ref{fig:waveformCreation}) and compare them to nominal photodetector responses to estimate the position of interaction and proton recoil energy in the pillar.

We used the Broyden-Fletcher-Goldfarb-Shanno (BFGS) bounded minimization to estimate scintillation position within the bounds of the pillar\cite{Broyden1970, Fletcher1970, Goldfarb1970, Shanno1970, Byrd1995, Zhu1997}. BFGS implements the quasi-Newtonian method of Broyden, Fletcher, Goldfarb and Shanno using only first derivatives of the cost function surface. The ``bounded'' implementation uses a form of the minimization method that approximates the inverse Hessian using a few vectors and ``bound box'' constraints on variables. We used BFGS to minimize a cost function to estimate position of scintillation and proton recoil energy; the cost function is described in detail in the next section. We explored two different approaches for finding the best estimate of position and energy. The first approach obtained the best estimate of scintillation position followed by an estimate of proton recoil energy. The second approach fit both position and energy simultaneously. Both approaches resulted in similar RMS errors for both scintillation position and proton recoil energy. 

\subsection{Cost Function}
We used a negative Poisson log likelihood (NLL) cost function to determine the best estimate of the scintillation position and proton recoil energy. We compared the observed outputs from the photodetectors to nominal responses predicated on assumed scintillation position and proton recoil energy. The negative log likelihood cost function assumed the charge carrier counts were Poisson distributed. Stating with the Poisson distribution, we can derive a NLL cost function. The Poisson probability of observing $n$ counts when the expected number of counts $m$ is given by

\begin{align}
P(n;m)&= \dfrac{m^n}{n!} e^{-m} \\
LL = \ln P &= \ln (m^n e^{-m}) = n \ln(m) - m
\end{align}

\noindent
Therefore, the log likelihood is 
\begin{align}
NLL &= -(L_{response} + R_{response}) \\
{L}_{response} &= n_{L} \cdot ln(m_{L}) - m_{L} \nonumber\\
{R}_{response} &= n_{R} \cdot ln(m_{R}) - m_{R} \nonumber
\end{align}

\noindent
where $LL$ is the log likelihood value, $n$ is an observed response, $m$ is the expected response, $NLL$ is the negative log likelihood value, $n_{L}$ and $n_{R}$ are the observed photodetector output waveform from the left/right photodetector and $m_{L}$ and $m_{R}$ are the nominal response for the left/right photodetector. Log likelihood cost functions are minimized where the observed and nominal responses match. We used maximum likelihood estimation maximization (MLEM) to estimate the most probable scintillation position and proton recoil energy for a given observed response. 

\subsection{Initial Guess}
A good initial guess decreases the number of cost function evaluations to find the minimum. To calculate an initial guess for proton recoil energy, we estimated scintillation position using the light intensity on both photodetectors; the proton recoil energy initial guess used the estimated scintillation position, scintillator luminosity and the inverse of the neutron light output function.

\subsection{Estimation of Scintillation Position and Proton Recoil Energy}
\label{RMS description}
\begin{figure}[h!!]
\centering
  \includegraphics[width=\columnwidth]{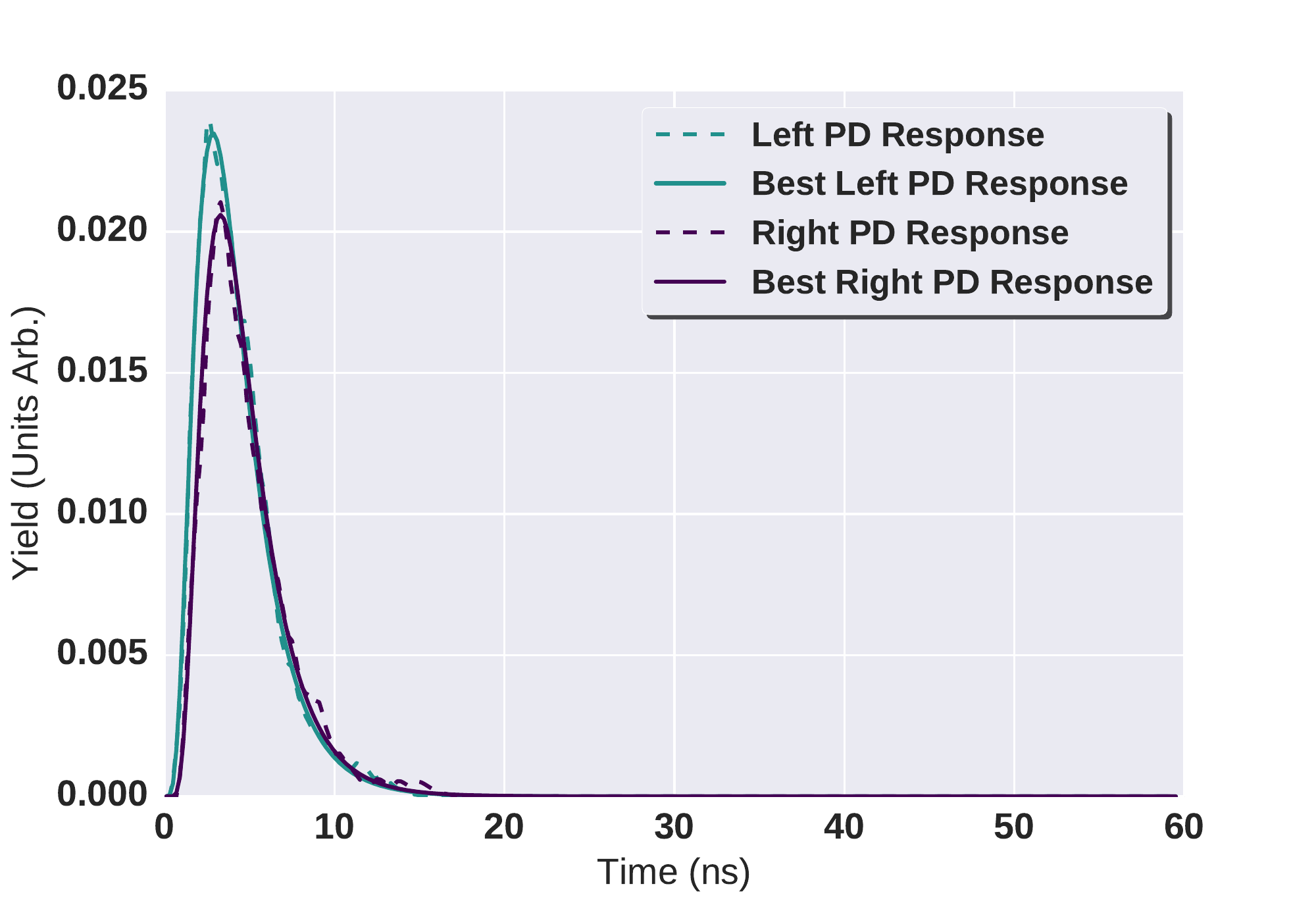}
  \caption{Observed waveforms are shown in dashed lines compared to the responses from the best estimate of scintillation position and proton recoil energy}
  \label{fig:EJ-232Q_sipm_fitWaves}
\end{figure}

Recall that we are trying to determine the best combination of pillar length, pillar width, scintillator, and photodetector that results in the smallest uncertainty in both position and proton recoil energy. For each combination, we simulated 10,000 events uniformly distributed along the pillar. We tabulated the true proton recoil energy, true position of interaction, best estimate of proton recoil energy and best estimate of position. An example of a best fit pair of waveforms is shown in Figure \ref{fig:EJ-232Q_sipm_fitWaves}.

To evaluate the precision of reconstruction, we computed sample RMS errors for each combination for a specified proton recoil energy. We calculated position and proton recoil RMS errors for 12 different combinations at 1 MeV and 2 MeV proton recoil energies.

\begin{figure}[h!!]
\centering
  \includegraphics[width=\columnwidth]{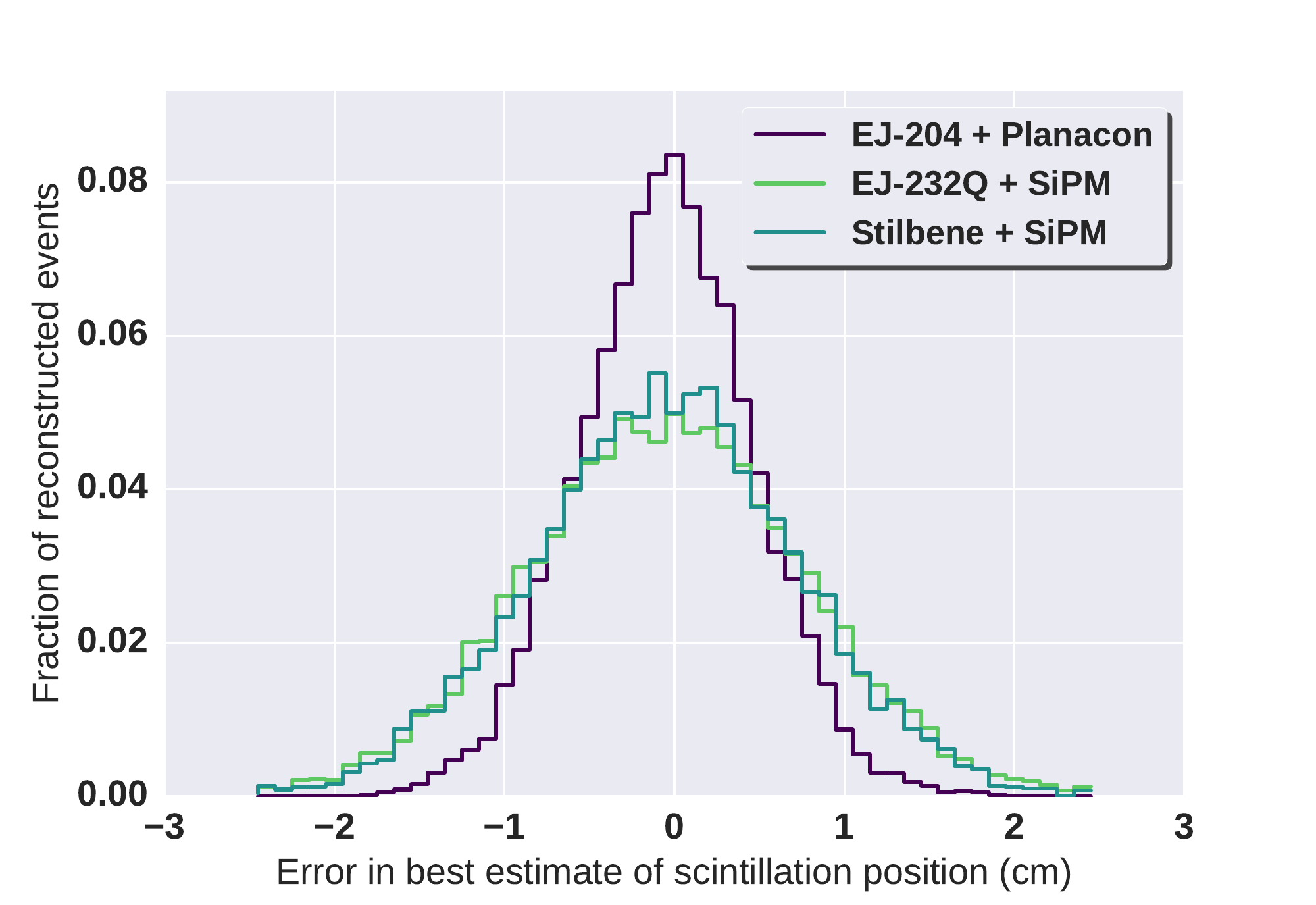}
  \caption{Error in position reconstruction for 10,000 events uniformly distributed throughout the pillar for 2 MeV proton recoil energy}
  \label{fig:rmsErrorPosHist}
\end{figure}

We histogramed the error for the best estimate of scintillation position for three selected combinations in Figure \ref{fig:rmsErrorPosHist}. This figure shows best estimates for a 2 MeV proton recoil energy for a 1 cm x 1 cm x 20 cm scintillator pillar with a 1 mm air gap using an ESR film on the pillar housing walls. Examining the figure, the EJ-204/MCP-PM combination has the best estimate of scintillation position of all three combinations. The position RMS errors are shown in Table \ref{table:threeResults}. Recall that the luminosity of EJ-204 is equivalent to that of stilbene (approximately 68\% of anthracene luminosity). The photon detection efficiency of SiPMs is almost double the quantum efficiency of the MCP-PM photocathode.  Even with better photostatistics, the stilbene/SiPM combination still has a slightly poorer position RMS error due to the long response time of both the scintillator and photodetector. Spreading the charge in time results in a more imprecise estimate of scintillation position. The fast light emission of EJ-232Q combined with an SiPM has a comparable position RMS error even though it produces one third the amount of scintillation light for the same proton recoil energy. Fast scintillation times are desirable; spreading charge over time produces poorer estimates of scintillation position.

\begin{table*}[t]
\centering
\caption{RMS errors for position and proton recoil energy for three scintillator/photodetector combinations}
\label{table:threeResults}
\begin{tabular}{ccc}
             & Position  & Energy  \\
Combination  & RMS Error  & RMS Error  \\
\hline
EJ-204/MCP-PM &  0.52 cm & 43 keV \\ 
Stilbene/SiPM &  0.74 cm & 35 keV \\ 
EJ-232Q/SiPM  &  0.82 cm & 188 keV \\ 
\end{tabular}
\end{table*}

\begin{figure}[h!!]
\centering
  \includegraphics[width=\columnwidth]{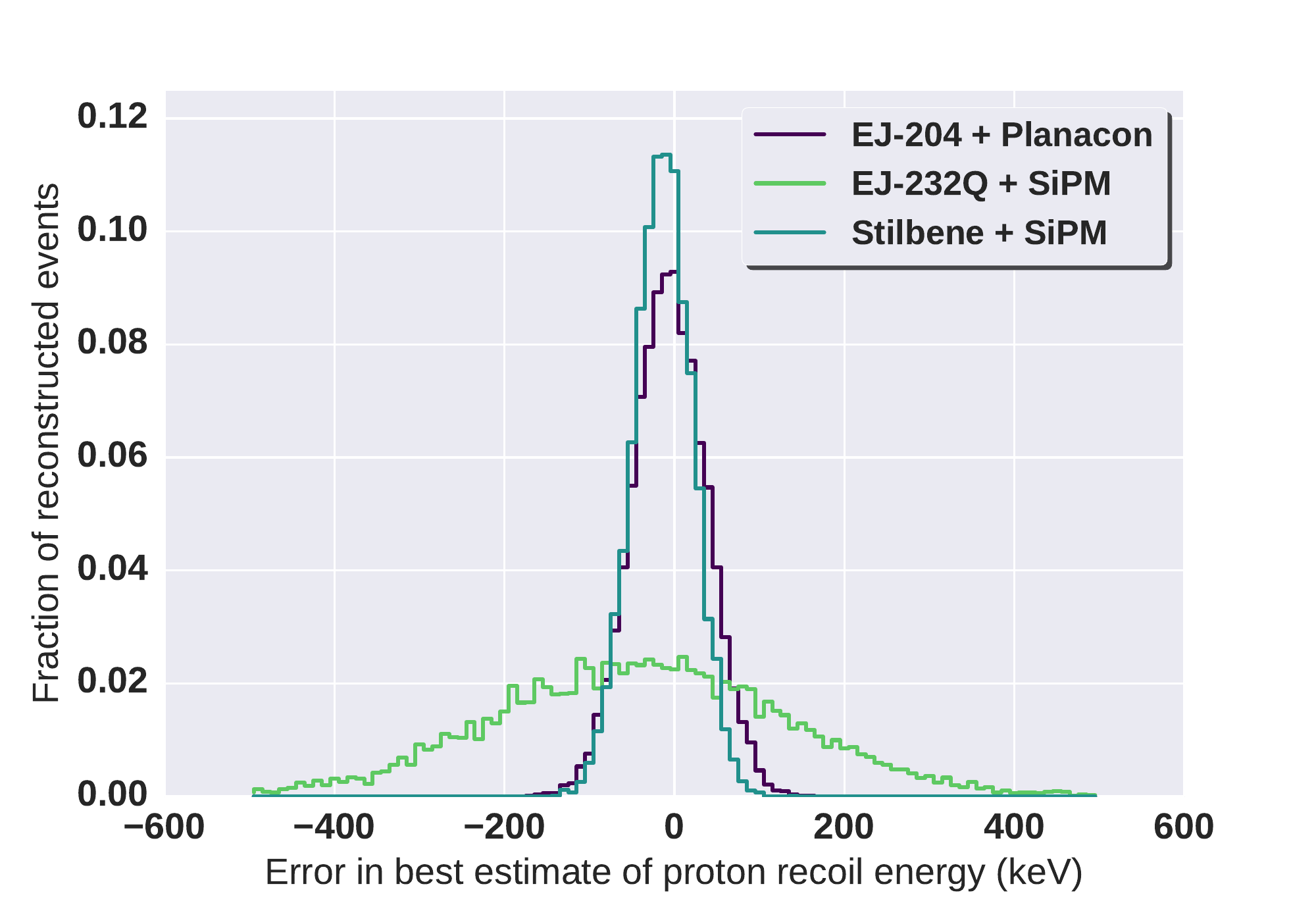}
  \caption{ Error in proton recoil energy reconstruction for 10,000 events uniformly distributed throughout the pillar for 2 MeV proton recoil energy}
  \label{fig:rmsErrorEnergyHist}
\end{figure}

We histogramed the error in the best estimate of proton recoil energy for the same three combinations, geometry, and proton recoil energy; the results are shown in Figure \ref{fig:rmsErrorEnergyHist}. The combination of stilbene and SiPM outperformed the others with a proton recoil RMS error of 35 keV (1.8\% error). Other RMS errors were 43 keV (2.2\% error) for EJ-204 + MCP-PM and 188 keV (9.9\% error) for the quenched plastic also shown in Table \ref{table:threeResults}. Better photostatistics is the main driving force in reconstructing proton recoil energy. The Stilbene/SiPM combination only slightly outperformed the EJ-204 + MCP-PM combination from photostatistics alone. EJ-232Q significantly underperformed the other combinations due to its small self-attenuation length of 8 cm. 

The proton recoil energy  RMS errors are smaller than expected using counting statistics of the number of charge carriers produced in the photodetectors. In the simulation, we estimated the amount of light produced in the interaction and calculated the corresponding proton recoil energy using an inverted form of the light output function. This transformation from brightness to proton recoil energy inherently reduces the RMS error; for a detailed description of this effect, see \ref{RMS reduction section}.

\subsection{Estimation of Scintillation Time}

Up to this point, we have assumed we knew the scintillation time. We showed nominal responses and pillar response functions starting at the true scintillation time. This assumption results in more precise position reconstruction estimates. Real world applications will use a lower level discriminator (LLD) to determine if an interaction occurred. If the LLD is exceeded, the waveform is digitized. Lower level discriminators are subject to large variations in arrival time from  amplitude walk. Consequently, the scintillation time will have to be estimated.

We recalculated position RMS error while trying to estimate scintillation position and scintillation time simultaneously for a 1 cm x 1 cm x 20 cm EJ-204 pillar using an MCP-PM. Current fast digitizers sample waveforms every 200 ps. Therefore, we binned our observed waveform in 200 ps intervals. We calculated the initial guess of scintillation time using a 20\% constant fraction discrimination level. 

\begin{figure}[h!!]
\centering
  \includegraphics[width=\columnwidth]{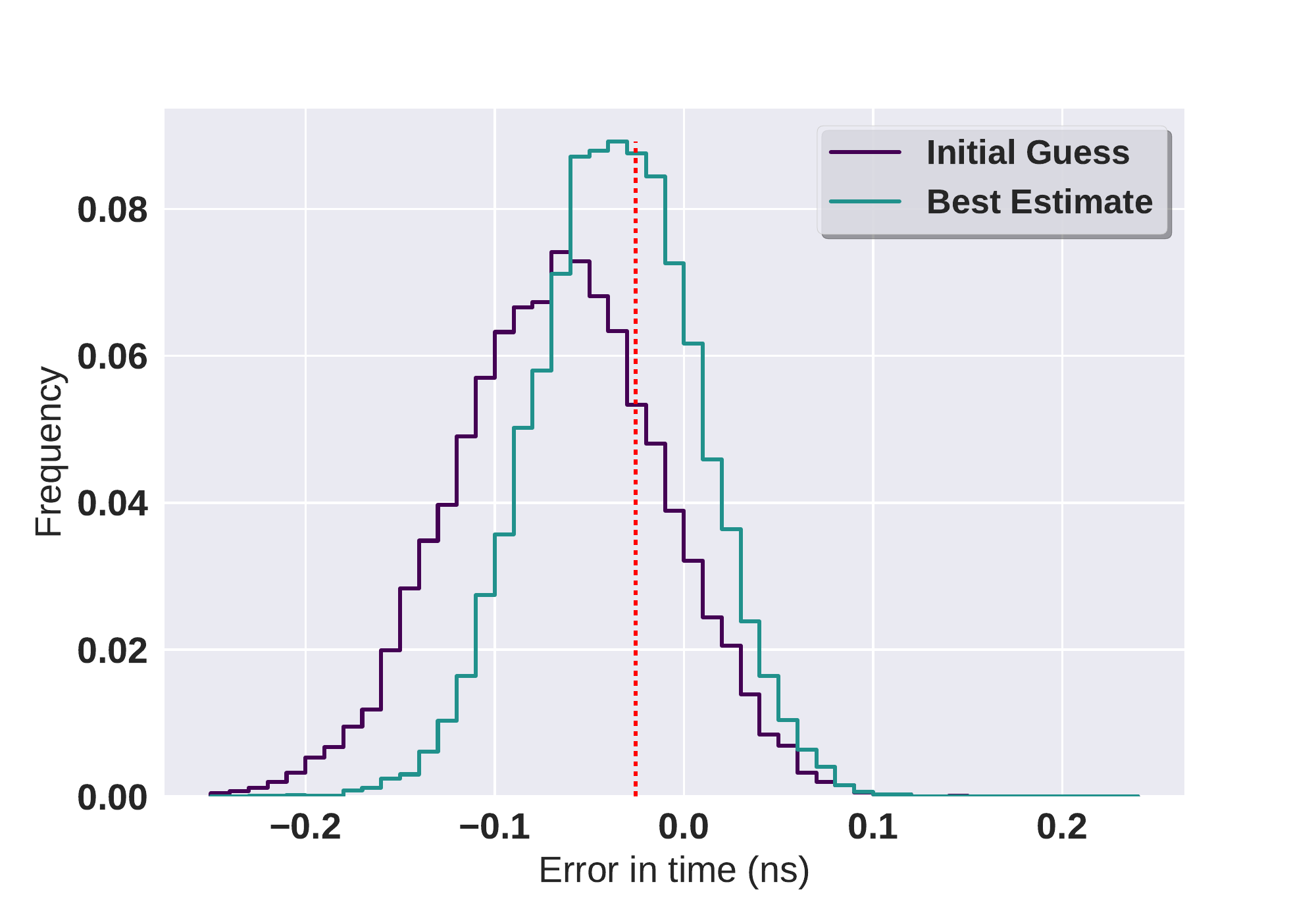}
  \caption{Initial guess using CFD and best estimate of scintillation time using MLEM for 10,000 uniformly distributed 2 MeV proton recoils throughout the pillar.}
  \label{fig:fitTiming}
\end{figure}

CFD timing does not estimate the true scintillation time; it includes a timing offset. We calculated an average offset of 1.2 ns using the nominal responses shown in Figure \ref{fig:idealResponses}. We calculated the initial guess for scintillation time by subtracting the average offset from the average of 20\% CFD times from both observed waveforms. The results of estimating the true scintillation time in the fit is shown in Figure \ref{fig:fitTiming}.

The simulation consisted of 10,000 events uniformly distributed throughout the pillar at 2 MeV proton recoil energy. True scintillation time varied from 5 ns to 20 ns. We calculated error in scintillation time by subtracting the estimated time from the true scintillation time. Both the initial guess and the best estimates of scintillation time exhibited a bias towards later scintillation times. The best estimate exhibited a 30 ps bias whereas CFD timing resulted in a bias of approximately 80 ps. We obtained 50.1 ps RMS error for the best estimate of scintillation time. 

Fitting scintillation time and scintillation position concurrently resulted in a 5\% increase in position RMS error from 0.52 cm to 0.54 cm.  At 1 MeV proton recoil energy, we observed an increase of just under 1 mm for position RMS error and approximately 100 ps RMS error for the best estimate of scintillation time. Both results show an increase of less than 10\% at 1 and 2 MeV proton recoil energy.  

\subsection{Cost Function Surface Analysis}

\begin{figure*}[!htb]
\centering
\begin{subfigure}[t]{0.45\textwidth}
  \centering
  \includegraphics[width=\columnwidth]{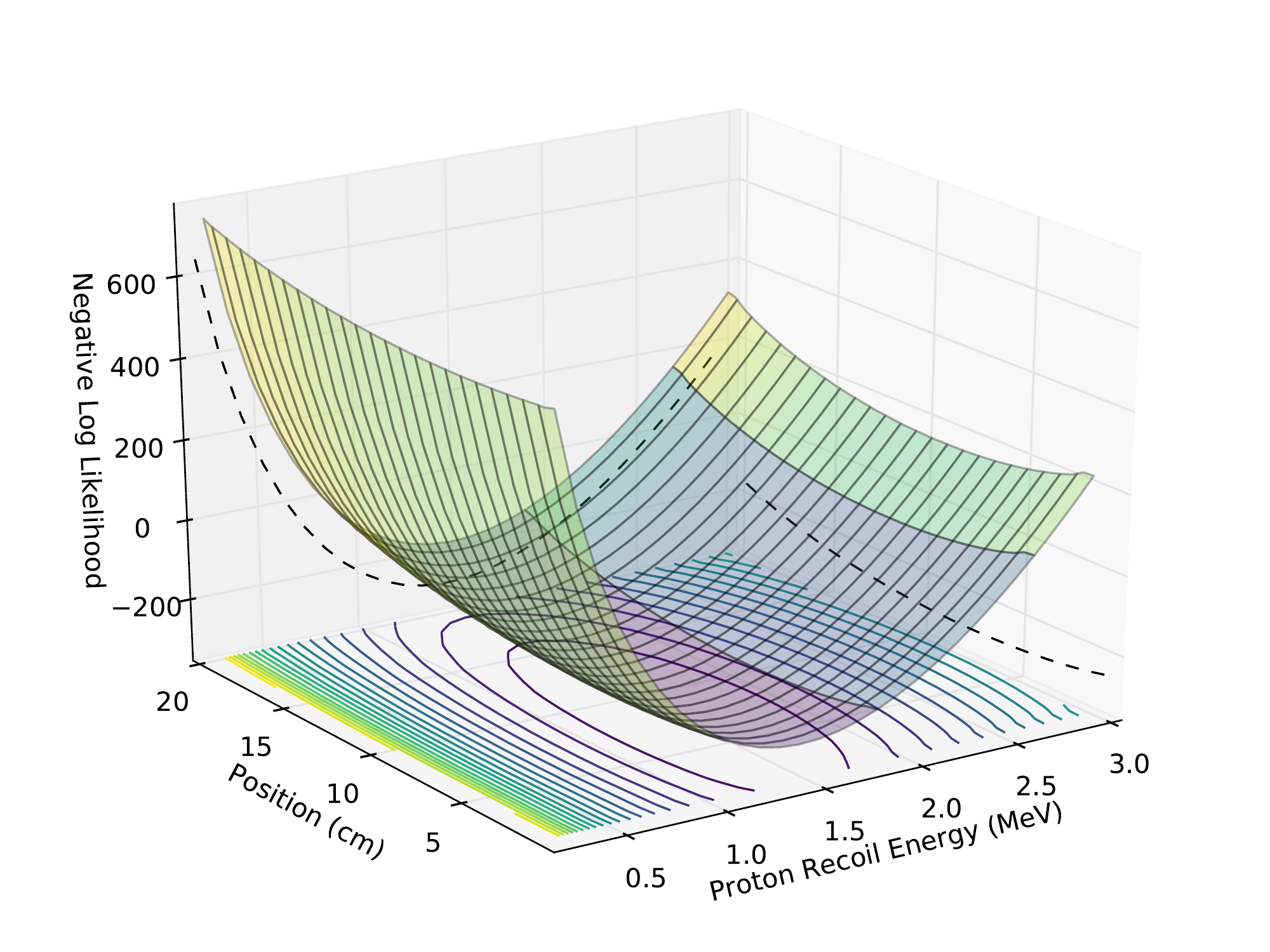}
  \caption{EJ-204/MCP-PM combination}
  
\label{fig:3d_ej204_planacon}
\end{subfigure}
\hspace{2em}
\begin{subfigure}[t]{0.45\textwidth}
  \centering
  \includegraphics[width=\columnwidth]{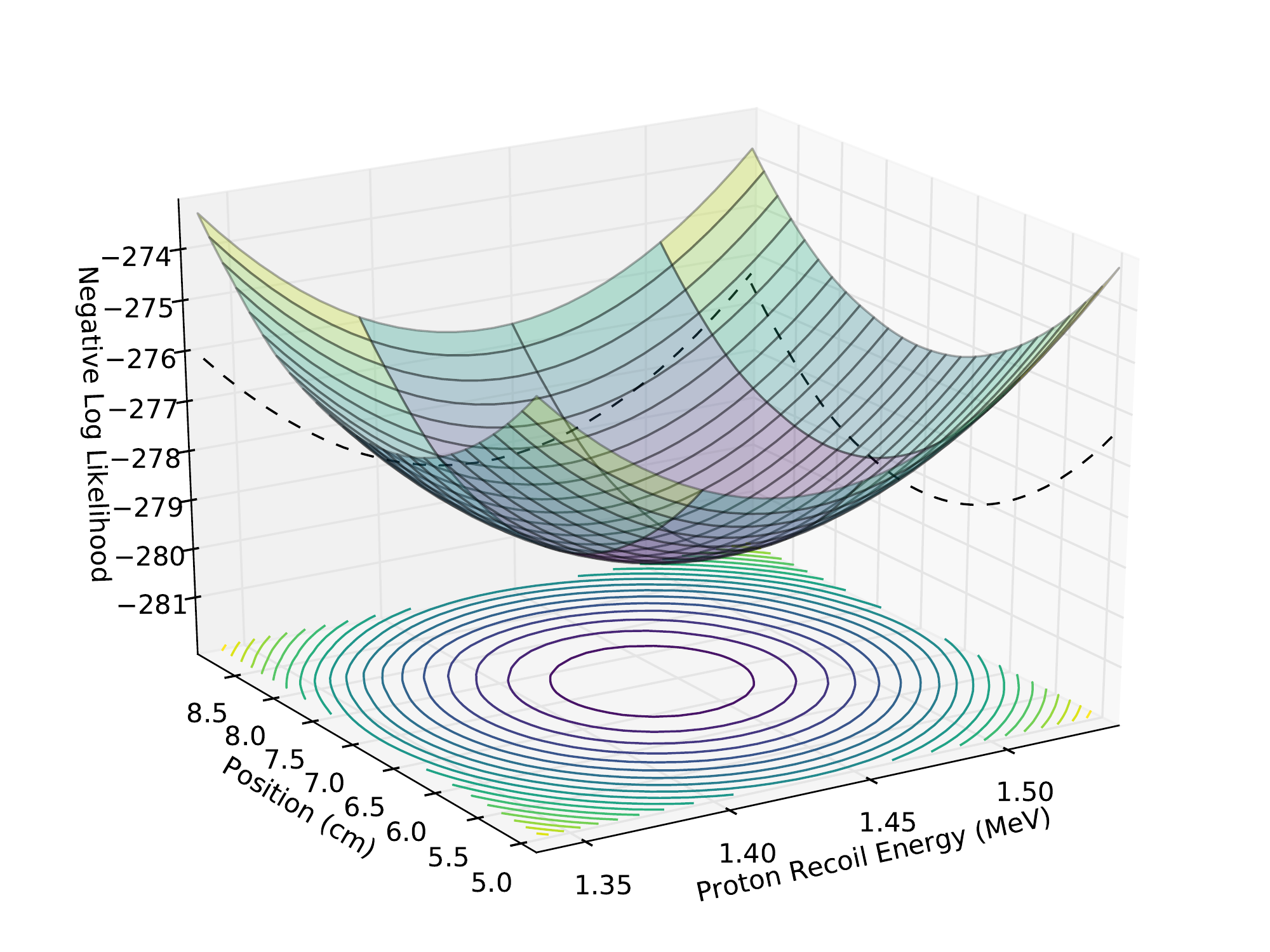}
  \caption{EJ-204/MCP-PM combination zoomed in on optimal solution}
  
\label{fig:3d_ej204_planacon_zoom}
\end{subfigure}\par\medskip
\begin{subfigure}[t]{0.45\textwidth}
  \centering
  \includegraphics[width=\columnwidth]{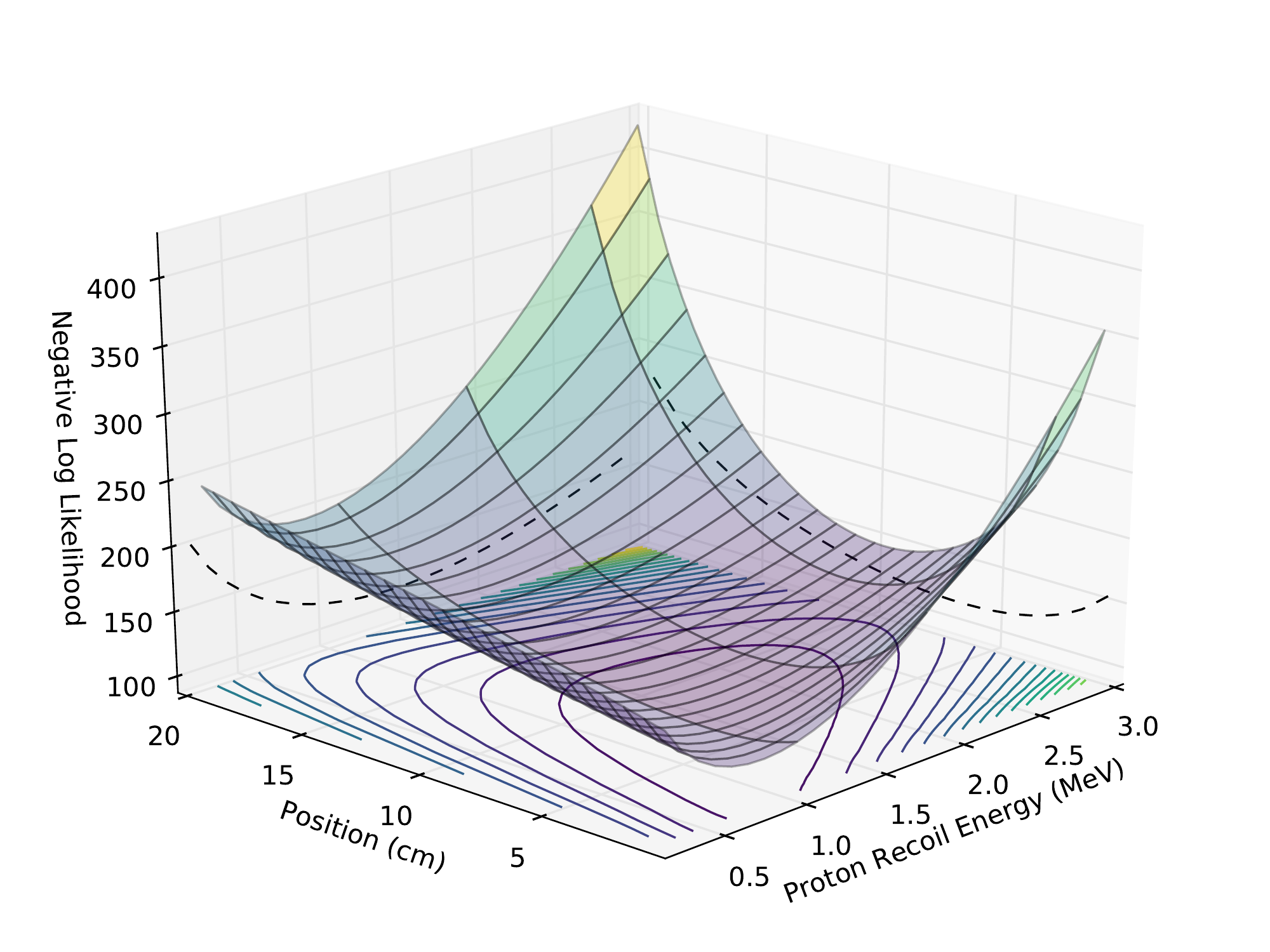}
  \caption{EJ-232Q/SiPM combination}
  
\label{fig:3d_ej232q_sipm}
\end{subfigure}
\hspace{2em}
\begin{subfigure}[t]{0.45\textwidth}
  \centering
  \includegraphics[width=\columnwidth]{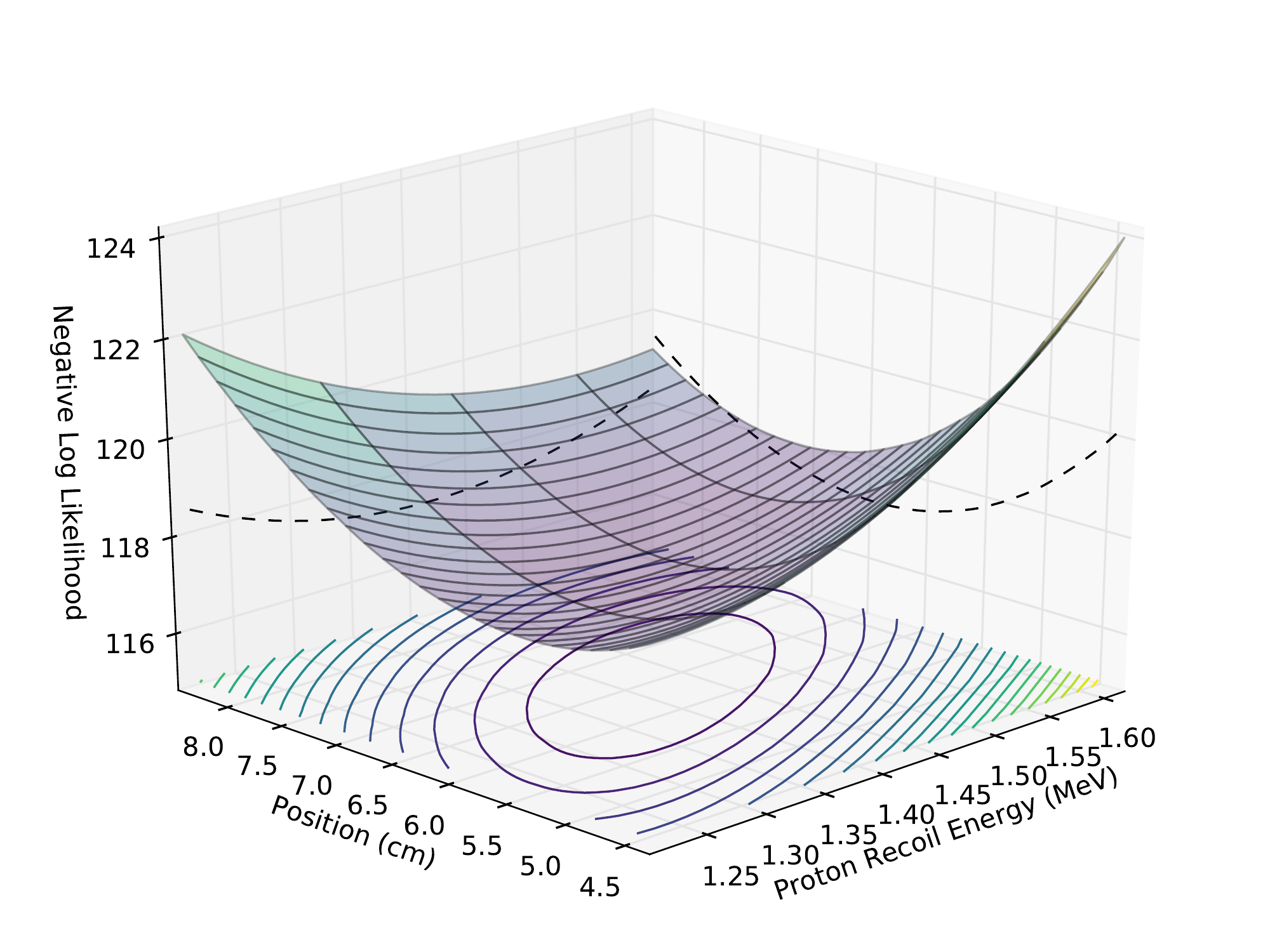}
  \caption{EJ-232Q/SiPM combination zoomed in on optimal solution}
\label{fig:3d_ej232q_sipm_zoom}
\end{subfigure}
\begin{subfigure}[t]{0.45\textwidth}
  \centering
  \includegraphics[width=\columnwidth]{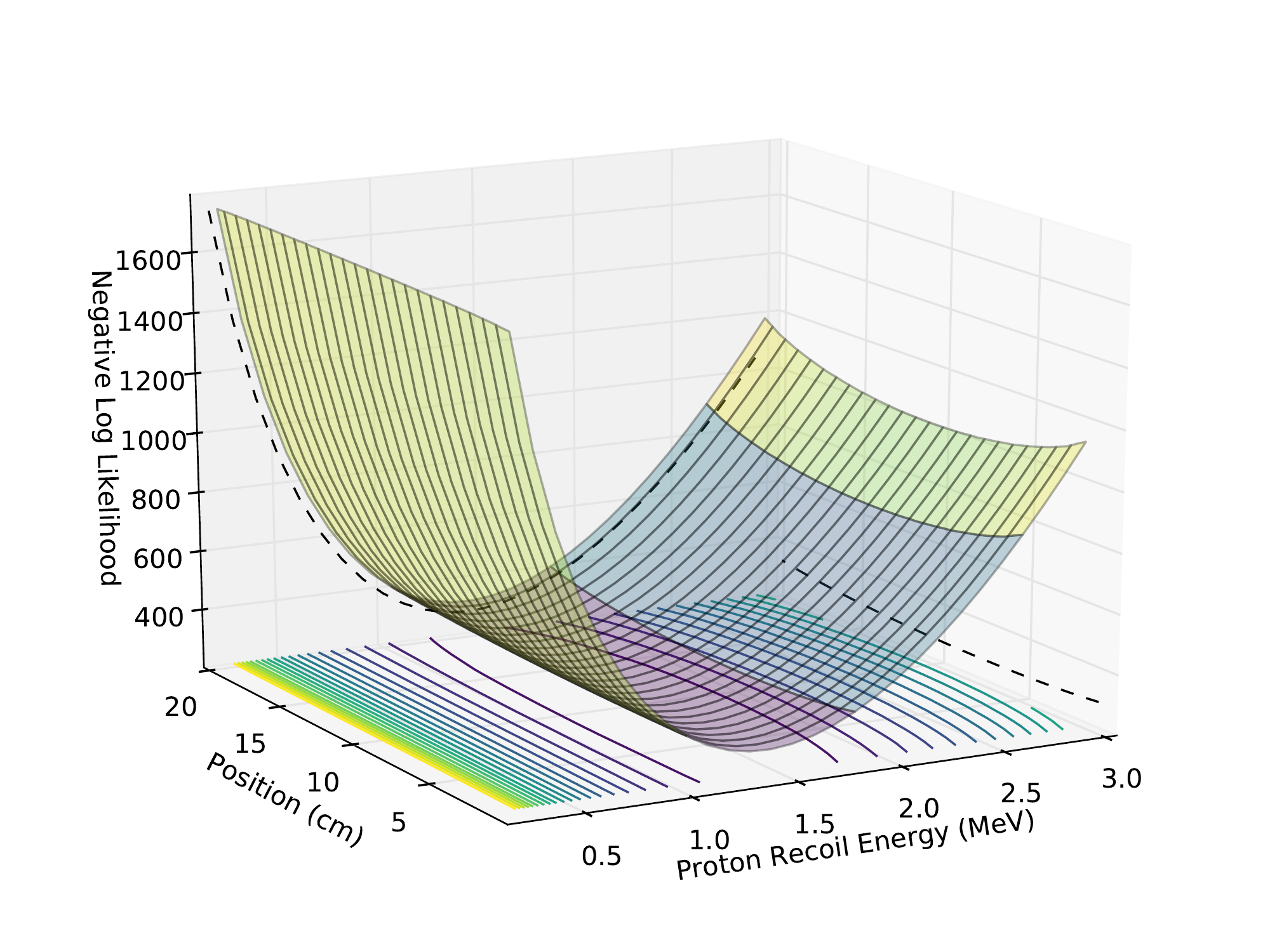}
  \caption{Stilbene/SiPM combination}
\label{fig:3d_stilbene_sipm}
\end{subfigure}
\begin{subfigure}[t]{0.45\textwidth}
  \centering
  \includegraphics[width=\columnwidth]{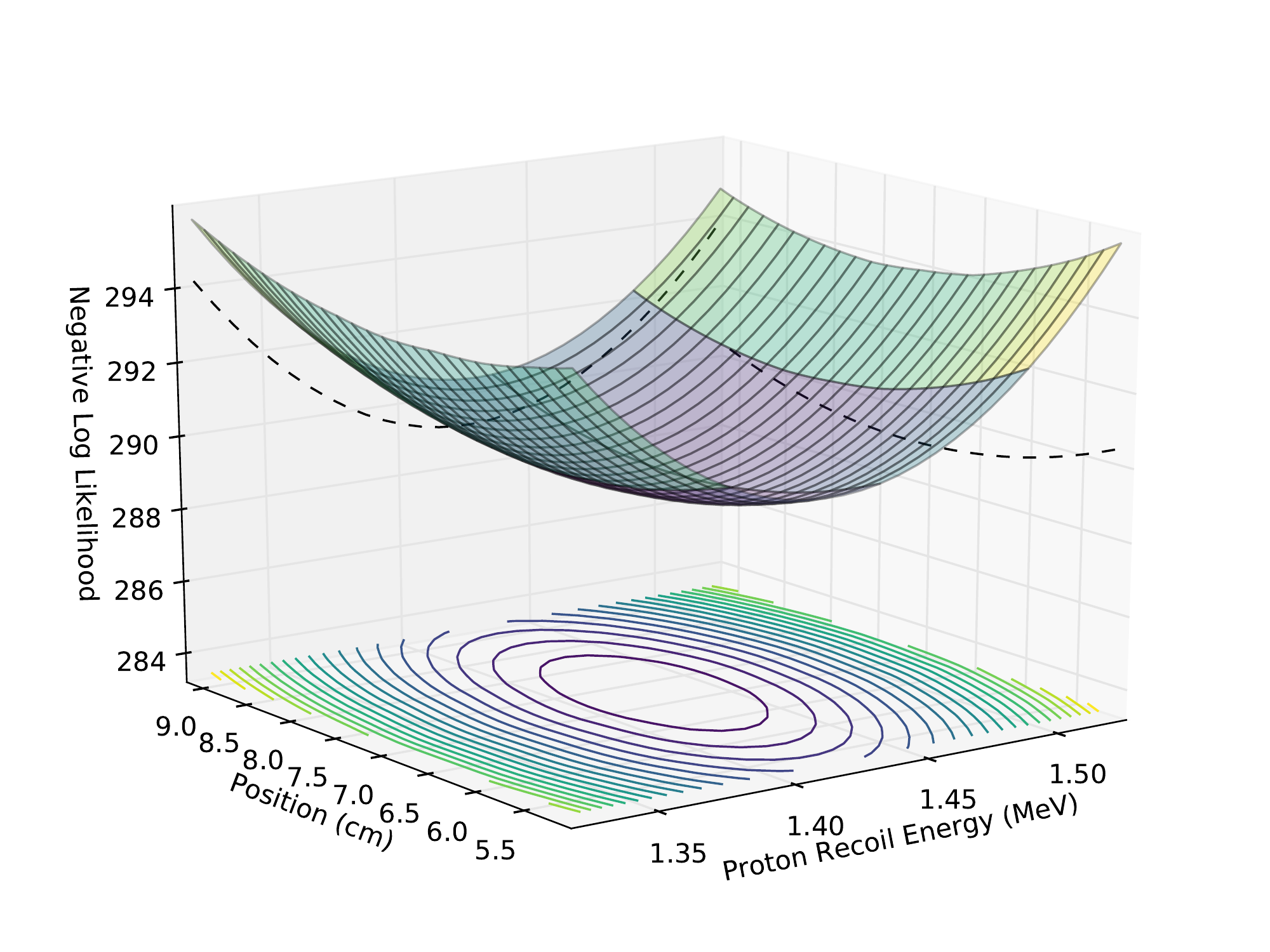}
  \caption{Stilbene/SiPM combination zoomed in on optimal solution}
\label{fig:3d_stilbene_sipm_zoom}
\end{subfigure}
\caption[]{3-D NLL surfaces created by iterating through scintillation position and proton recoil energy for a 1 cm x 1 cm x 20 cm pillar. Subfigures show different scintillator/photodetector combinations. For all combinations, the cost function surface exhibits a single global minimum with no local minima. Proton recoil energy estimation is more precise due to the steep decent towards the optimal solution; however, we observe a near-optimal solution (shallow slope) for estimating scintillation position which produces less precise estimates of scintillation position.}
\label{fig:NLL_plots}
\end{figure*}

We performed a more in-depth analysis of how each combination behaves by examining the cost function surfaces associated  with each case starting with the best combination of EJ-204/MCP-PM in Figure \ref{fig:3d_ej204_planacon}. For all NLL surfaces in this section, the true position is at 7 cm with a proton recoil energy of 1.4 MeV.

The figure shows the cost function surface NLL at different positions and proton recoil energies throughout the pillar. The dashed lines on the x-z and y-z planes are 2-D NLL values that pass through the best estimate of position and proton recoil energy. These iso-lines show the behavior of the cost function NLL when iterating over one variable, keeping the other constant.  Examining the curvature of the iso-lines through the best estimate, we arrive at a best estimate of proton recoil energy quickly (i.e., in few iterations) as the descent to the best estimate is steep. Conversely, we observe a shallow slope of the cost function surface as we vary position. The shallow slope indicates we have a near-optimal solution for a large range of scintillation positions. A nearly flat surface increases the number of cost function evaluations to find the minimum.

The overall behavior of the cost function surface exhibits a single global minimum with no local minima. The NLL value varies rapidly with changing proton recoil energy. Conversely, the NLL value varies more slowly with changing scintillation position. 


Figure \ref{fig:3d_ej204_planacon_zoom} shows a zoomed-in image of Figure \ref{fig:3d_ej204_planacon} in the vicinity of best fit. It shows a well-defined, quadratic surface near the best fit for both proton recoil energy and position. This indicates that a short scintillation pulse width and fast photodetector impulse response lead to a well behaved negative log likelihood surface. This event reconstructed scintillation position to about 7.5 cm and 1.45 MeV proton recoil energy.

The NLL cost function surface for the EJ-232Q/SiPM combination is shown in Figure \ref{fig:3d_ej232q_sipm}. The behavior is substantially different from the previous combination of EJ-204/MCP-PM. EJ-232Q has a much faster scintilltor time response, but it produces less than one third of scintillation light compared to the other scintillators, and it has an attenuation length of 8 cm. This results in tens of charge carriers created at the photodetector rather than hundreds.
A shallow and nearly optimal solution exists for a range of scintillation positions and proton recoil energies. This demonstrates that position and proton recoil energy have comparable effects on measured waveforms at a very low number of charge carriers. Recall the true scintillation position and proton recoil energy is 7 cm and 1.4 MeV respectively. Lower charge carrier counts on both photodetectors made it difficult to estimate scintillation position and proton recoil energy precisely.

Figure \ref{fig:3d_ej232q_sipm_zoom} shows a zoomed in view of the NLL surface near the optimal solution. This combination results in a different behavior near the global minimum compared to Figure \ref{fig:3d_ej204_planacon_zoom}. The NLL surface results in larger uncertainty in both position and proton recoil energy. This event's optimal fit was approximately 6.5 cm and 1.45 MeV.

The third and last combination of scintillator and photodetector uses stilbene/SiPM shown in Figure \ref{fig:3d_stilbene_sipm}. The resulting NLL surface looks similar to the first combination of EJ-204/MCP-PM. Stilbene has a luminosity similar to EJ-204, but it is more widely distributed in time. Also recall that SiPMs have a higher PDE resulting in more charge carriers created in the photodetector. Previous results have shown this enables slightly more precise proton recoil energy estimation, but less precise position estimation. Examining the dashed 2-D cost function contours in Figure \ref{fig:3d_stilbene_sipm}, the dashed line for proton recoil energy has a steep slope to the minimum value of the NLL cost function. In contrast, the NLL is nearly flat with position, which is why the position can only be estimated imprecisely using stilbene/SiPM.

Zooming in on the optimal fit location in Figure \ref{fig:3d_stilbene_sipm_zoom}, we see a very distinct minimum in proton recoil energy with a more elongated basin for estimating the best position. This is expected due to the time spread of charge from both stilbene and the SiPM.

\subsection{Precision of Position Dependent Reconstruction}

\begin{figure}[h!!]
\centering
  \includegraphics[width=\columnwidth]{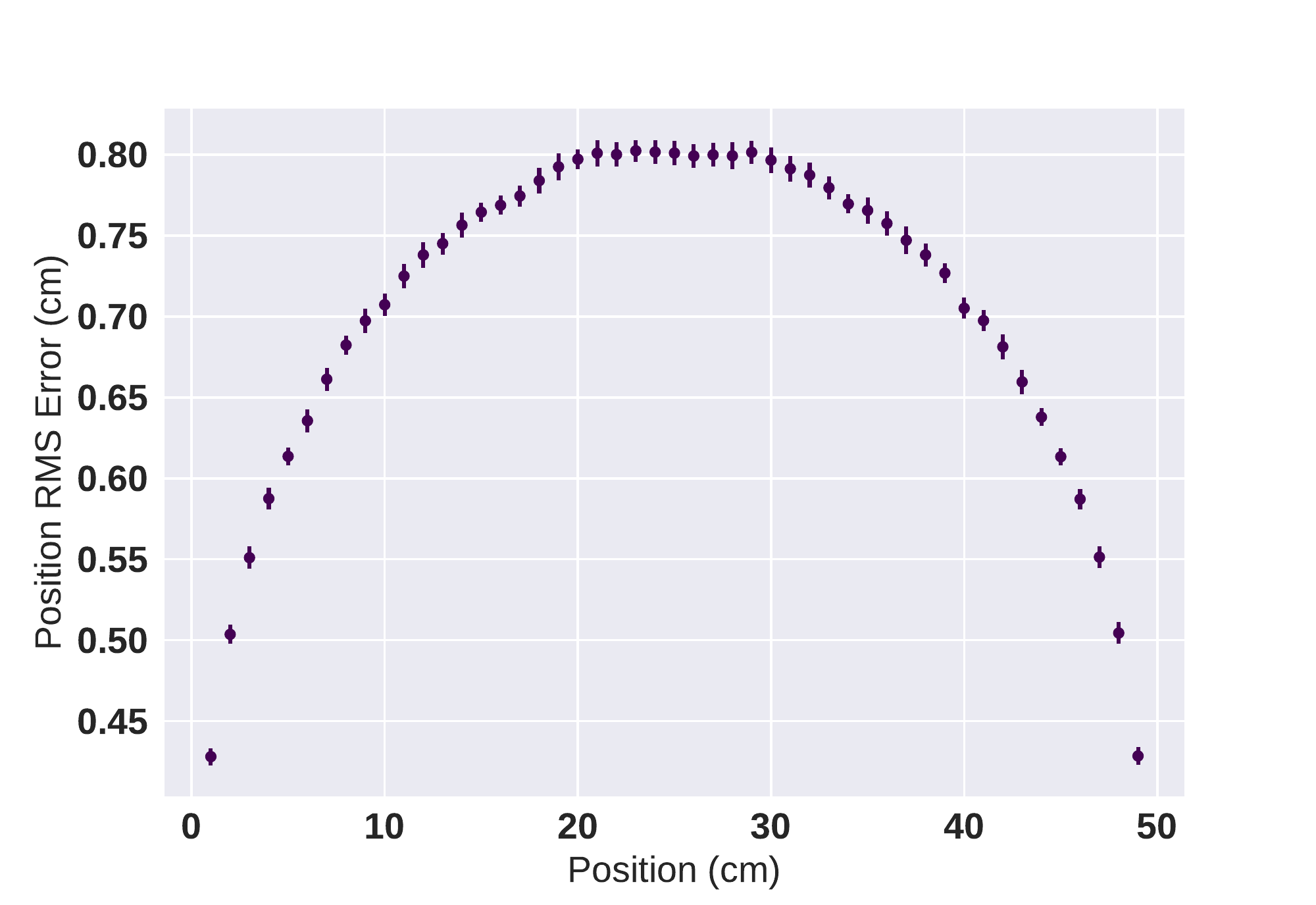}
  \caption{ Position RMS errors across the length of the pillar for a 1 cm x 1 cm x 50 cm pillar. RMS errors decrease near the photodetectors due to an increase in light collection efficiency in conjunction with edge effects.}
  \label{fig:acrossBarRMS}
\end{figure}

Light collection efficiency is highest when scintillation occurs near a photodetector, and it is lowest in the center of the pillar. To determine if this affects our estimate of scintillation position,  we calculated position RMS errors as a function of position in the pillar. We used a 1 cm x 1 cm x 50 cm EJ-204 pillar with a 2 MeV proton recoil energy for 10,000 events in 1 cm increments. These results are displayed in Figure \ref{fig:acrossBarRMS}. Recall that we bounded scintillation position reconstruction to within our tabulated pillar time spread histograms 0.5 cm away from the pillar ends. This reduced the position reconstruction error near the edges of the pillar. To reduce this edge effect, we simulated a longer pillar to examine if collection efficiency influences position reconstruction. Reconstruction results are shown in Figure \ref{fig:acrossBarRMS}.

 As expected, the position RMS error near the photodetectors is lowest due to edge effects and higher collection efficiencies. The RMS error in the center of the pillar is roughly double the error near the ends of the pillar. RMS error is symmetric about the center of the pillar due to the symmetry of the pillar geometry. Neutron  scatters near photodetectors will have more accurate pointing vectors and scintillation position estimates resulting in more accurate back-projected cones. 

\subsection{Reconstruction Results}

\begin{figure}[h!!]
\centering
  \includegraphics[width=\columnwidth]{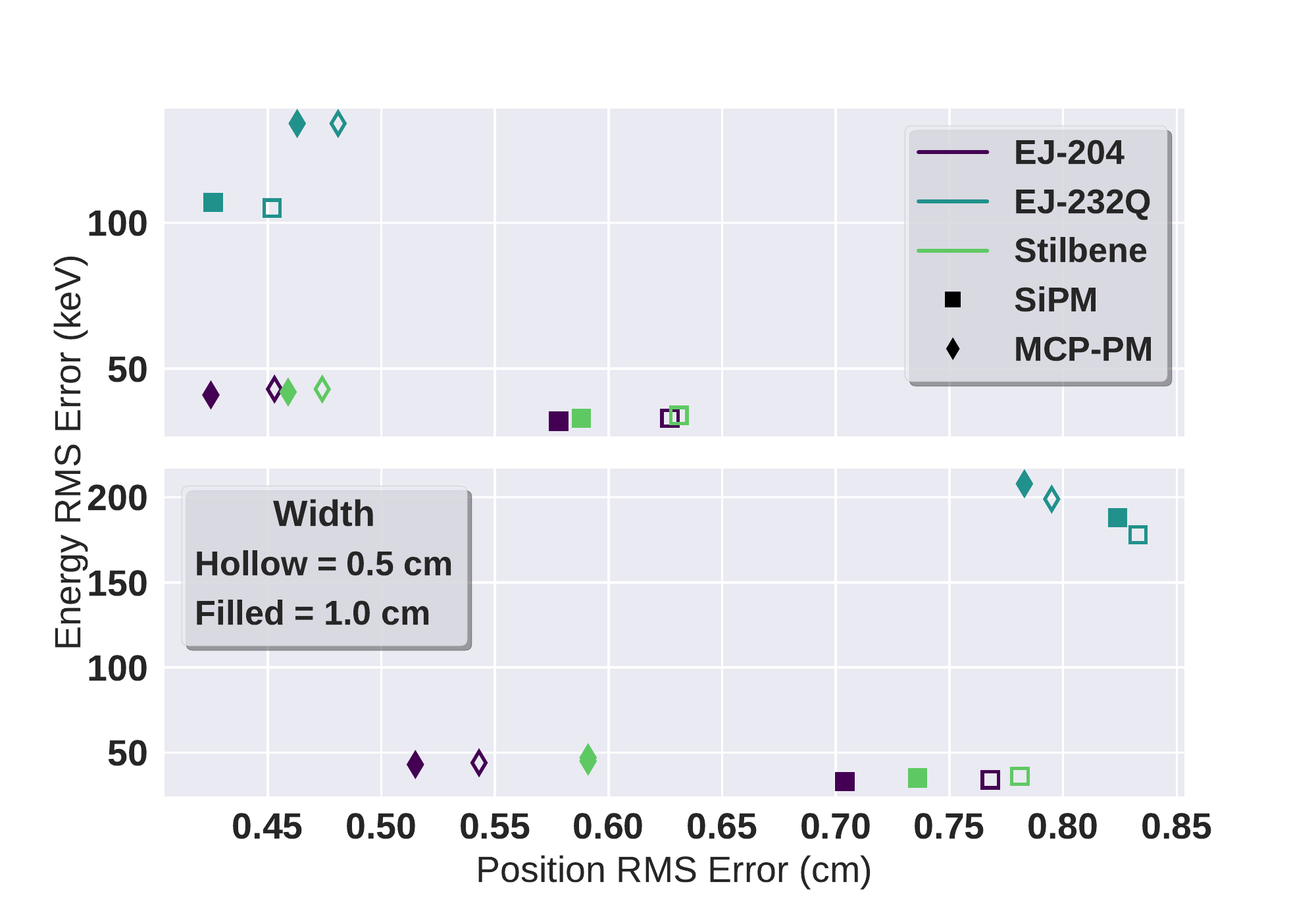}
  \caption{Reconstructed position and proton recoil energy RMS errors for a 2 MeV proton recoil energy. The top of the figure shows results from a 10 cm pillar length and the bottom shows results from a 20 cm pillar length. The same data is displayed in Tables \ref{table:10cm_2MeV} and \ref{table:20cm_2MeV}}
  \label{fig:2MeV_tableFig}
\end{figure}

\begin{figure}[h!!]
\centering
  \includegraphics[width=\columnwidth]{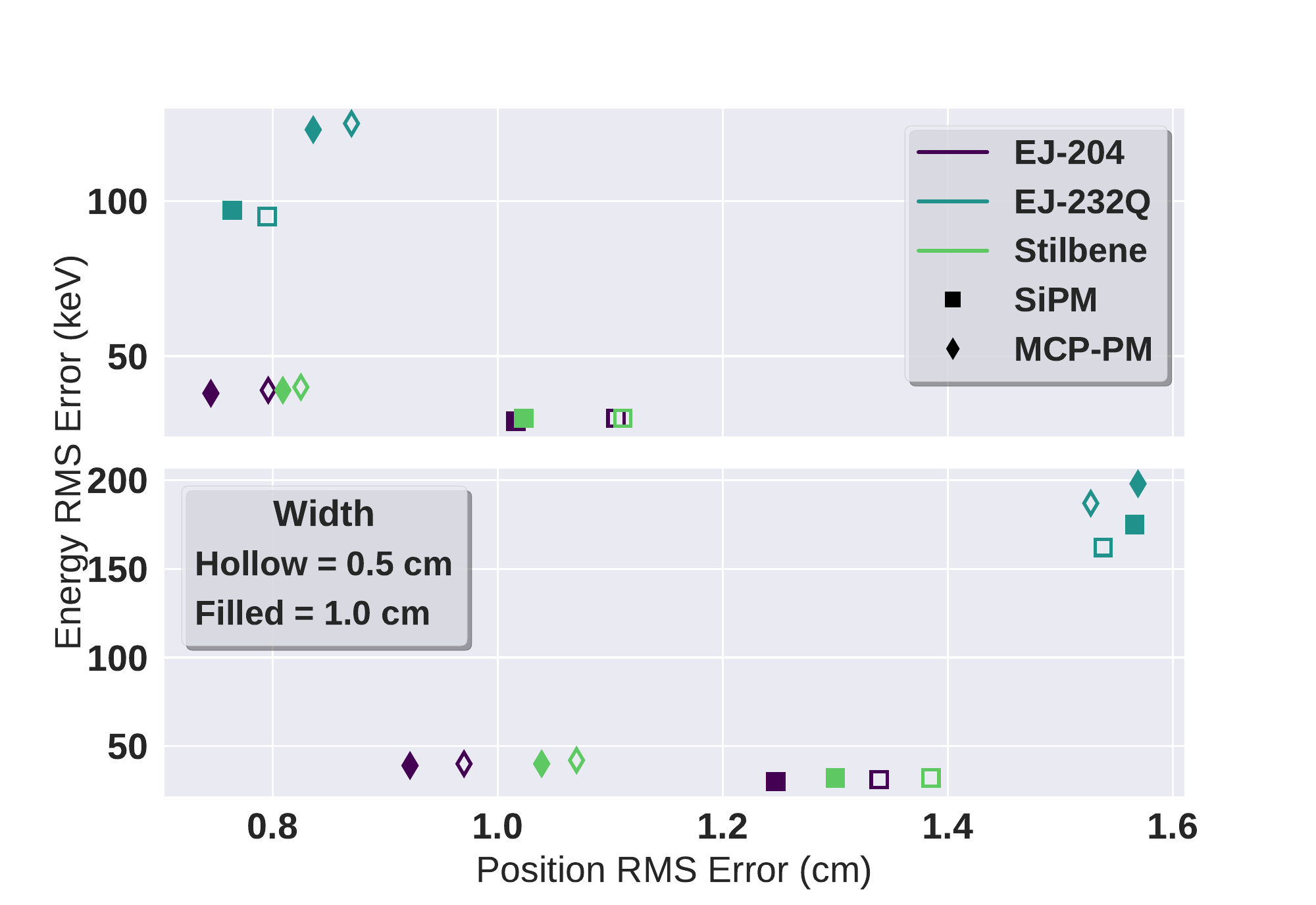}
  \caption{Reconstructed position and proton recoil energy RMS errors for a 1 MeV proton recoil energy. The top of the figure shows results from a 10 cm pillar length and the bottom shows results from a 20 cm pillar length. The same data is displayed in Tables \ref{table:10cm_1MeV} and \ref{table:20cm_1MeV}}
  \label{fig:1MeV_tableFig}
\end{figure}

We tabulated position and proton recoil energy RMS errors for 1 MeV and 2 MeV proton recoil energies with 12 different scintilltor/photodetector combinations. The results are tabulated in Appendix A in Tables \ref{table:10cm_1MeV}, \ref{table:20cm_1MeV}, \ref{table:10cm_2MeV} and \ref{table:20cm_2MeV}. Results are also displayed, for ease of viewing, in Figures \ref{fig:1MeV_tableFig} and \ref{fig:2MeV_tableFig} for 1 MeV and 2 MeV proton recoil energies respectively. The results shown used 10,000 scintillation events uniformly distributed throughout the pillar length.

We will first discuss the brighter events in Figure \ref{fig:2MeV_tableFig}. The top of the figure shows the position and energy RMS errors for a 10 cm pillar, and the bottom shows results for a 20 cm pillar. Examining both lengths, the width of the scintillator pillar does not have a substantial impact on the ability to reconstruct scintillation position and energy precisely. The effect observed in Figure \ref{fig:sideSizeChannelTimeSpread} produces less than a 10\% increase in scintillation location reconstruction. For cone back-projection, we assumed the (x,y) position of scintillation was located at the center of the illuminated pillar. Increasing pillar width increases the uncertainty in position in (x,y). This negatively impacts the angular resolution of the image reconstruction. 

Examining the top of Figure \ref{fig:2MeV_tableFig} for a 10 cm pillar length, all combinations using the MCP-PM resulted in similar position RMS errors around 0.45 cm. Even though EJ-232Q emits much less light than the other two scintillators, the fast scintillator time response compensates  to allow more accurate position reconstruction. Consequently, with poor photostatistics, the proton recoil energy reconstruction is about 3.5 times more imprecise for the quenched plastic. Examining EJ-204 and stilbene combinations, they produce similar energy RMS error with the SiPM very slightly outperforming with approximately a 30\% decrease in RMS error. The position RMS error is larger, by almost 50\%, due to the temporal spread characteristic of an SiPM. 

For the 20 cm pillar length on the bottom of Figure \ref{fig:2MeV_tableFig}, we see a similar trend for most combinations of scintillator and photodetector. Overall, there is about a 20\% increase in position RMS error and 10\% increase in proton recoil energy RMS error.  However, we see a substantial discrepancy in reconstruction precision for all combinations using EJ-232Q. Instead of achieving one of the best position RMS errors, it has the poorest. This is due to the short self-attenuation length of the scintillator. Increased photon transit distance substantially reduces the number of charge carriers created in the photodetectors, producing larger uncertainties in both reconstructed variables.

We also investigated the performance of all combinations at 1 MeV proton recoil energy to determine if reconstruction results varied as a function of proton recoil energy. Results are shown in Figure \ref{fig:1MeV_tableFig}. First, notice the almost 100\% increase in position RMS error for all combinations due to poor photostatistics. It is important to note that the energy RMS error stayed approximately constant for both energies. However, percent RMS error is inversely proportional to proton recoil energy. For example, an RMS error of 40 keV for a 1 MeV proton recoil results in a 4\% uncertainty in the proton recoil energy. If we have a similar 40 keV RMS error for a 2 MeV proton recoil energy, our uncertainty is halved to 2\% uncertainty. We observed relatively insubstantial changes to the overall trend of all combinations.  Again, 1 cm pillar widths slightly outperformed the 0.5 cm widths. EJ-232Q reconstructed position precisely in a 10 cm pillar, but very imprecisely in a 20 cm pillar. This is a good indicator that the relative performance of the scintillator and photodetector combinations does not vary as a function of proton recoil energy.

\begin{figure}[h!!]
\centering
  \includegraphics[width=\columnwidth]{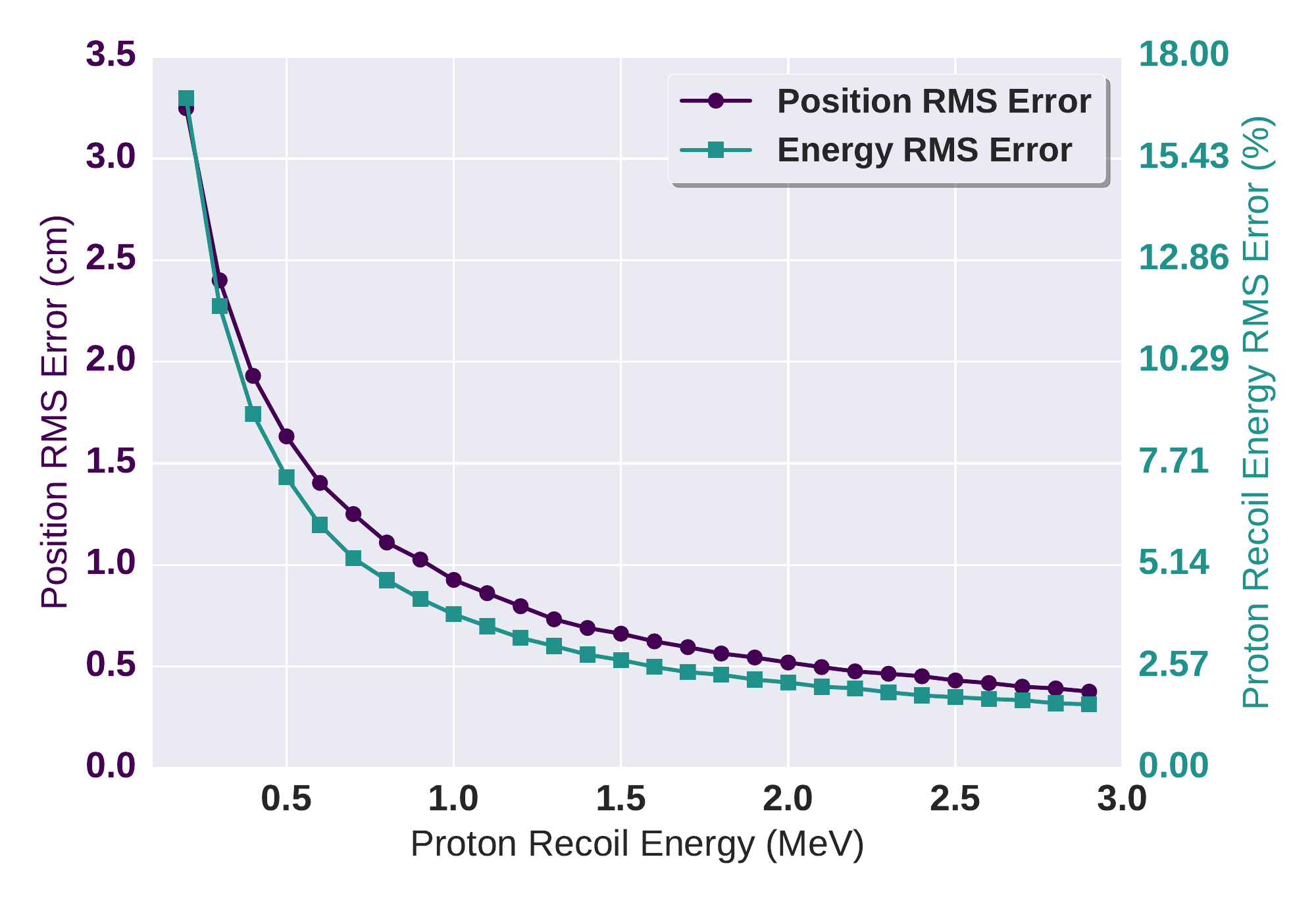}
  \caption{Reconstructed position and proton recoil energy RMS error as a function of recoil energy. Proton recoil energy RMS error stays approximately constant at 40 keV over the range investigated.}
  \label{fig:protonRecoilEnergyStudy}
\end{figure}

In Figure \ref{fig:EffComparison}, we simulated SVSC-PiPS and NSC devices. The results exhibited an order of magnitude increase in efficiency for neutrons incident on an SVSC-PiPS device. However, the uncertainty in scintillation position increased for all combinations when we reduced proton recoil energy from 2 MeV to 1 MeV for the SVSC-PiPS device. To quantify this increase, we computed RMS errors as a function of proton recoil energy for the best combination of an EJ-204 scintillator and MCP-PM. Pillar dimensions were set to 1 cm x 1 cm x 20 cm for this simulation. Figure \ref{fig:protonRecoilEnergyStudy} shows the results of this study using 10,000 uniformly distributed scintillation events for each proton recoil energy. 

Proton recoil energies ranged from 0.2 MeV through 2.9 MeV in 0.1 MeV increments. At 0.2 MeV proton recoil energy, we calculated a position RMS error of 3.25 cm. This gives us an uncertainty on the scintillation position of approximately one-third the length of the pillar. We also calculated a proton recoil RMS error at 18\%; this is equivalent to 34 keV. Position RMS error decreases with diminishing returns around 1.5 MeV. This saturation effect may be caused by the neutron light output function. Around 1.5 MeV proton recoil, neutron light production becomes approximately linear as a function of proton recoil energy. At the highest proton recoil energy of 2.9 MeV, we calculated a position RMS error of 0.35 cm with a proton recoil energy under 2\%.

\subsection{Position, Energy and Time Reconstruction Conclusions}
Figures \ref{fig:1MeV_tableFig} and \ref{fig:2MeV_tableFig} and Tables \ref{table:10cm_1MeV} - \ref{table:20cm_2MeV} reiterate the same conclusions shown in Figures \ref{fig:rmsErrorPosHist} - \ref{fig:NLL_plots}. The most substantial impact on estimating scintillation position is the temporal spread of charge carriers. Large spreads result in increased uncertainty in scintillation position. The number of charge carriers in the photodetector governs proton recoil energy reconstruction precision. 

Calculation of scintillation position using light intensity of opposing photodetectors resulted in an RMS error of approximately 2.3 cm and 1.3 cm for 1 MeV and 2 MeV proton recoils (shown in Figure \ref{fig:lightIntensityEstimate}). A comparison of best estimated scintillation position utilizing MLEM produced RMS errors of 0.92 and 0.52 cm for 1 and 2 MeV proton recoils. Estimating scintillation position using light intensity result in 250\% larger RMS error. 

Allowing scintillation time to be a free floating variable in the fit produced less than 10\% increase in position RMS error with a timing RMS error of less than 100 ps for proton recoil energies above 1 MeV.

\section{Source Localization}
In this section we will estimate the image resolution of the device using the best scintillator and photodetector combination of EJ-204/MCP-PM with MCNPX-PoliMi. Back-projected images are characteristic of each imaging device and is a good comparison metric between different imagers. Back-projection image resolution is not an estimate of the resolution on the location of a point source. The position RMS error improves at higher proton recoil energies; therefore, we estimated the best back-projection image resolution at a threshold of 1.5 MeV (for both neutron scatters) of $\sigma_{\theta}=19.51^\circ$ and $\sigma_{\phi}=15.67^\circ$. Using light intensity to estimate scintillation position produced a poorer detector resolution.

\subsection{MCNPX-PoliMi Simulation}
To calculate the resolution of an SVSC-PiPS device, we simulated a Cf-252 point source at multiple locations surrounding a 20 cm x 20 cm x 20 cm EJ-204 SVSC-PiPS device. We placed a Cf-252 source at four different locations in polar and azimuth angle ($\theta$, $\phi$): (-45$^\circ$, 0$^\circ$), (0$^\circ$, 45$^\circ$), (0$^\circ$, 0$^\circ$), (-30$^\circ$, 50$^\circ$).

\noindent
These simulations consisted of fission energy neutrons emitted isotropically 1 m away from the center of the SVSC-PiPS device. We fixed pillar dimensions at 1 cm x 1 cm x 20 cm resulting in 400 pillars. We used MCNPX-PoliMi to simulate the transport of neutrons throughout the detector. We simulated the EJ-204/MCP-PM combination based on the results from the previous section. 


As previously described, MCNPX-PoliMi records neutron interaction state variables within specified detector ``cells''. We specified each pillar as a detector cell to return all collision information within the pillars. We used the collision information recorded by MCNPX-PoliMi and the optical photon transport methods described in Section~\ref{reconstructionMethod}. Neutrons that interacted multiple times in the same pillar were not used for image back-projection.  We used true proton recoil energy and location of interaction within the pillar to create randomly sampled observed photodetector waveforms. Subsequently, we found the best estimate of scintillation position and proton recoil energy using MLEM by comparing the nominal responses to the observed waveforms as shown in Section~\ref{positionReconstruction}.

\subsection{Back-Projected Images}
For clarity, we will reiterate the information needed to back-project cones. We need the first and second location of scatter in (x,y,z). We also need the proton recoil energy of the first scatter. Finally, we require the time of flight of between scintillation events. The back-projected cone has an apparent 'thickness' correlated to the uncertainty of the input variables: (x,y,z) position of interaction for both neutron scatters, time of interaction for both scatters and proton recoil energy from the first scatter. To quantify this thickness, we need to quantify the uncertainty of all inputs. Uncertainties for all quantities are given as one standard deviation, or 1$\sigma$.

Recall the (x,y) position cannot be estimated within a pillar; we assume all interactions occur in the center of the pillar in the (x,y) plane. Therefore, the uncertainty in (x,y) position is that of a uniform distribution shown in Equation \ref{stdXY}

\begin{align}
\label{stdXY}
\sigma_{x,y} = \sqrt{\dfrac{w^2}{12}}
\end{align}

\noindent
where $w$ is the width of the  pillar.
 
Uncertainty in z-position estimates depend upon proton recoil energy (shown in Figure \ref{fig:protonRecoilEnergyStudy}). We created a look up table to estimate the uncertainty as a function of proton recoil energy. We fixed proton recoil energy uncertainty to 40 keV; proton recoil RMS errors were nearly constant over a wide range of energies. Timing uncertainty for cone back-projection was set to a constant 100 ps for all proton recoil energies.

\begin{figure}[h!!]
\centering
  \includegraphics[width=\columnwidth]{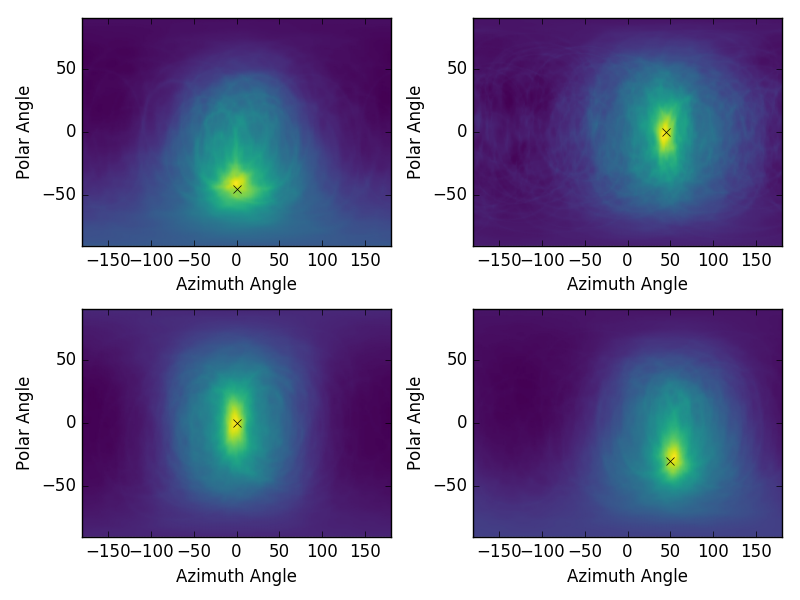}
  \caption{Plots of the sum of back-projected cones from a Cf-252 neutron source simulated at 1 meter distances from the center of the SVSC-PiPS device.  True source positions are indicated with an 'X' in each plot}
  \label{fig:backprojections}
\end{figure}

We propagated the uncertainty of the input parameters to the opening angle of the cone. Cone back-projections include interactions with incorrect pointing vectors. This includes events that first scatter off carbon then subsequently double scatter off hydrogen-1, and events where the neutron first scatters on hydrogen, then carbon, then hydrogen again. We set the detection threshold at 0.5 MeV proton recoil energy. Each back-projected scenario is shown in Figure \ref{fig:backprojections}. For all four source locations, the peak of the back-projected image coincide with the true source position. 
 
\subsection{Imager Resolution}

\begin{figure}
\centering
  \includegraphics[width=\columnwidth]{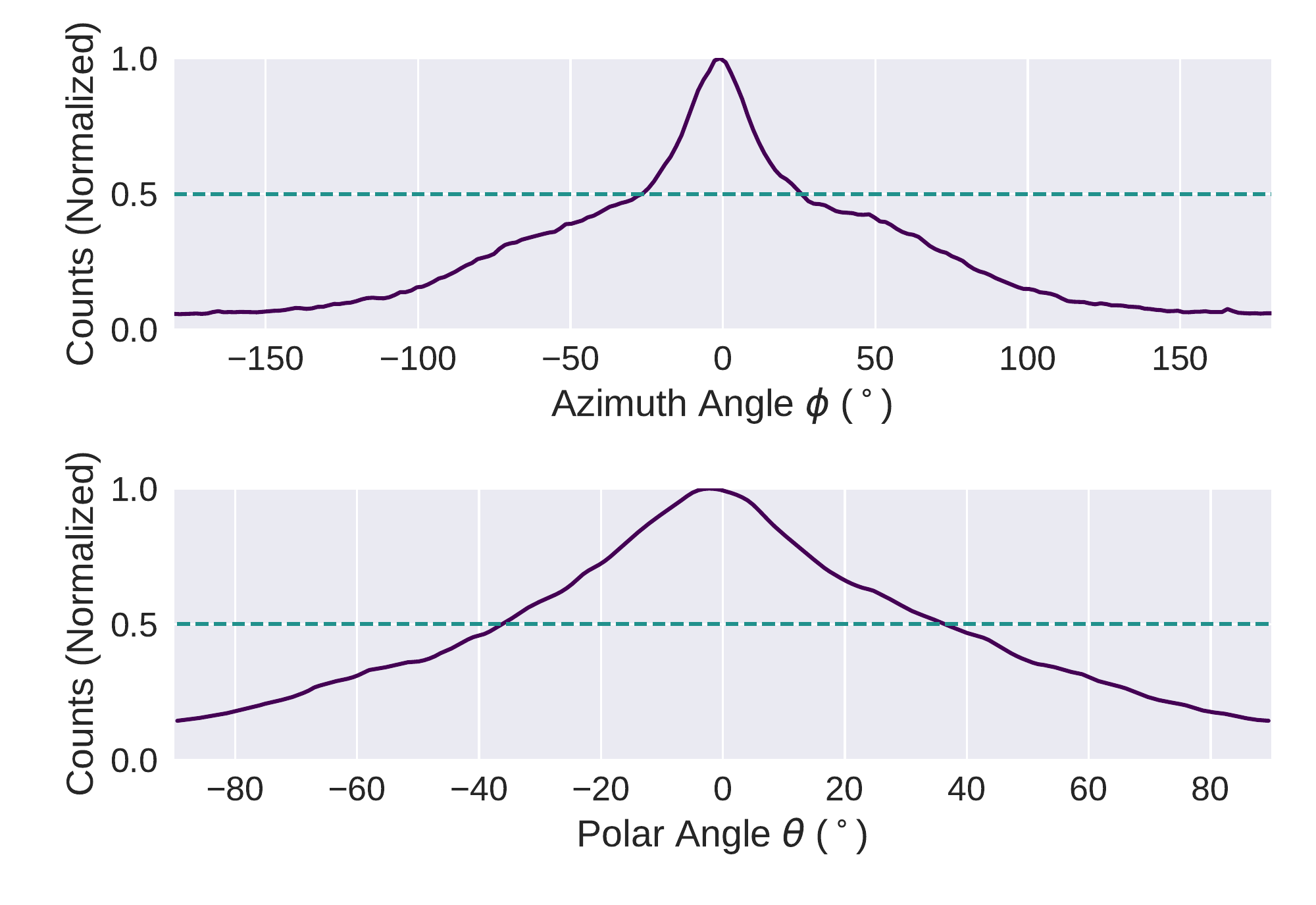}
  \caption{Normalized 2-D profiles of the back-projected image for a Cf-252 source located at (0$^\circ$, 0$^\circ$) a distance of 1 m from the center of the SVSC-PiPS device. }
  \label{fig:2dProfile}
\end{figure}

Using the back-projected cone image, we calculated the back-projected image resolution of a SVSC-PiPS device for a Cf-252 point source at (0$^\circ$, 0$^\circ$) at a distance of 1 m. We normalized the histogram to the maximum back-projected bin value. Then, we took a cut through the maximum bin in both polar and azimuthal directions to create a two dimensional profile of the back-projected image. Profiles are shown in Figure \ref{fig:2dProfile}. For each profile, we calculated the full width at half-maximum (FWHM) to estimate a 1$\sigma$ resolution for polar and azimuth directions. We assumed the profiles follow a normal distribution to directly compare to the Sandia National Laboratories neutron scatter camera back-projected image resolution of 10$^\circ$; $\sigma$ can be approximated using Equation \ref{eqn:FWHM}.

\begin{align}
\label{eqn:FWHM}
FWHM \approx 2.355 \sigma
\end{align}

\begin{figure}[h!!]
\centering
  \includegraphics[width=\columnwidth]{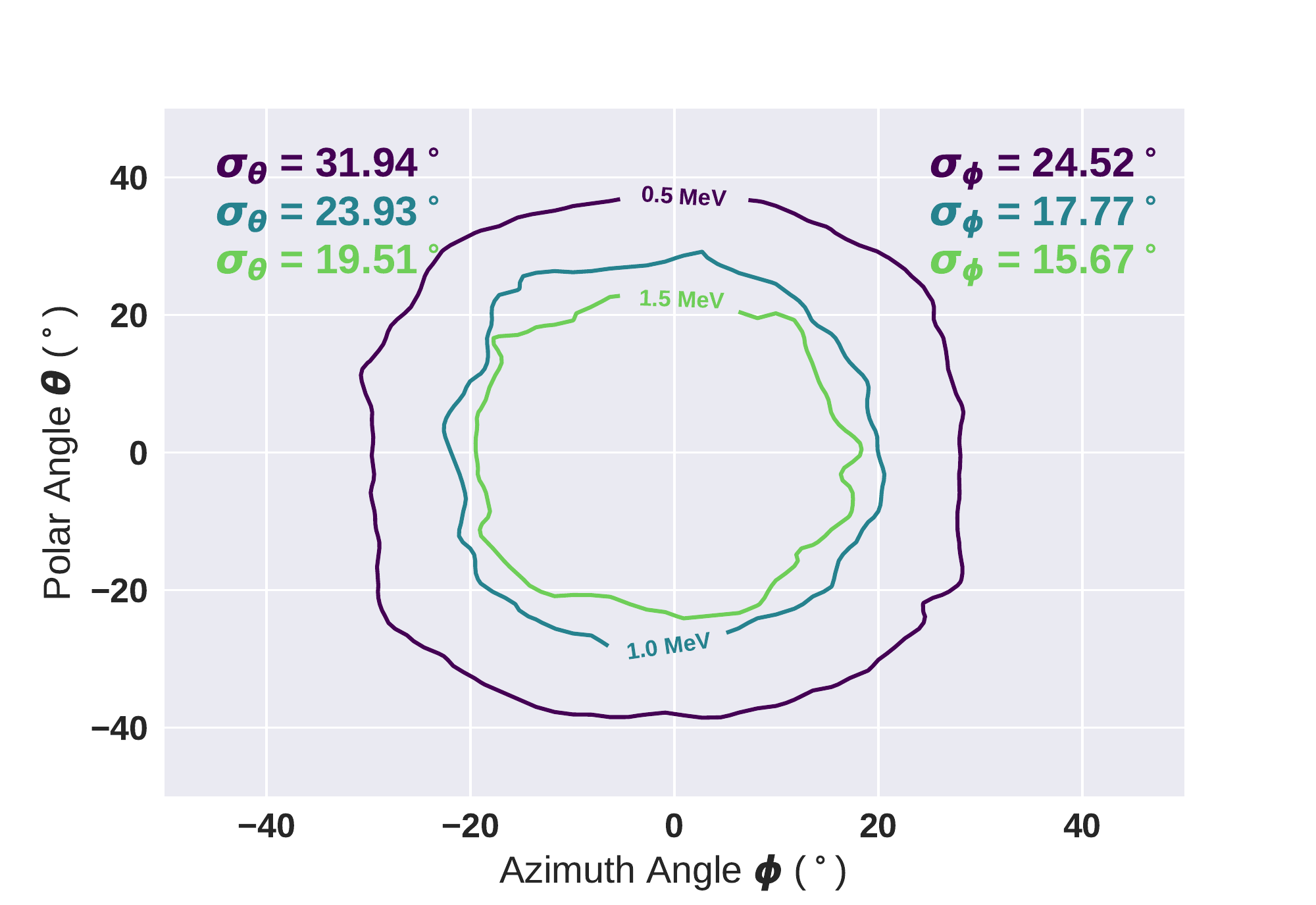}
  \caption{SVSC-PiPS FWHM contour lines for back-projected images using thresholds of 0.5, 1.0 and 1.5 MeV. Polar and azimuth resolutions for each threshold are shown in the figure}
  \label{fig:resolutions}
\end{figure}

Cone thickness is dependent upon proton recoil energy. There is less uncertainty in high energy proton recoil events. Therefore, high proton recoil energy events generate a sharper back-projected image. We can obtain a higher resolution at a reduced efficiency if we require both scatters to have higher energy proton recoils. An FWHM contour line for 0.5, 1 and 1.5 MeV proton recoil energy thresholds is shown in Figure \ref{fig:resolutions}. 

For Figure \ref{fig:resolutions}, we simulated $10^8$ isotropic neutrons with fission spectrum energy. We used the same reconstructed cone histogram for each threshold cut. The number of back-projected cones for each threshold is displayed in Table \ref{table:cones}.

\begin{table*}[t]
\centering
\caption{Number of cones back-projected for each image when estimating detector resolution}
\label{table:cones}
\begin{tabular}{cccc}
Threshold (MeV)   & Number of Cones & $\sigma_{\theta}$ & $\sigma_{\phi}$\\
\hline
0.5 & 32,337 & 31.94$^\circ$ & 24.52$^\circ$\\ 
1.0 &  8,112 & 23.93$^\circ$ & 17.77$^\circ$\\ 
1.5 &  2,367 & 19.51$^\circ$ & 15.67$^\circ$\\ 
\end{tabular}
\end{table*}

 Fewer cones than the numbers shown in Table~\ref{table:cones} are needed to estimate incident neutron direction; nonetheless, to obtain a similar number of back-projected cones at a 1~m source distance, $10^8$ neutrons are emitted from an IAEA significant quantity of plutonium (8 kg) in approximately 5 minutes.

 For the highest threshold of 1.5 MeV, we back-projected 2,367 cones of possible incident neutron direction. Using the FWHM, we calculated an SVSC-PiPS back projected image resolution of $\sigma_{\theta}=19.51^\circ$ and $\sigma_{\phi}=15.67^\circ$.
 
The limit on detector resolution is not dictated by the back-projected image. Image resolution can be improved using maximum likelihood estimation maximization to determine the most probable source location given a back projected image. An example of this process can be found in \cite{Brennan2010}.

\subsection{Back-Projected Images Using Light Intensity Position Estimation}

\begin{figure}
\centering
  \includegraphics[width=\columnwidth]{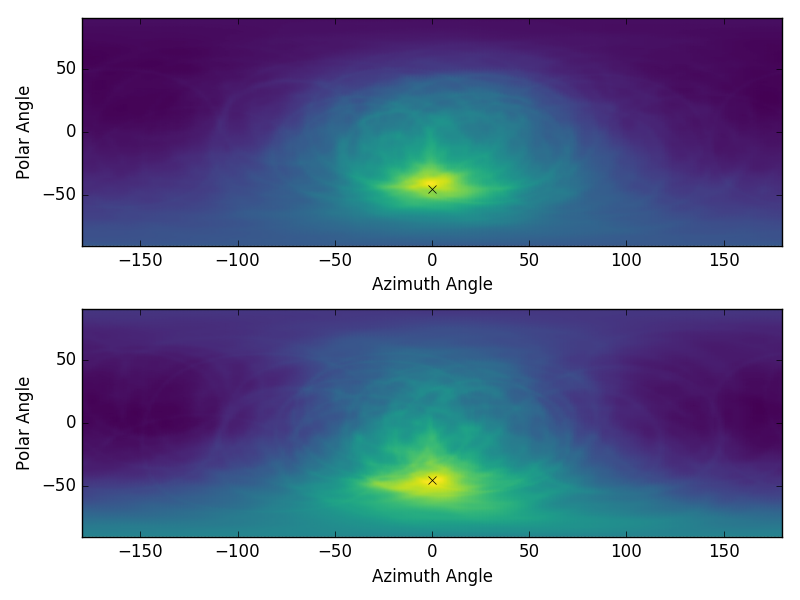}
  \caption{Comparison of cone back-projection images. The bottom half of the figure shows cone back-projection using the intensity of light on opposing photodetectors to estimate scintillation position. The top half of uses the scintillation estimation technique described in this paper. True source positions are indicated with an 'X' in each plot}
  \label{fig:intensityBackPro}
\end{figure}

We directly compared back-projected images using two different scintillation position reconstruction techniques. The first technique was shown in Figure \ref{fig:lightIntensityEstimate}. To reiterate, this technique estimates scintillation position using the ratio of light intensity on opposing photodetectors. The second technique, earlier in the paper, compares observed photodetector responses to nominal responses using MLEM. These results are shown in Figure \ref{fig:intensityBackPro}. The top of the figure uses the MLEM technique and bottom of the figure uses light intensity to estimate scintillation position. We created both images using the same collision data log from MCNPX-PoliMi. Consequently, we back-projected the same number of cones.


We performed a similar simulation shown in Figure \ref{fig:protonRecoilEnergyStudy} to ascertain the uncertainty in scintillation position using the ratio of light intensity on opposing photodetectors. We tabulated scintillation position RMS errors to create a look up table at different proton recoil energies. We used the look up table as an input parameter to cone back-projections for z-position estimates. Examining Figure \ref{fig:intensityBackPro}, larger uncertainties in scintillation location (using light ratio estimates) produces a smeared out back-projected image. This is illustrated in the bottom figure. We calculated a poorer resolution for both polar and azimuth angles when using light intensity to estimate scintillation position. Light intensity estimates produced $\sigma_{\theta} \approx 40^\circ$ and $\sigma_{\phi}\approx 47^\circ$, an increase in back-projected image resolution of 9$^\circ$ and 23$^\circ$ for polar and azimuth angle respectively.

\section{Conclusions}
In this paper, we modeled the performance of a compact fast neutron scatter camera comprising pillars of organic scintillator. Using a contiguous active volume increases detection efficiency by an order of magnitude compared to the multivolume neutron scatter camera developed at Sandia National Laboratories. We estimated scintillation position and proton recoil energy using the intensity vs. time history of charge carriers emitted by the photodetectors at each end of the pillar.

This pillar-based design resolves some of the inherent difficulties in implementing a monolithic single volume scatter camera. The monolithic design comprises a ``block'' of scintillator with pixelated photodetectors attached to (at least) two opposing sides of the cube. When the scintillator ``block'' is contiguous, all photodetector pixels must be digitized to estimate position (in three dimensions), scintillation time, and intensity; this requires large data throughput. Scintillation light from the first interaction overlaps in time with light from the second interaction and subsequent interactions. This creates difficulty in identification of which elastic scatter the scintillation light originated. 

In contrast, the optically segmented design presented confines the light to a single pillar for each interaction. Optical isolation enables independent analysis for each elastic scatter; As stated previously, the user only needs to estimate the z-position (along the length of the pillar) of interaction rather than (x, y, z). Additional scatters can be ignored. Optical isolation significantly reduces the number of channels digitized to just four, the two opposing anode channels for both illuminated pillars. The anode signals from the photodetectors can be sent through a switchyard of discriminators to determine which four channels need to be digitized.

A single volume scatter camera has the inherent difficulty of neutron/gamma discrimination. One could use a pulse shape discrimination capable scintillator to discriminate on particle type or require a minimum separation between interactions and use time of flight to discriminate between particles.

We used the light transport module in Geant4 to evaluate light collection efficiency for different reflectors. ESR film was the best choice for the reflector material with light collection efficiency of approximately 45\%  when scintillation light was emitted 5 cm from one end of the pillar. A 1 mm air gap surrounding the scintillator pillar enabled for total internal reflection of scintillation light.

 Iterating through combinations of pillar dimensions, scintillator type and photodetector type, we observed that scintillation position reconstruction is highly dependent on the time spread of charge carriers in the photodetector; we also found that position reconstruction is more precise near the ends of the pillar. The pillar exhibited higher collection efficiencies near photodetectors (lowest collection efficiency at the center of the pillar).  The combination of EJ-204 and an MCP-PM resulted in the lowest reconstructed position RMS errors for all detector sizes simulated. This design produced a position RMS error of 0.52 cm for 2 MeV proton recoil energy and a 20 cm pillar; Small changes in pillar width proved negligible when estimating scintillation position and proton recoil energy. 

Reconstructed proton recoil energy RMS error is dependent upon the number of charge carriers produced in the photodetectors. All detector/scintillator combinations (except for ones including EJ-232Q) reconstructed proton recoil energy very precisely at approximately 40 keV RMS error over a wide range of proton recoil energies; short scintillator attenuation lengths, like that of EJ-232Q, do not perform well in this imager design.
 
 The RMS errors for different detector/scintillator combinations relative to each other did not change when we halved the proton recoil energy from 2 MeV to 1 MeV. However, scintillation position reconstruction is dependent upon proton recoil energy. Lower proton recoil energies produce less light in the scintillator and poorer photostatistics. This produces larger uncertainties in position reconstruction.

We located a simulated Cf-252 source by back-projecting cones of probable incident neutron direction. When cones overlap, they will reveal the source location. We approximated the back-projected image resolution of the SVSC-PiPS device at $\sigma_{\theta}=19.5^\circ$ and $\sigma_{\phi}=15.7^\circ$ compared to a 10$^\circ$ for the Neutron Scatter Camera developed at Sandia National Labs.

Using light intensity to estimate scintillation position degraded position estimation precision by 250\%  compared to using the intensity vs. time history of charge carriers emitted by the photodetectors at each end of the pillar. The precision will be poorer at all proton recoil energies, especially low energies. Scintillation position estimation using only light intensity ratios produced a 50-100\% poorer detector resolution.

\section{Acknowledgements}
Sandia National Laboratories is a multimission laboratory managed and operated by National Technology and Engineering Solutions of Sandia, LLC., a wholly owned subsidiary of Honeywell International, Inc., for the U.S. Department of Energy’s National Nuclear Security Administration under contract DE-NA0003525.

This work was funded in-part by the Consortium for Verification Technology under Department of Energy National Nuclear Security Administration award number DE-NA0002534.

\section{References}
\setcounter{table}{0}
\setcounter{figure}{0}
\renewcommand{\thetable}{A\arabic{table}}






\bibliography{trimLibrary}

\begin{thebibliography}{10}
\expandafter\ifx\csname url\endcsname\relax
  \def\url#1{\texttt{#1}}\fi
\expandafter\ifx\csname urlprefix\endcsname\relax\def\urlprefix{URL }\fi
\expandafter\ifx\csname href\endcsname\relax
  \def\href#1#2{#2} \def\path#1{#1}\fi

\bibitem{knoll_textbook}
G.~Knoll, {Radiation Detection and Measurement}, John Wiley {\&} Sons, 2010.

\bibitem{Grannan1972}
R.~T. Grannan, et~al., {A Large Area Detector for Neutrons Between 2 and 100
  MeV}, Nuclear Instruments and Methods 3 (1972) 99--108.

\bibitem{Preszler1972}
A.~M. Preszler, G.~M. Simnett, R.~S. White, {Earth albedo neutrons from 10 to
  100 MeV}, Physical Review Letters 28~(15) (1972) 982--985.
\newblock \href {http://dx.doi.org/10.1103/PhysRevLett.28.982}
  {\path{doi:10.1103/PhysRevLett.28.982}}.

\bibitem{Preszler1974}
A.~M. Preszler, G.~M. Simnett, R.~S. White, {Angular distribution and altitude
  dependence of atmospheric neutrons from 10 to 100 MeV}, Journal of
  Geophysical Research 79~(1) (1974) 17--22.
\newblock \href {http://dx.doi.org/10.1029/JA079i001p00017}
  {\path{doi:10.1029/JA079i001p00017}}.

\bibitem{Zych1975}
A.~D. Zych, et~al., {Large Area Double Scattering Telescope for Balloon-Borne
  Studies of Neutrons and Gamma Rays}, IEEE Transactions on Nuclear Science
  (1975) 605--610.

\bibitem{Walker1986}
S.~E. Walker, A.~M. Preszler, W.~A. Millard, {Double scatter neutron
  time‐of‐flight spectrometer as a plasma diagnostic}, Review of Scientific
  Instruments 57~(8) (1986) 1740--1742.
\newblock \href {http://dx.doi.org/10.1063/1.1139167}
  {\path{doi:10.1063/1.1139167}}.

\bibitem{Ryan1992}
J.~M. Ryan, et~al., {COMPTEL as a Solar Gamma Ray and Neutron Detector},
  Springer US, Boston, MA, 1992, pp. 261--270.

\bibitem{Ryan1993}
J.~Ryan, et~al., {Comptel Measurements of Solar Flare Neutrons}, Adv. Space
  Res. 13~(9) (1993) 255--258.

\bibitem{Moser2005}
M.~R. Moser, et~al., {A fast neutron imaging telescope for inner heliosphere
  missions}, Advances in Space Research 36~(8) (2005) 1399--1405.
\newblock \href {http://dx.doi.org/10.1016/j.asr.2005.03.037}
  {\path{doi:10.1016/j.asr.2005.03.037}}.

\bibitem{Bravar2006}
U.~Bravar, P.~J. Bruillard, E.~O. Fluckiger, J.~R. Macri, M.~L. McConnell,
  M.~R. Moser, J.~M. Ryan,
  \href{http://link.aip.org/link/?PSI/6213/62130G/1}{{FNIT: the fast neutron
  imaging telescope for SNM detection}}, Non-Intrusive Inspection Technologies
  6213~(1) (2006) 62130G.
\newblock \href {http://dx.doi.org/10.1117/12.666119}
  {\path{doi:10.1117/12.666119}}.
\newline\urlprefix\url{http://link.aip.org/link/?PSI/6213/62130G/1}

\bibitem{Bravar2006a}
U.~Bravar, et~al., {Design and testing of a position-sensitive plastic
  scintillator detector for fast neutron imaging}, IEEE Transactions on Nuclear
  Science 53~(6) (2006) 3894--3903.
\newblock \href {http://dx.doi.org/10.1109/TNS.2006.886046}
  {\path{doi:10.1109/TNS.2006.886046}}.

\bibitem{Bravar2009}
U.~Bravar, et~al., {Calibration of the fast neutron imaging telescope (FNIT)
  prototype detector}, IEEE Transactions on Nuclear Science 56~(5) (2009)
  2947--2954.
\newblock \href {http://dx.doi.org/10.1109/TNS.2009.2028025}
  {\path{doi:10.1109/TNS.2009.2028025}}.

\bibitem{Woolf2009}
R.~S. Woolf, et~al., {Advanced characterization and simulation of SONNE: a fast
  neutron spectrometer for Solar Probe Plus}, Proceedings of SPIE 7438 (2009)
  74380S--74380S--12.
\newblock \href {http://dx.doi.org/10.1117/12.826425}
  {\path{doi:10.1117/12.826425}}.

\bibitem{Marleau2007}
P.~Marleau, et~al., {Advances in Imaging Fission Neutrons with a Neutron
  Scatter Camera}, in: IEEE Nuclear Science Symposium Conference Record, 2007,
  pp. 170--172.

\bibitem{Mascarenhas2008a}
N.~Mascarenhas, et~al., {Results with the neutron scatter camera}, IEEE Nuclear
  Science Symposium Conference Record 56~(3) (2008) 3368--3371.
\newblock \href {http://dx.doi.org/10.1109/NSSMIC.2008.4775064}
  {\path{doi:10.1109/NSSMIC.2008.4775064}}.

\bibitem{NSC_patent1}
N.~Mascarenhas, et~al., {Method for Improving the Angular Resolution of a
  Neutron Scatter Camera} (2012).
\newblock \href {http://arxiv.org/abs/arXiv:1011.1669v3}
  {\path{arXiv:arXiv:1011.1669v3}}, \href
  {http://dx.doi.org/10.1074/JBC.274.42.30033.(51)}
  {\path{doi:10.1074/JBC.274.42.30033.(51)}}.

\bibitem{Poitrasson-Riviere2014}
A.~Poitrasson-Rivi{\`{e}}re, et~al., {Dual-particle imaging system based on
  simultaneous detection of photon and neutron collision events}, Nuclear
  Instruments and Methods in Physics Research, Section A: Accelerators,
  Spectrometers, Detectors and Associated Equipment 760 (2014) 40--45.
\newblock \href {http://dx.doi.org/10.1016/j.nima.2014.05.056}
  {\path{doi:10.1016/j.nima.2014.05.056}}.

\bibitem{Pelowitz2011}
D.~Pelowitz, {MCNPX Users Manual Version 2.7.0} (2011).

\bibitem{ej204}
{EJ-204 Data Sheet}, Eljen Technology.

\bibitem{Zaitseva2015}
N.~Zaitseva, et~al., {Scintillation properties of solution-grown trans-stilbene
  single crystals}, Nuclear Instruments and Methods in Physics Research,
  Section A: Accelerators, Spectrometers, Detectors and Associated Equipment
  789 (2015) 8--15.
\newblock \href {http://dx.doi.org/10.1016/j.nima.2015.03.090}
  {\path{doi:10.1016/j.nima.2015.03.090}}.

\bibitem{ej232q}
{EJ-232Q Data Sheet}, Eljen Technology.

\bibitem{Kuchnir1968}
F.~T. Kuchnir, F.~J. Lynch, {Time Dependence of Scintillations and the Effect
  on Pulse-Shape Discrimination}, IEEE Transactions on Nuclear Science 15
  (1968) 107--113.
\newblock \href {http://dx.doi.org/10.1109/TNS.1968.4324923}
  {\path{doi:10.1109/TNS.1968.4324923}}.

\bibitem{Schuster2016}
P.~Schuster, E.~Brubaker, {Characterization of the scintillation anisotropy in
  crystalline stilbene scintillator detectors}, Nuclear Instruments and Methods
  in Physics Research, Section A: Accelerators, Spectrometers, Detectors and
  Associated Equipment\href {http://arxiv.org/abs/1606.06715}
  {\path{arXiv:1606.06715}}, \href
  {http://dx.doi.org/10.1016/j.nima.2016.11.016}
  {\path{doi:10.1016/j.nima.2016.11.016}}.

\bibitem{Stilbene1910}
{Scintinel Data Sheet}, InradOptics.

\bibitem{planacon_xp85012}
{Planacon XP-85012 Data Sheet}, Photonis.

\bibitem{sipm_Jseries}
{J-Series Silicon Photomultiplier Data Sheet}, SensL.

\bibitem{SensL2011}
SensL, {Introduction to the SPM (Silicon Photomultiplier)} (2011) 1--8.

\bibitem{Grigoryev2016a}
V.~A. Grigoryev, et~al., {Study of the Planacon XP85012 photomultiplier
  characteristics for its use in a Cherenkov detector}, Journal of Physics:
  Conference Series 675~(4) (2016) 042016.
\newblock \href {http://dx.doi.org/10.1088/1742-6596/675/4/042016}
  {\path{doi:10.1088/1742-6596/675/4/042016}}.

\bibitem{Agostinelli2003a}
S.~Agostinelli, et~al., {GEANT4 - A simulation toolkit}, Nuclear Instruments
  and Methods in Physics Research, Section A: Accelerators, Spectrometers,
  Detectors and Associated Equipment 506 (2003) 250--303.
\newblock \href {http://dx.doi.org/10.1016/S0168-9002(03)01368-8}
  {\path{doi:10.1016/S0168-9002(03)01368-8}}.

\bibitem{Allison2006}
J.~Allison, et~al., {Geant4 developments and applications}, IEEE Transactions
  on Nuclear Science 53~(1) (2006) 270--278.
\newblock \href {http://dx.doi.org/10.1109/TNS.2006.869826}
  {\path{doi:10.1109/TNS.2006.869826}}.

\bibitem{Janecek2008}
M.~Janecek, W.~W. Moses, {Optical reflectance measurements for commonly used
  reflectors}, IEEE Transactions on Nuclear Science 55~(4) (2008) 2432--2437.
\newblock \href {http://dx.doi.org/10.1109/TNS.2008.2001408}
  {\path{doi:10.1109/TNS.2008.2001408}}.

\bibitem{Janecek2012}
M.~Janecek, {Reflectivity spectra for commonly used reflectors}, IEEE
  Transactions on Nuclear Science 59~(3) (2012) 490--497.
\newblock \href {http://dx.doi.org/10.1109/TNS.2012.2183385}
  {\path{doi:10.1109/TNS.2012.2183385}}.

\bibitem{Loignon-houle2016}
F.~Loignon-houle, C.~M. Pepin, S.~A. Charlebois, R.~Lecomte, {Reflectivity
  Quenching of ESR Multilayer Polymer Film Reflector in Optically Bonded
  Scintillator Arrays}, Submitted to Nuclear Instruments and Methods in Physics
  Research Section A 851~(October 2016) (2016) 62--67.
\newblock \href {http://dx.doi.org/10.1016/j.nima.2017.01.051}
  {\path{doi:10.1016/j.nima.2017.01.051}}.

\bibitem{Brubaker2009a}
E.~Brubaker, et~al., {Calibration and Simulation of a Coded Aperture Neutron
  Imaging System}, IEEE Nuclear Science Symposium Conference Record (2009)
  920--924.

\bibitem{Leo1993}
L.~Leo, {Techniques for Nuclear and Particle Physics Experiments}, 1993.

\bibitem{3M2010}
3M, {Vikuiti Enhanced Specular Reflector (ESR) - ESR Sales Literature} (2010).

\bibitem{Naeem2013}
S.~F. Naeem, S.~D. Clarke, S.~A. Pozzi, {Validation of Geant4 and MCNPX-PoliMi
  simulations of fast neutron detection with the EJ-309 liquid scintillator},
  Nuclear Instruments and Methods in Physics Research, Section A: Accelerators,
  Spectrometers, Detectors and Associated Equipment 714 (2013) 98--104.
\newblock \href {http://arxiv.org/abs/arXiv:1508.06575v1}
  {\path{arXiv:arXiv:1508.06575v1}}, \href
  {http://dx.doi.org/10.1016/j.nima.2013.02.017}
  {\path{doi:10.1016/j.nima.2013.02.017}}.

\bibitem{Enqvist2012}
A.~Enqvist, C.~Lawrence, T.~Massey, S.~Pozzi, {Neutron light output functions
  measured for EJ-309 liquid scintillation detectors}, in: Proceedings of the
  INMM 53rd Annual Meeting, 2012.

\bibitem{Nemallapudi2016}
M.~Nemallapudi, S.~Gundacker, P.~Lecoq, E.~Auffray, {Single photon time
  resolution of state of the art SiPMs}, Journal of Instrumentation 11~(10)
  (2016) P10016--P10016.
\newblock \href {http://dx.doi.org/10.1088/1748-0221/11/10/P10016}
  {\path{doi:10.1088/1748-0221/11/10/P10016}}.

\bibitem{Ritt2010}
S.~Ritt, R.~Dinapoli, U.~Hartmann, {Application of the DRS chip for fast
  waveform digitizing}, Nuclear Instruments and Methods in Physics Research,
  Section A: Accelerators, Spectrometers, Detectors and Associated Equipment
  623~(1) (2010) 486--488.
\newblock \href {http://arxiv.org/abs/1011.0226} {\path{arXiv:1011.0226}},
  \href {http://dx.doi.org/10.1016/j.nima.2010.03.045}
  {\path{doi:10.1016/j.nima.2010.03.045}}.

\bibitem{Bitossi2016}
M.~Bitossi, R.~Paoletti, D.~Tescaro, {Ultra-fast sampling and data acquisition
  using the drs4 waveform digitizer}, IEEE Transactions on Nuclear Science
  63~(4) (2016) 2309--2316.
\newblock \href {http://dx.doi.org/10.1109/TNS.2016.2578963}
  {\path{doi:10.1109/TNS.2016.2578963}}.

\bibitem{Oberla2014}
E.~Oberla, et~al., {A 15 GSa/s, 1.5 GHz bandwidth waveform digitizing ASIC},
  Nuclear Instruments and Methods in Physics Research, Section A: Accelerators,
  Spectrometers, Detectors and Associated Equipment 735 (2014) 452--461.
\newblock \href {http://arxiv.org/abs/arXiv:1309.4397v1}
  {\path{arXiv:arXiv:1309.4397v1}}, \href
  {http://dx.doi.org/10.1016/j.nima.2013.09.042}
  {\path{doi:10.1016/j.nima.2013.09.042}}.

\bibitem{Broyden1970}
C.~G. Broyden, {The Convergence of a Class of Double-rank Minization
  Algorithms}, Journal of the Mathematics and its Applications 6~(September)
  (1970) 76--90.
\newblock \href {http://dx.doi.org/10.1093/imamat/6.3.222}
  {\path{doi:10.1093/imamat/6.3.222}}.

\bibitem{Fletcher1970}
R.~Fletcher, {A New Approach to Variable Metric Algorithms}, The Computer
  Journal 13~(3) (1970) 317--322.

\bibitem{Goldfarb1970}
D.~Goldfarb, {A family of variable-metric methods derived by variational
  means}, Mathematics of Computation 24~(109) (1970) 23--23.
\newblock \href {http://dx.doi.org/10.1090/S0025-5718-1970-0258249-6}
  {\path{doi:10.1090/S0025-5718-1970-0258249-6}}.

\bibitem{Shanno1970}
D.~F. Shanno, {Conditioning of quasi-Newton methods for function minimization},
  Mathematics of Computation 24~(111) (1970) 647--647.
\newblock \href {http://arxiv.org/abs/arXiv:1011.1669v3}
  {\path{arXiv:arXiv:1011.1669v3}}, \href
  {http://dx.doi.org/10.1090/S0025-5718-1970-0274029-X}
  {\path{doi:10.1090/S0025-5718-1970-0274029-X}}.

\bibitem{Byrd1995}
R.~Byrd, P.~Lu, J.~Nocedal, C.~Zhu, {A Limited Memory Algorithm for Bound
  Constrained Optimization}, Society for Industrial and Applied Mathematics
  16~(5) (1995) 1190--1208.

\bibitem{Zhu1997}
C.~Zhu, R.~H. Byrd, P.~Lu, J.~Nocedal, {Algorithm 778: L-BFGS-B: Fortran
  subroutines for large-scale bound-constrained optimization}, ACM Transactions
  on Mathematical Software 23~(4) (1997) 550--560.
\newblock \href {http://dx.doi.org/10.1145/279232.279236}
  {\path{doi:10.1145/279232.279236}}.

\bibitem{Brennan2010}
J.~Brennan, et~al., {Applying the neutron scatter camera to treaty verification
  and warhead monitoring}, IEEE Nuclear Science Symposium Conference
  Record~(cm) (2010) 691--694.
\newblock \href {http://dx.doi.org/10.1109/NSSMIC.2010.5873848}
  {\path{doi:10.1109/NSSMIC.2010.5873848}}.

\bibitem{Roemer2009}
K.~Roemer, et~al., {A technique for measuring the energy resolution of low-Z
  scintillators}, 2009 IEEE Nuclear Science Symposium Conference Record
  (NSS/MIC) (2009) 6--11\href {http://dx.doi.org/10.1109/NSSMIC.2009.5401909}
  {\path{doi:10.1109/NSSMIC.2009.5401909}}.

\end{thebibliography}
\bibliographystyle{elsarticle-num} 

\appendix
\section{}
\label{RMS reduction section}
In the simulation, we estimated proton recoil energy using the number of charge carriers produced in the left and right photodetectors. If we assume

\begin{enumerate}
  \item A typical organic scintillator luminosity for a 1 MeV and 2 MeV recoil proton is about 1,500 scintillation photons and 5,000 photons respectively
  \item A light collection efficiency of 35\% for a single photodetector
  \item The quantum efficiency of a MCP-PM photocathode is about 20\%
\end{enumerate}

\noindent
we can calculate the approximate number of photoelectrons produced in the left photocathode using Equation \ref{eqn:createdPE}. For a 1 MeV proton recoil, the left photodetector produces about 100 photoelectrons; this will produce a relative uncertainty of 10\% or 100 keV (235 keV FWHM). We can calculate the total uncertainty ($\sigma_T$) by adding the left and right photodetector uncertainties ($\sigma_L, \sigma_R$) in quadrature.

\begin{align}
\sigma_{T}^2 &= \sigma_L^2 + \sigma_R^2
\end{align}

\noindent
We want to examine the total relative uncertainty so we divide by the total number of charge carriers ($\rho_T$)
\begin{align} \label{eqn:relUncertainty}
\frac{\sigma_{T}^2}{\rho_T^2} &= \frac{\sigma_L^2}{\rho_T^2} + \frac{\sigma_R^2}{\rho_T^2} \\
\end{align}

\noindent
where the total number of charge carriers is the sum of both left and right photodetector charge carriers

\begin{align} \label{eqn:center1}
\rho_T &= \rho_L + \rho_R
\end{align}

\noindent
If we assert that the scintillation occurred at the center of the pillar, we can assume

\begin{align} \label{eqn:center2}
\sigma_R &\approx \sigma_L 
\end{align}

\noindent
Substituting in Equations \ref{eqn:center1} and \ref{eqn:center2} into Equation \ref{eqn:relUncertainty} and simplifying results in Equation \ref{eqn:relUncertainty2}.

\begin{align} \label{eqn:relUncertainty2}
\frac{\sigma_{T}}{\rho_T} = \frac{1}{\sqrt{2}} \frac{\sigma_L}{\rho_L}
\end{align}

Using Equation \ref{eqn:relUncertainty2}, we calculate a total relative uncertainty of 7.1\% or 71 keV (167 keV FWHM) for a 1 MeV proton recoil.

If we scale up to a 2 MeV proton recoil using the same efficiencies, we estimate approximately 340 photoelectrons produced in the left and right photodetectors each with a relative uncertainty of 5.4\%. If we again account for both photodetectors, we estimate a total relative uncertainty of 3.8\% or 76 keV (179 keV FWHM).

Using the previous FWHM values of 167 keV and 179 keV, if we divide by the original proton recoil energy, we calculate a resolution of 16.7\% and 9\% for 1 MeV and 2 MeV proton recoil energies respectively. These resolutions are similar to plastic scintillator resolution measurements of a similar plastic scintillator (EJ-200), published in \cite{Roemer2009}.

So far, using simple photon counting statistics, we have estimated an uncertainty of 70-80 keV for proton recoil energies between 1 and 2 MeV. However, our modeled results in Figures \ref{fig:2MeV_tableFig} and \ref{fig:1MeV_tableFig} indicate uncertainties below 50 keV. This can be understood as a consequence of the non-linearity of the scintillator light output function, as follows. Let $L(E_p)$ be the light output function which converts proton recoil energy ($E_p$) to electron equivalent energy ($LO$) and $L^{-1}(LO)$ be the inverse function that transforms electron equivalent energy to proton recoil energy as shown  
\begin{align}
L(E_P) &= LO \\
L^{-1}(LO) &= E_p
\end{align}

\noindent
If we introduce a small perturbation in the light output, we find
\begin{align}
L^{-1}(LO + \Delta LO) &= E_p + \Delta E_p
\end{align}

\noindent
Setting $\Delta LO = \sigma_{LO}$ and $ \Delta E_p = \sigma_{E_p}$

\begin{align}
L^{-1}(LO + \sigma_{LO}) &= E_p + \sigma_{E_p}
\end{align}

\noindent
If we use a first order expansion of $L^{-1}$, we obtain Equation \ref{eqn:ffa}
\begin{align} \label{eqn:ffa}
L^{-1}(LO + \sigma_{LO}) &\approx L^{-1}(LO) + \sigma_{LO} L'^{-1}(LO)
\end{align}
\noindent
where $L'^{-1}$ is the derivative of the inverse light output function given by
\begin{align}
L'^{-1}(LO) = \frac{\delta L^{-1}}{dLO}
\end{align}

\noindent
Using the first order expansion, we can equate

\begin{align} \label{test}
\frac{\sigma_{E_p}}{E_p} &= \frac{\sigma_{LO}L'^{-1}(LO)}{L^{-1}(LO)}
\end{align}

\begin{figure}[h!!]
\centering
  \includegraphics[width=\columnwidth]{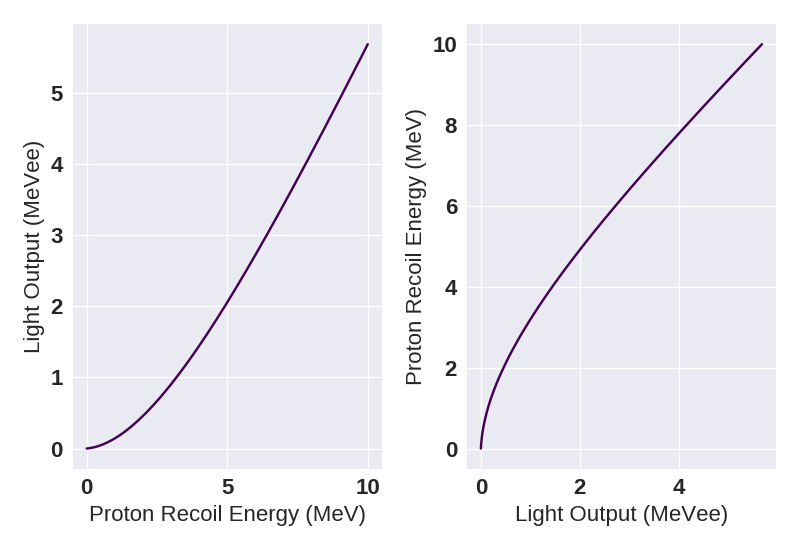}
  \caption{Left: Light output function, $L(E_p)$, for EJ-309 as described in \cite{Naeem2013} and \cite{Enqvist2012}. Right: Inverse EJ-309 light output function, $L^{-1}(LO)$}
  \label{fig:LOandInverse}
\end{figure}

In the simulation, we estimate the amount of light produced in 10,000 simulated scintillation events and calculate the corresponding proton recoil energy using $L^{-1}(LO)$; the distribution of light outputs is transformed into a distribution of proton recoil energies using $L^{-1}$. When performing the transformation, the relative uncertainty of the proton recoil energy will be reduced if 
\begin{align} \label{eqn:ineq}
\frac{\sigma_{E_p}}{E_p} &< \frac{\sigma_{LO}}{LO}
\end{align}


\noindent
If we substitute Equation \ref{test} into Equation \ref{eqn:ineq} and simplify, we see
\begin{align} \label{eqn:result}
L'^{-1}(LO) &< \frac{L^{-1}(LO)}{LO}
\end{align}

\noindent
This shows that we will observe a reduction in the relative uncertainty in the proton recoil energy if the slope of the inverted light output function is less than that of the proton recoil energy divided by the light produced at that proton recoil energy.

As an example, if the light output function is linear with respect to the proton recoil energy (similar to light production from gamma interactions) then

\begin{align}
L(E_p) &= a E_p \\
L^{-1}(LO) &= LO/a \\
L'^{-1}(LO) &= 1/a  
\end{align}

\noindent
If we substitute these results into Equation \ref{eqn:result}, we get
\begin{align}
\frac{1}{a} &< \frac{LO/a}{LO} 
\end{align}

\noindent
where both sides simplify to unity and the inequality does not hold true, therefore there is no reduction in relative uncertainty.

We used numerical methods to estimate the inequality for Equation \ref{eqn:result} since the light output function we used (shown in Figure \ref{fig:LOandInverse}) cannot be analytically inverted. The inequality holds true for all proton recoil energies. The results show a reduction of uncertainty by a factor of $\approx$1.65 for proton recoil energies from 1-2 MeV with diminishing returns as proton recoil energy increases.   

\section{}
\begin{table*}[h!!]
\centering
\caption{10 cm length pillars simulation results, 1 MeV proton recoil energy}
\label{table:10cm_1MeV}
\begin{tabular}{ccccc}
Width (cm)   & Scintillator  & Photodetector   & Position RMS Error (cm) & Energy RMS Error (MeV) \\
\hline
1.0 & EJ-204 & MCP-PM & 0.745 & 0.038 \\ 
1.0 & EJ-232Q & SiPM & 0.764 & 0.097 \\ 
0.5 & EJ-232Q & SiPM & 0.795 & 0.095 \\ 
0.5 & EJ-204 & MCP-PM & 0.796 & 0.039 \\ 
1.0 & Stilbene & MCP-PM & 0.809 & 0.039 \\ 
0.5 & Stilbene & MCP-PM & 0.825 & 0.040 \\ 
1.0 & EJ-232Q & MCP-PM & 0.836 & 0.123 \\ 
0.5 & EJ-232Q & MCP-PM & 0.870 & 0.125 \\ 
1.0 & EJ-204 & SiPM & 1.016 & 0.029 \\ 
1.0 & Stilbene & SiPM & 1.023 & 0.030 \\ 
0.5 & EJ-204 & SiPM & 1.105 & 0.030 \\ 
0.5 & Stilbene & SiPM & 1.111 & 0.030 \\ 
\end{tabular}
\end{table*}

\begin{table*}[h!!]
\centering
\caption{20 cm length pillars simulation results, 1 MeV proton recoil energy}
\label{table:20cm_1MeV}
\begin{tabular}{ccccc}
Width (cm)   & Scintillator  & Photodetector  & Position RMS Error (cm) & Energy RMS Error (MeV) \\
\hline
1.0 & EJ-204 & MCP-PM & 0.922 & 0.039 \\ 
0.5 & EJ-204 & MCP-PM & 0.970 & 0.040 \\ 
1.0 & Stilbene & MCP-PM & 1.039 & 0.040 \\ 
0.5 & Stilbene & MCP-PM & 1.070 & 0.042 \\ 
1.0 & EJ-204 & SiPM & 1.247 & 0.030 \\ 
1.0 & Stilbene & SiPM & 1.300 & 0.032 \\ 
0.5 & EJ-204 & SiPM & 1.339 & 0.031 \\ 
0.5 & Stilbene & SiPM & 1.385 & 0.032 \\ 
0.5 & EJ-232Q & MCP-PM & 1.527 & 0.187 \\ 
0.5 & EJ-232Q & SiPM & 1.538 & 0.162 \\ 
1.0 & EJ-232Q & SiPM & 1.566 & 0.175 \\ 
1.0 & EJ-232Q & MCP-PM & 1.569 & 0.198 \\ 
\end{tabular}
\end{table*}

\begin{table*}[h!!]
\centering
\caption{10 cm length pillars simulation results, 2 MeV proton recoil energy}
\label{table:10cm_2MeV}
\begin{tabular}{ccccc}
Width (cm)   & Scintillator  & Photodetector  & Position RMS Error (cm) & Energy RMS Error (MeV) \\
\hline
1.0 & EJ-204 & MCP-PM & 0.425 & 0.041 \\ 
1.0 & EJ-232Q & SiPM & 0.426 & 0.107 \\ 
0.5 & EJ-232Q & SiPM & 0.452 & 0.105 \\ 
0.5 & EJ-204 & MCP-PM & 0.453 & 0.043 \\ 
1.0 & Stilbene & MCP-PM & 0.459 & 0.042 \\ 
1.0 & EJ-232Q & MCP-PM & 0.463 & 0.134 \\ 
0.5 & Stilbene & MCP-PM & 0.474 & 0.043 \\ 
0.5 & EJ-232Q & MCP-PM & 0.481 & 0.134 \\ 
1.0 & EJ-204 & SiPM & 0.578 & 0.032 \\ 
1.0 & Stilbene & SiPM & 0.588 & 0.033 \\ 
0.5 & EJ-204 & SiPM & 0.627 & 0.033 \\ 
0.5 & Stilbene & SiPM & 0.631 & 0.034 \\ 
\end{tabular}
\end{table*}

\
\begin{table*}[h!!]
\centering
\caption{20 cm length pillars simulation results, 2 MeV proton recoil energy}
\label{table:20cm_2MeV}
\begin{tabular}{ccccc}
Width (cm)   & Scintillator  & Photodetector  & Position RMS Error (cm) & Energy RMS Error (MeV) \\
\hline
1.0 & EJ-204 & MCP-PM & 0.515 & 0.043 \\ 
0.5 & EJ-204 & MCP-PM & 0.543 & 0.044 \\ 
1.0 & Stilbene & MCP-PM & 0.591 & 0.045 \\ 
0.5 & Stilbene & MCP-PM & 0.591 & 0.047 \\ 
1.0 & EJ-204 & SiPM & 0.704 & 0.033 \\ 
1.0 & Stilbene & SiPM & 0.736 & 0.035 \\ 
0.5 & EJ-204 & SiPM & 0.768 & 0.034 \\ 
0.5 & Stilbene & SiPM & 0.781 & 0.036 \\ 
1.0 & EJ-232Q & MCP-PM & 0.783 & 0.208 \\ 
0.5 & EJ-232Q & MCP-PM & 0.795 & 0.199 \\ 
1.0 & EJ-232Q & SiPM & 0.824 & 0.188 \\ 
0.5 & EJ-232Q & SiPM & 0.833 & 0.178 \\ 
\end{tabular}
\end{table*}

These tables list the scintillation position and proton recoil energy reconstruction RMS errors for all combinations of scintillator, photodetector and pillar geometry using the method described in this paper.


\end{document}